\RequirePackage{fix-cm}
\documentclass[smallextended,natbib]{svjour3}       
\smartqed  
\usepackage{graphicx}
\usepackage[bookmarks=false]{hyperref}
\usepackage{mathptmx}      
\usepackage{color}
\usepackage{datetime}

\def\HI{H{\,\small I}}

\newcommand{\kms}{$\,$km$\,$s$^{-1}$}
\newcommand{\ergs}{$\,$erg$\,$s$^{-1}$}

\newcommand{\mJybeam}{mJy beam$^{-1}$}
\newcommand{\msun}{{${M}_\odot$}}
\newcommand{\msunyr}{{${ M}_\odot$ yr$^{-1}$}}

\newcommand{\tspin}{$T_{\rm s}$}
\newcommand{\tc}{$T_{\rm c}$}
\newcommand{\tb}{$T_{\rm b}$}
\newcommand{\cf}{$c_{\rm f}$}

\newcommand{\cmtwo}{cm$^{-2}$}
\newcommand{\nhi}{$N_{\rm HI}$}

\def\HI{H{\,\small I}}
\def\NaI{Na{\,\small I}}
\def\NaII{Na{\,\small II}}

\def\titHI{H{\,\normalsize I}}

\def\OIII{[O{\,\small III}]}

\def\emph#1{{\sl #1}}

\newcommand{\atlas}{{ATLAS$^{\rm 3D}$}} 

\begin{document}

\title{The interstellar and circumnuclear medium of active nuclei traced by \titHI\ 21-cm absorption
}

\titlerunning{H{\scriptsize\,I}\ 21-cm absorption in AGN}       

\author{Raffaella Morganti        
 \and
       Tom Oosterloo
}

\institute{R. Morganti, T.A. Oosterloo  \at
              ASTRON, the Netherlands Institute for Radio Astronomy,  Postbus 2, 7990 AA, Dwingeloo, \\The Netherlands \\
              and \\
              Kapteyn Astronomical Institute, University of Groningen,  Postbus 800,
9700 AV Groningen, \\The Netherlands \\
              Tel.: +31(0)521-595100\\
              \email{morganti@astron.nl, oosterloo@astron.nl}           
}


\date{Version  \today}
\maketitle

\tableofcontents
\newpage

\begin{abstract}
This review summarises what we have learnt in the last two decades based on  \HI\ 21-cm absorption observations about the cold interstellar medium (ISM) in the central regions of active galaxies and about the interplay between this gas and the active nucleus (AGN).  \HI\ absorption is a powerful tracer on all  scales, from the parsec-scales close to the central black hole to structures of many tens of kpc tracing interactions and mergers of galaxies.   Given the strong radio continuum emission  often associated with the central activity, \HI\ absorption observations can  be used to study the \HI\  near an active nucleus  out to much higher redshifts than is possible by using \HI\ emission. 
In this way,  \HI\  absorption has been  used to characterise in detail the general ISM in active galaxies, to trace the fuelling of radio-loud AGN, to study the  feedback occurring between the energy released by the active nucleus and the ISM, and the impact of such interactions on the evolution of  galaxies and of their AGN.  
In the last two decades, significant progress has been made in all these areas. It is now well established that many radio loud AGN are surrounded by small, regularly rotating gas disks that contain a significant fraction of \HI.
The structure of these disks has been traced down to parsec scales by VLBI observations. Some groups of objects, and in particular young and recently restarted radio galaxies, appear to have a particularly high detection rate of \HI. This is interesting in connection with  the evolution of these AGN and their impact on the surrounding ISM.
This is further confirmed by an important  discovery, made thanks to technical upgrades of  radio telescopes, namely the presence of fast, AGN-driven outflows of cold gas which give a direct view of the impact of the energy released by AGN on the evolution of galaxies (AGN feedback). In addition, evidence has been collected that clouds of cold gas can play a role in fuelling the nuclear activity.
This review ends by briefly describing the upcoming large, blind \HI\  absorption surveys planned for the new radio telescopes which will soon become operational. These  surveys will allow to significantly expand existing work, but will also allow to explore new topics, in particular the evolution of the cold ISM in AGN.

\end{abstract}

\section{Introduction}
\label{intro}


As  is well known, hydrogen is   the most common element in the Universe and it is observed in structures which span from the largest cosmological scales to the small, pc-scale of the centres of galaxies. Depending on the physical conditions, hydrogen  occurs in all possible phases, ranging from atomic and molecular gas in the colder   interstellar medium (ISM; with temperatures up to a few thousand kelvin),  warm ionised gas (with temperatures of the order of $10^4$ K) in the warmer ISM and in intragalactic space,  to very hot ionised gas (with temperatures of $10^6$ K or even higher) in the circum-galactic medium around galaxies, and in groups and clusters of galaxies \citep[e.g.,][]{Wolfire10}. Obviously, by using different types of observations of the different phases of hydrogen,  a very  large variety of astronomical phenomena can be studied. 

In this context, since the prediction of the hyperfine transition of atomic hydrogen  (\HI)  at a rest-frame frequency of 1.4204 GHz, or wavelength $\lambda = 21.1$ cm, by Henk van de Hulst in 1944 and the successive first detections of the \HI\ line in 1951 by Ewen and Purcell in the USA \citep{Ewen51}, Muller and Oort in the Netherlands \citep{Muller51}, and  Christiansen and Hindman in Australia \citep{Christiansen52}, studies based on radio observations of atomic hydrogen have indeed made very significant contributions to the understanding of a very wide range of astronomical problems.

From the start, studies using the \HI\  line in  {\sl emission} have played a key role in the understanding of the  kinematics, structure and evolution  of our own Milky Way \citep{Oort58}, as well as of other galaxies in general. In particular, such work has played a central role in uncovering the properties of dark matter  in galaxies, and in showing the importance for the evolution of galaxies of  gas accretion/circulation as well  of interactions between galaxies with their environment (see \citealt{Sanders14} and \citealt{Sancisi08} for reviews of these topics and \citealt{Giovanelli2015} for a review on recent \HI\ emission surveys).  The relevance of \HI\ studies for understanding the formation and  evolution of galaxies, and even for studying the evolution of the Universe as a whole (e.g., the Epoch of Reionisation), is still very high and is one of the main scientific drivers for the construction of the Square  Kilometre Array (SKA) and its pathfinders and precursors \citep{Bourke14,Staveley15}.

However,  observations of the \HI\ line in {\sl absorption} due to gas in front of a bright radio continuum source have also proven to be very powerful because they allow to investigate processes and conditions in the ISM on very different spatial scales and in objects at much higher redshift compared to what is possible with emission line studies.

There are two  settings for \HI\ absorption line studies. One  is {\sl intervening absorption} line work where one  observes gas in a foreground Galactic or extragalactic object along the  line-of-sight to a distant, {\sl unrelated} continuum source. Such observations are used, for example, to study the detailed properties of the ISM in our Galaxy as well as in  distant galaxies, well beyond the reach of emission line observations \citep{Kanekar04}.

The other setting  is to observe \HI\  where the absorbing gas is located in the same extragalactic object  as where the background radio continuum is produced (usually an AGN)  and where  there is an active relation between the gas and the continuum source. This is known as {\sl associated absorption} work and  is the topic of this review. The ISM of a galaxy with an AGN is complex with a large variety of physical conditions tracing a similarly large variety of processes.  For all these situations, \HI\  is a powerful tracer on all  scales, from the pc-scales close to the central black hole to scales of many tens of kpc tracing interactions and mergers of galaxies. 

Because the absorbing gas and the bright continuum source are located in the same galaxy, associated absorption studies are generally focussing on AGN and their  interplay with the ISM of the same object.  One of the main topics of study is   the general structure and kinematics of the ISM of galaxies harbouring an AGN and how this varies among the different types of AGN. In the literature, AGN are often associated with galaxy interactions and mergers, but the real connection, if present, is still unclear. The statistics of the shapes and widths of \HI\ absorption lines gives information on whether the ISM is likely to be in a regularly rotating structure or whether there are indications for more irregular gas distributions and thus indirectly on the role of galaxy interactions. Moreover, such studies also give information on to what extent different kinds of large- and small-scale gas structures are found in different types of AGN which, in turn, tells something about the evolution of the different types of objects. 
The main results from recent work on these issues are discussed in  \S \ref{sec:kinematics}.

Another main topic, and which has seen major developments in recent years, is the detection of fast AGN-driven outflows of cold gas, detected as broad, blue-shifted wings in the \HI\ absorption profiles.  Such outflows   are a manifestation of  AGN feedback which is an important ingredient in current models for galaxy evolution and of the evolution of their central super massive black hole.  \HI\ observations of such outflows provide very useful diagnostics on how the energy released by an AGN interacts with the ISM which can be used to  help to understand the physics of these interactions. These \HI\ outflows are discussed in \S \ref{sec:kinematicallydisturbed}. 

The third main topic we discuss is what \HI\ absorption observations tell us about   the gaseous circumnuclear regions around the super massive black hole (SMBH). Important questions are to what extent such gas  can represent the fuel reservoir   for the SMBH  and whether it can make the difference between an SMBH being active or quiescent. It has often been postulated that  \HI\ profiles that are redshifted with respect to the systemic velocity could be signs of this feeding process, but from an observational point of view, the situation is still not clear (see \S \ref{sec:fuelling}).

Before summarising the main results obtained from observations of associated \HI\ absorption, we highlight the observational differences  between  \HI\  emission and absorption studies,  we discuss how the intrinsic properties of the absorbing \HI\ column can be derived from observations, and discuss some other technical aspects.

\section{Technical aspects}

\subsection{Observing \HI\ in emission and absorption: some  differences}

Although for emission and for absorption studies one uses exactly the same type of observations (and indeed for nearby galaxies emission and absorption can be detected in the same object, see e.g. \ref{fig:cena}, compared to emission line work, \HI\ absorption studies have both advantages and disadvantages. 
\HI\ emission line observations are column density limited and because  column density sensitivity scales as $\sigma/\theta^{2}$ (with $\sigma$ the noise level of the data and $\theta$ the spatial resolution), this means that, given a  noise level,  extended \HI\ gas can only be detected with spatial resolutions larger than a certain limit. With the sensitivity of present-day radio interferometers, in practise this means that \HI\ observations can trace gas structures with column densities down to a few times $10^{18}$ \cmtwo\ at spatial resolutions of about 1 arcminute (corresponding to 60 kpc for an object at $z=0.05$) to  just below $10^{21}$ \cmtwo\ at a few arcseconds resolution (corresponding to a few kiloparsec at the same redshift). For a given spatial resolution, sources beyond a given distance will not be spatially resolved anymore and because of  beam dilution, the effective column density detection limit will increase with increasing distance. Given that the typical column densities of \HI\ are mostly in the range $10^{19}$ -- $10^{21}$ \cmtwo, this means that, unless very long observing times are used,  \HI\ emission line studies with current telescopes are in general limited to the relatively local Universe  \citep[$z < 0.1$; for exceptions see, e.g.][]{Verheijen07,Fernandez16}.

\begin{figure}
\centering
  \includegraphics[angle=0,width=0.7\textwidth]{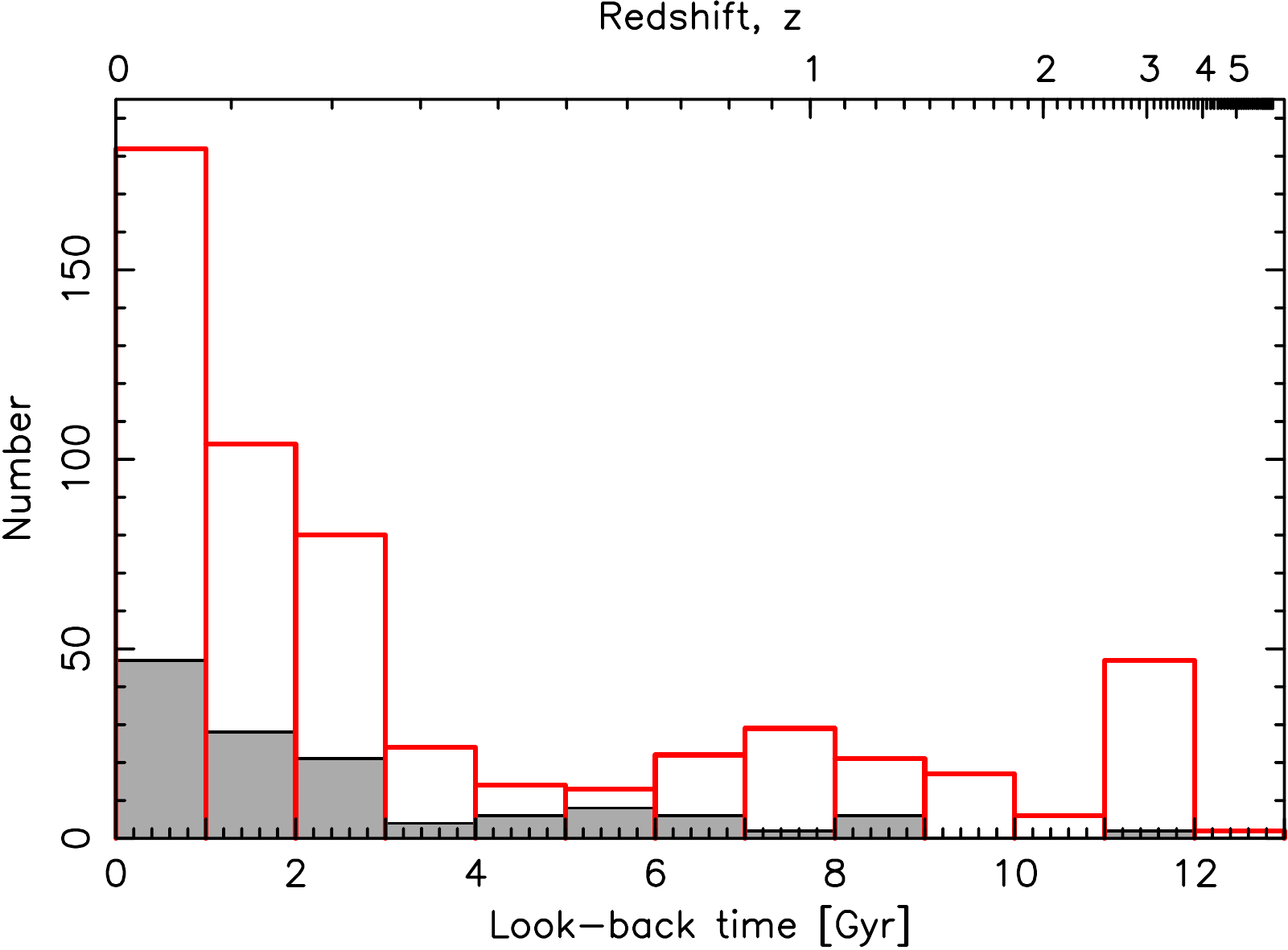}
\caption{ 
The distribution of the current state of detections of associated \HI\ absorption (shaded histogram) and non-detections (unshaded) as function of look-back time for a standard $\Lambda$CDM cosmology with $H_\circ$ = 71 \kms\ Mpc$^{-1}$, $\Omega_{\rm matter} =  0.27$ and $\Omega_{\Lambda} = 0.73$ (taken from \cite{Curran18}).
}
\label{fig:curranHist}       
\end{figure}

This is  different for \HI\ absorption observations. Because the depth in Jy of an \HI\ absorption line   depends on the strength of the background continuum source, the detectability of an \HI\ absorption line is,  everything else being equal, independent of redshift. As long as  bright continuum sources exist at high redshift, \HI\ absorption can in principle be detected as easily at high redshift as in the local Universe.  Therefore, \HI\ absorption observations are much better suited to address the evolution of the \HI\ properties of galaxies and associated issues than emission line studies.  
 However, in practice,  relatively few detections have been made  at higher redshift so far, as is shown in Fig.\ \ref{fig:curranHist} which gives the distribution of detections and non-detections as function of lookback time \citep[taken from the compilation given by][]{Curran18}. This may be due to different source properties at higher redshift (either intrinsic or due to selection effects), but technical limitations of current radio telescopes (frequency coverage, sensitivity) also play a role. The fact that large ranges in redshift are not accessible due to RFI is another reason why most sources detected are at lower redshift. The highest redshift detection of \HI\ absorption currently is $z = 3.4$ (towards TXS~0902+343, corresponding to a look-back time of 11.6 Gyr; \citealt{Uson91}). The GMRT is currently the main instrument for high-redshift absorption line work,  but with the improved low-frequency capabilities of the radio telescopes which are  becoming available,   (MeerKat, ASKAP, LOFAR, MWA and SKA1), the observational possibilities will much improve (see \S \ref{sec:future}).

A related advantage is that \HI\  absorption can be detected and studied at very high spatial resolutions.  The column density detection limit for absorption line work is inversely proportional to the surface brightness of the background continuum. Therefore, as long as the background source remains bright enough, \HI\ absorption can be imaged at very high spatial resolution, including the milli-arcsecond scales reached by  Very Long Baseline Interferometry (VLBI).   Such spatial resolutions are not accessible for \HI\ emission, and will not be so even with the upcoming facilities.   The spatial scales that can be observed with \HI\ absorption observations can be as small as a few tens of parsecs, even in high-redshift sources. Such resolutions match very well the scales of many processes occurring in and around AGN and  \HI\ absorption observations are therefore a particularly good way to probe them. As we will see later in this review,  at the much higher spatial resolutions of VLBI, \HI\  absorption observation  trace gas with similar column densities as are usually detected  in emission at much lower resolution.  
 
As every observing technique, also \HI\ absorption is affected by a number of limitations. 
Perhaps the main limitation is that one, and sometimes two, important parameters are ill constrained, namely the excitation temperature of the 21-cm transition (commonly referred to as the spin temperature  \tspin) and, in the case the continuum source is unresolved, the fraction of the background continuum that is covered by absorbing gas (the covering factor $c_{\rm f}$). This implies uncertainties in the derived \HI\ column densities and masses. Given the importance of this issue, we discuss this in more detail in \S \ref{sec:process} and \S \ref{sec:tspin}.

Moreover, intrinsic to the technique is that for the gas to be observed in absorption, it has to be located in front of the  radio continuum source and gas behind it will not be seen. This is obviously a limitation because we can only obtain information on the gas distribution and kinematics over the extent covered by the  continuum emission (which can be quite limited). Often this is a much smaller area than the full extent of the gas structures.  In addition, if \HI\ absorption is detected,  the  absorbing medium can be a combination of  a number of different gas clouds along the line of sight being located at different distances from the active nucleus, and one needs additional arguments, such as the width of the observed profile combined with the morphology of the background continuum,   to disentangle the structure and locations of the absorbing clouds.  A positive aspect of this is that  absorption allows to unambiguously disentangle the kinematics of the gas in terms of whether the gas is infalling or outflowing, something often not possible with emission line observations. As we will illustrate below, this property has led to a number of very relevant results.


\subsection{Basic theory of \HI\ absorption}
\label{sec:process}
The study of atomic hydrogen at radio frequencies is possible thanks to the fact that a hyperfine transition exists for  hydrogen atoms in the ground state ($1^2S_{\frac{1}{2}}$). Depending on the relative spins of the proton and the electron, there is a very small  difference in the energy of the ground state, with the state with parallel spins having a slightly higher energy than the anti-parallel state. The energy difference between these two states is very small (5.87 $\mu$eV) so a photon associated with this transition has a frequency in the radio domain 1420.405752 MHz (corresponding to 21.106 cm). The prediction that this should be an observable transition was made by van de Hulst in 1944 when he was investigating whether there were observable spectral lines in the radio domain, and the detectability of the line was soon after confirmed in 1951 by observations done by \cite{Ewen51}, by \cite{Muller51}, and  by \cite{Christiansen52}.
The theory of the 21-cm line can be found in any textbook on the physics of interstellar medium  \citep[e.g.,][]{Spitzer78,Field59,Verschuur88,Wilson13} and here we only summarise the essentials.

The relative populations of the two energy levels of the transition can be written as  $n_{\rm u}/n_{\rm l} = g_{\rm u}/g_{\rm l} \exp(-h\nu/kT_{\rm s})$  where $n_{\rm u}$ and  $n_{\rm l}$ are the number of atoms in the upper and lower level respectively and  $g_{\rm u}$ and  $g_{\rm l}$ the statistical weights of the two levels. \tspin\ is  the {\sl spin temperature}, the commonly used term for the excitation temperature of the transition. The statistical weights are $g_{\rm u}=3$ and $g_{\rm l}=1$ and because in general $h\nu \ll T_{\rm s}$, the relative populations are  $n_{\rm u}/n_{\rm l}\simeq3$.

Several processes, both radiative and collisional, play a role in determining the  \tspin\  of an interstellar cloud, depending on density, temperature and radiation field. We refer  the reader to \cite{Field59} for the classic discussion of  the details of this, but we will mention a few relevant aspects in \S \ref{sec:tspin}.  

From standard theory of radiative transfer one can easily derive that the observed spectrum \tb$(v)$\ of an \HI\ cloud with optical depth $\tau(v)$ and spin temperature \tspin, and which is partially covering (with covering factor \cf) a background continuum source with brightness temperature \tc, is the combination of absorbed emission from the continuum source, the emission from the part of the continuum source not covered by the cloud, and self-absorbed 21-cm line emission from the \HI\ cloud itself:
\[
T_{\rm b}(v) = c_{\rm f}T_{\rm c} e^{-\tau(v)} +(1-c_{\rm f})T_{\rm c} + T_{\rm s}(1-e^{-\tau(v)}).
\]

The brightness \tc\ can be found from the line-free parts of the spectrum so that one can obtain
\begin{equation}
\Delta T(v) = T_{\rm b}(v) - T_{\rm c} =  (T_{\rm s}-c_{\rm f}T_{\rm c})(1-e^{-\tau(v)}).
\end{equation}
Equation 1 shows that the spectral line will be seen either in emission or absorption depending on whether \tspin\ $>$ \cf\tc\ or not. In emission line studies, it is usually assumed that the optical depth is low so that for the case of pure line emission (\tc\ = 0)  
\[T_{\rm b}(v) = \tau(v) T_{\rm s}.
\]
The observed profile $T_{\rm b}(\nu)$ as function of frequency $\nu$ can be converted to a spectrum $T_{\rm b}(V)$ as function of velocity $V$ using $V=c \cdot(\nu_\circ-\nu)/\nu $ where $\nu_\circ$ is the rest frequency of the 21-cm transition. From atomic physics we know that $\tau \propto N_{\rm HI}/T_{\rm s}$, so we can  obtain the \HI\ column density from an emission line by integrating the observed emission  over velocity $V$:
\begin{equation}
N_{\rm HI} = 1.82\times 10^{18} \int T_{\rm b}(V)[\mathrm{K}]\ \ dV[\mathrm{km\ s}^{-1}],
\end{equation}
which is independent of \tspin.

If \tspin\ $<$ \cf\tc,   the absorption dominates over the emission of the 21-cm line and an absorption line will be detected. Except for lower resolution absorption observations of  objects in the relatively nearby Universe, for most \HI\ absorption observations, and certainly those done with VLBI, confusion of the absorption spectrum with emission is not an issue. In other words: \tspin\ $\ll$ \cf\tc\ and the observed absorption spectrum is  $\Delta T(V) = -c_{\rm f}T_{\rm c}(1-e^{-\tau(V)})$ and the optical depth of the \HI\ cloud can  be derived from the spectrum using
\begin{equation}
\tau(V) = -\ln\big(1+\frac{\Delta T(V)}{c_{\rm f}T_c}\big) 
\end{equation}
or, if the optical depth is low ($\tau \ll 1$ i.e.\ $|\Delta T(V)|\ll c_{\rm f}T_c$)
\begin{equation}
\tau(V) = \frac{|\Delta T(V)|}{c_{\rm f}T_c}. 
\end{equation}
 The \HI\ column density can then be found from
\begin{equation}
N_{\rm HI} = 1.82\times 10^{18}\  T_{\rm s}[\mathrm{K}]\int \tau(V)\ \ dV[\mathrm{\mathrm{km\ s}^{-1}}]
\end{equation}
which, for faint lines ($\tau \ll 1$) reduces to
\begin{equation}
N_{\rm HI} = 1.82\times 10^{18} \  \frac{T_{\rm s}}{c_{\rm f}T_c} \int {|\Delta T(V}|\ dV
\end{equation}
where all temperatures are in K and $v$ is in km s$^{-1}$.

The above equation shows  an important difference with emission line work namely that  the  column density derived from an absorption spectrum {\sl does} depend on the spin temperature \tspin. The reason for the difference is that in the case of optically thin 21-cm emission line, the emission is due to spontaneous transitions. Because $n_{\rm u}/n_{\rm l}\simeq3$, basically independent of \tspin,  one can directly translate the number of detected photons into number of atoms present. The strength of an absorption line, however, depends on the detailed balance between absorbing and emitting photons along the line of sight and this does depend on \tspin.

In addition, the derived column densities also depend on the covering factor \cf\ which often is unknown. This dependence of $N_{\rm HI}$ on \tspin\ and on \cf\  is a significant complication for many absorption line studies.
We will briefly discuss this in the following section.

\subsection{Spin temperature and covering factor} 
\label{sec:tspin}

The discussion of the previous section shows that there are two main parameters that introduce uncertainties in the estimate of 
 the \HI\ column density derived from absorption observations: the spin temperature \tspin\ and the covering factor   $c_f$.  
 
The spin temperature \tspin\ is often the one more difficult  to constrain. A number of factors control the spin temperature of atomic hydrogen  \citep[see, e.g.,][]{Field59}: absorption of photons from the radiation field, emission stimulated by such  photons, collisions of  H atoms with other particles,  and pumping by Ly$\alpha$ photons. 
 In the normal ISM of a galaxy, collisions dominate the excitation of the line. Radiative effects are particularly important in the vicinity of an AGN and can give rise to high spin temperatures, depending on gas density.  Under such conditions the spin temperature can be as large as a few thousand K, up to  8000 K  \citep{Bahcall69,Maloney96} and there is observational evidence that this indeed occurs.

The  \HI\ in the normal ISM of a normal galaxy is formed by a mix of phases, each with different densities and temperatures. Such ISM 
is typically  divided in a colder, denser phase (the Cold Neutral Medium; CNM) and a warmer, less dense phase (the Warm Neutral Medium; WNM). The typical temperatures of the CNM are around 100 K while those of the WNM are in the range 1000 -- 7000 K. In the normal ISM of galaxies, the spin temperature approaches the kinetic temperature  because the transition is collisionally excited. The existence of two phases of \HI\ in the typical ISM means that there are also two regimes for \tspin.

However, since $\tau \propto T_{\rm s}^{-1}$, \HI\ absorption observations are more sensitive to the colder components of the  gas and the effective \tspin\ of an absorption line due to a number of intervening clouds  with a different values for \tspin\ is the harmonic mean of the \tspin\  of all intervening clouds. In many \HI\ absorption studies a canonical \tspin\ = 100  K is assumed, but there are many situations where a higher \tspin\ is more appropriate. For example, for \HI\ clouds in the the Milky Way, a relation between \tspin\ and \nhi\  has been observed in the sense that a threshold exists (around \nhi\ $\simeq 2\times 10^{20}$ \cmtwo) below which the \tspin\ of absorbing clouds is high (800 -- 7000 K) while for clouds with \nhi\ above this threshold, \tspin\ is only a few hundred K \citep{Kanekar11}. This threshold is thought to arise because inefficient self-shielding against ultraviolet photons at lower column densities prevents the formation of the CNM.

Low spin temperatures have been found in some radio sources. One case is Centaurus A where \HI\ absorption is detected against the Northern and beginning of the Southern lobe, i.e.\ absorption on scales of a few hundred parsec to one kiloparsec and not in the vicinity of the central AGN. By comparing the strength of the absorption with that of emission from neighbouring positions, a spin temperature of \tspin\ $\sim$100 K was found \citep{Struve10b}.  However, this absorption occurs quite far  from the central AGN and may not be not representative for absorption by clouds closer to the AGN.

An example of higher \tspin\ is the young radio source  PKS~1549--79 \citep{Holt06} for which an  independent estimate of the neutral column towards  this source was made based on the optical  extinction derived from   modelling  the optical continuum of the quasar. By comparing this estimate with the \HI\ absorption data, it was found that the spin temperature  must be in the range $3000 <$ \tspin\ $< 6000$ K. However, given its properties, this object could be an extreme example. Other methods for putting  limits to the \tspin, e.g.\ using X-ray observations, have given less extreme values, see \S \ref{sec:OtherPhasesGas}.

\begin{figure}
\centering
  \includegraphics[angle=0,width=0.45\textwidth]{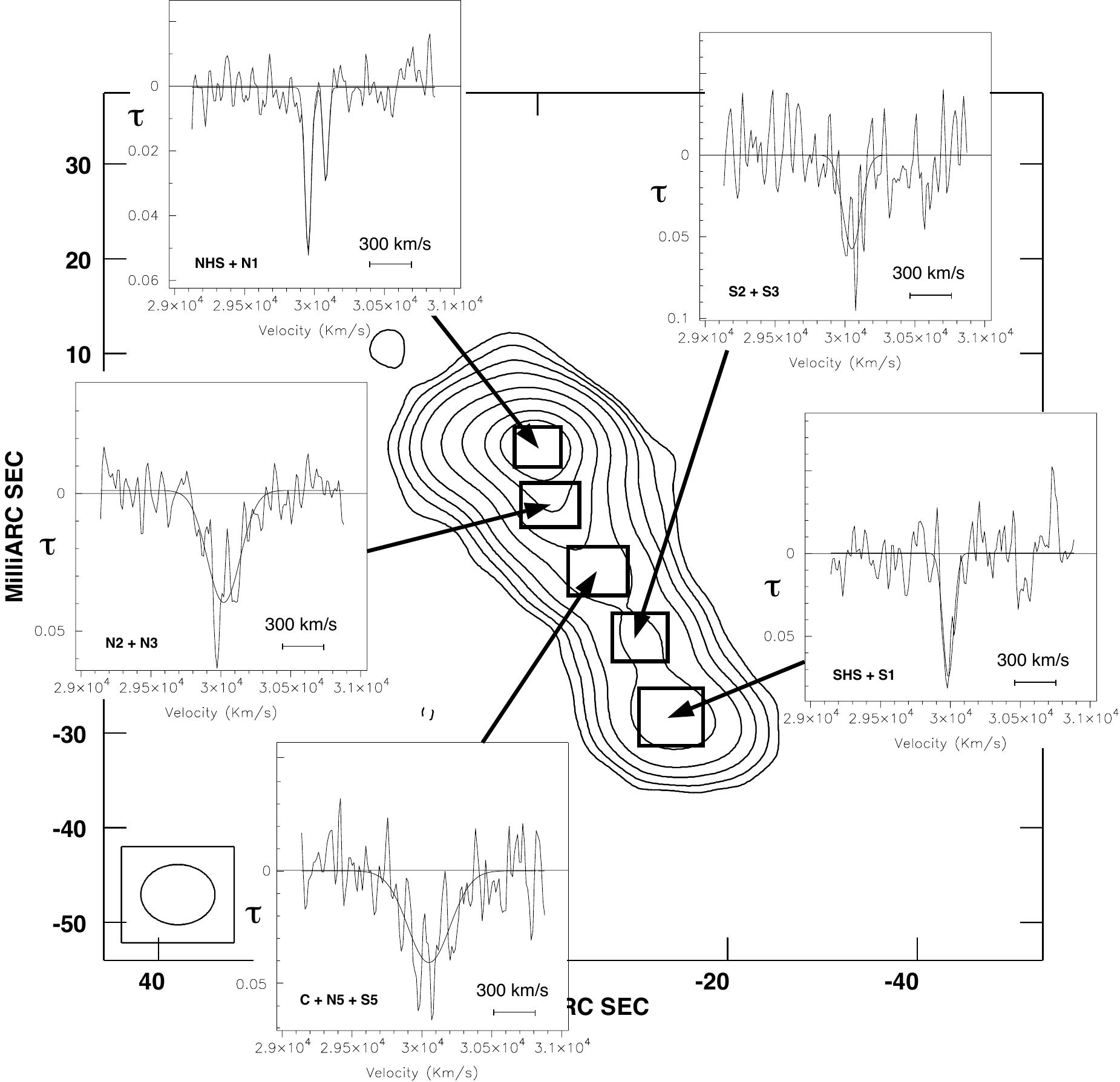}
  \includegraphics[angle=0,width=0.45\textwidth]{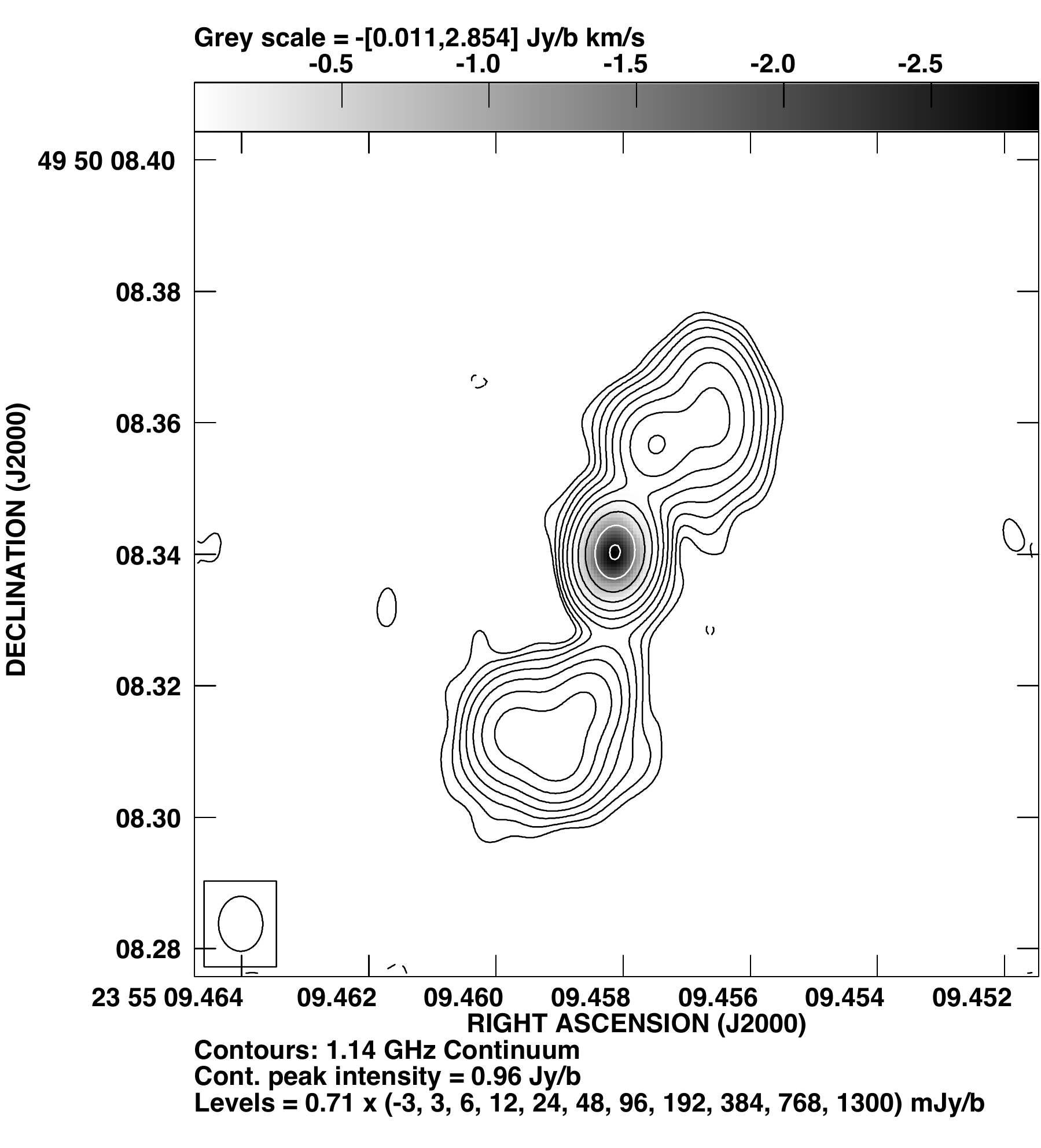} 
  \includegraphics[angle=0,width=0.45\textwidth]{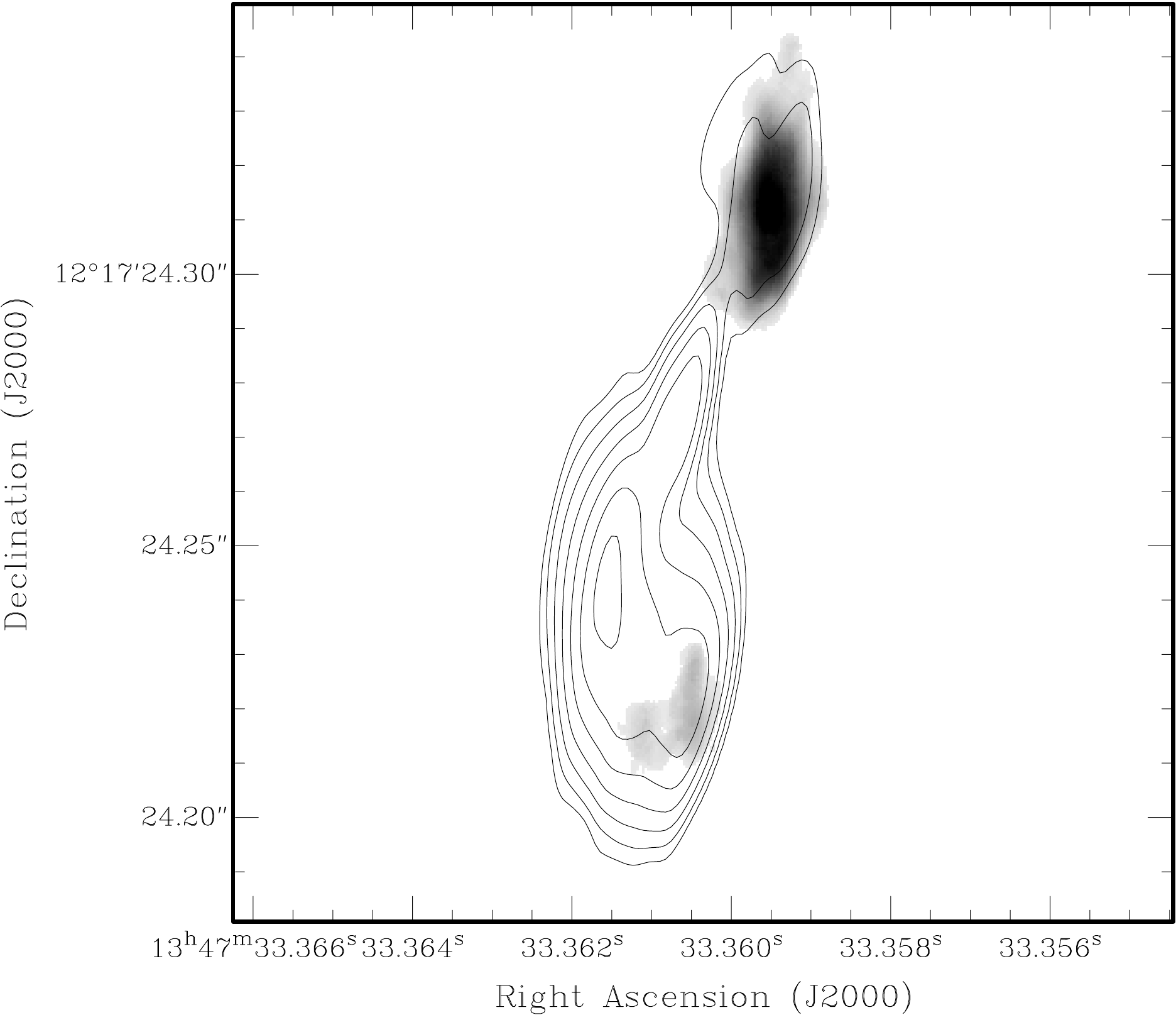}    
\caption{Three examples showing extreme cases for the covering factor. For PKS~1946+708 shown in the top-left panel \citep[taken from][]{Peck99}, absorption is detected across the entire source and thus \cf\ = 1. In  the source in the top-right panel (B2352+495), absorption is only detected against the core, resulting in a low covering factor of \cf\ $\sim$ 0.2 \citep[taken from][]{Araya10}, while the bottom panel shows the opposite situation with a similar covering factor, but where absorption is only detected against part of the radio lobes \citep[4C12.50; data taken from][]{Morganti13}.}
\label{fig:covFac}       
\end{figure}

The covering factor \cf\ is often assumed to be unity, mostly due to lack of    information that can be used to motivate another value.  \citet{Curran13a} pointed out that  apparent systematic differences could exist between the covering factor of compact and of extended sources, simply for geometrical reasons due to differences in size between the two classes of sources, even if there is no difference in the intrinsic properties of the absorbing gas in the two types of sources. However, VLBI observations suggest that their model may be too simple (see Fig.\ \ref{fig:covFac}).  For example, the covering factor is close to \cf\ = 1 in the compact source PKS~1946+708 \citep{Peck99}, but examples of compact sources with much lower covering factor  also exist (e.g.,  \cf\ $\sim 0.2$ in B2352+495, \citealt{Araya10}, or in 4C~12.50, \citealt{Morganti13}).  Extended sources with  a large covering factor are also known (e.g., \cf\ = 1 in the re-started source 3C~293 \citealt{Beswick04}). However a covering factor of a few percent is measured in the extended source NGC 315 \citep{Morganti09}, while \cf\ = 0.5 is found for the extended source 3C~305 \citep{Morganti05b}.  These examples suggest that size is not always a good predictor for the covering factor (see also \S  \ref{sec:young})

\subsection{Current and future telescopes and their limitations} 
\label{sec:TelescopesLimitations}

\HI\ absorption observations have been carried out with all  available radio telescopes. First observations were done using single dish telescopes 
(such as Arecibo, Parkes, and the Green Bank Telescope (GBT)),  but most of the recent work has been done with  interferometers: VLA, ATCA, WSRT, GMRT and Very Long Baseline Interferometry network (mainly European VLBI Network (EVN) and Very Long Baseline Array (VLBA)). Interferometers provide high  spatial resolution which is a distinct   advantage, given that associated absorption is typically found in the very central regions of galaxies. They also have better spectral stability of the observing band and are less affected by RFI.  The recent upgrades of many radio telescopes have further improved the capabilities and performance for \HI\ absorption observations.  

A number of limitations have affected  \HI\ absorption observations in the past. One is the limited velocity range that could be covered by an observation while maintaining good spectral resolution. This was due to the relatively small width of the observing bands available at the time combined with limited correlator capacity.  In many objects, the width of the absorption profile  exceeds 1000 \kms\  which in the past was similar, or even larger, than the velocity range covered by the observations. Due to upgrades, all relevant radio telescopes now allow observations with very large bandwidth with good velocity resolution and this  has resolved this issue. Parallel with this development has been the improved spectral stability of the instruments through the use of digital systems. While 15--20 years ago it was difficult to obtain a spectral dynamic range better than 1:1000, current instruments perform at least a factor 10 better. This means that in particular  broad but faint absorption components can now be detected much more easily.  The combination of broader observing bands and better spectral dynamic range has provided a major step forward and new discoveries have resulted (see \S \ref{sec:outflows}).

\begin{figure}
\centering
  \includegraphics[angle=0,width=0.52\textwidth]{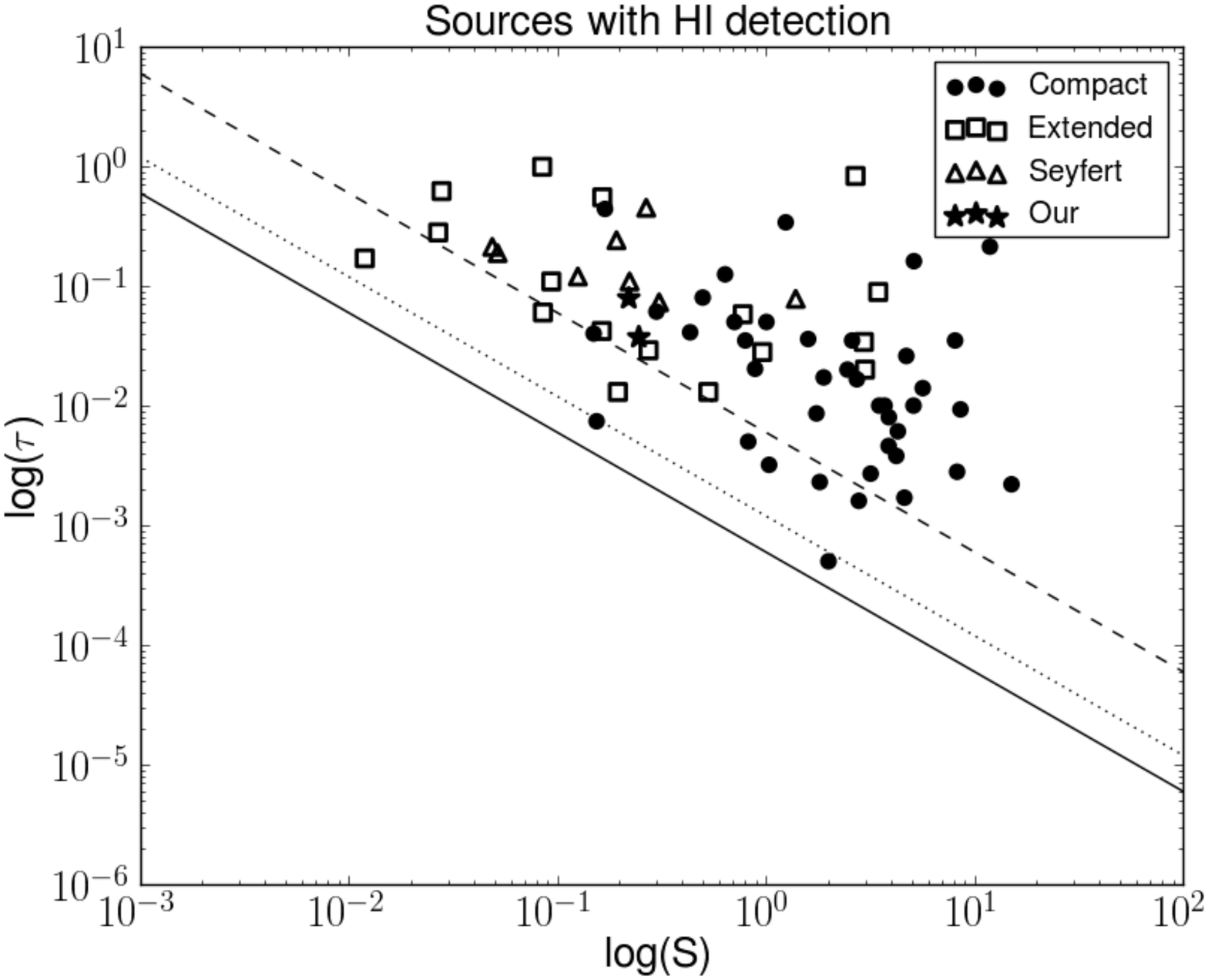}
  \includegraphics[angle=0,width=0.45\textwidth]{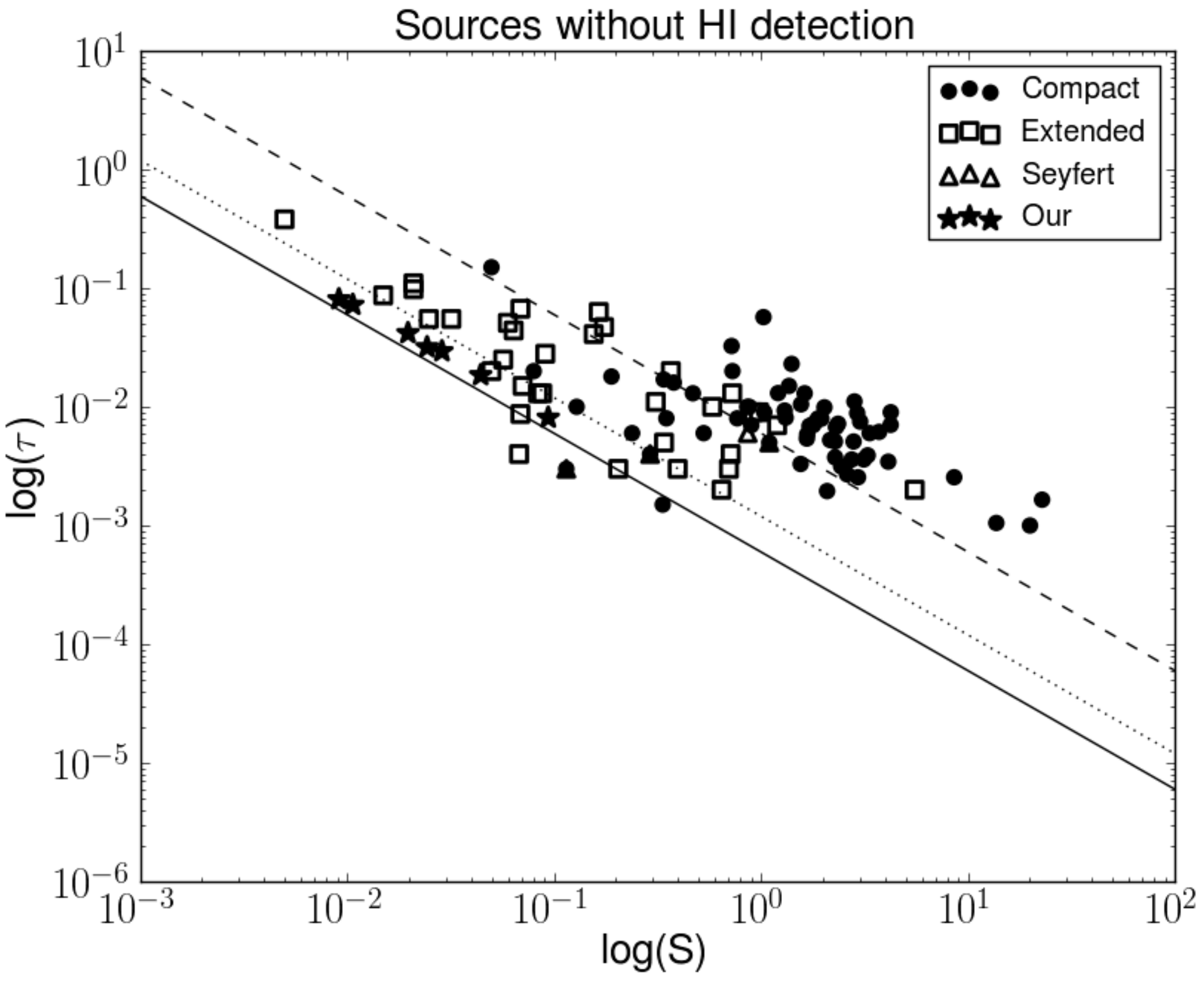}  
\caption{Optical depth vs flux density for \HI\ detections (top) and non-detections (bottom) collected from the literature. The solid, dotted and hashed line represent the $3 \sigma$ limits in the absorption line signal corresponding in order to $1 \sigma = 0.2, 0.4, 2$ mJy. Taken from \cite{Gereb14a}. }
\label{fig:tau}       
\end{figure}

Another main limitation has been, and to some extent still is, the limited coverage and sensitivity at lower frequencies which makes it harder to study high-redshift objects.  Only a few  radio telescopes are fitted with receivers suitable for observing \HI\ at redshifts higher than 0.2--0.4 and  the few receivers that were available  in the past, for example the UHF receivers mounted on the WSRT until a few years ago, had relatively high system noise. The situation has improved with the recent upgrade of the GMRT which offers continuous frequency coverage with good sensitivity, bandwidth and spectral resolution up to high redshifts  \citep{Gupta17b}. An important problem is, of course,  that many frequency ranges below 1300 MHz are very significantly affected by RFI so that many redshift ranges are not observable. The possibilities for high-$z$ observations will   improve with advent of MeerKat, which is located at a relatively RFI-poor site in South Africa and which will have very good sensitivity up to redshifts of 1.4,  and in particular with ASKAP in Western Australia which can observe up to $z = 1$ and for which the RFI situation for redshifts higher than about 0.4 is particularly good. Ultimately SKA1-MID  will allow studies up to $z \sim 3$ and with SKA1-LOW for even higher redshifts.

This new generation of radio telescopes will also offer an important improvement (a factor 10 or more) in sensitivity, even at low redshifts, which  obviously  is currently  another limitation for  \HI\ absorption observations.  To give an example, for a  velocity resolution of 10-20 \kms\ (which is   commonly used  because it provides a compromise between velocity resolution and sensitivity, i.e.\ observing time),  a telescope like the VLA can reach - for an observation of a several hours - a typical  rms noise per channel of $\sim 0.3$ \mJybeam. This means that for a radio source of 100 mJy one can detect a peak optical depth of  $\tau \sim 0.015$ (i.e.\  the flux absorbed by the \HI\ is a few percent of the radio continuum of the source). This opacity level   is where  currently the bulk of the detections have been found, while  objects with optical depth above 0.1 are  rare. Figure \ref{fig:tau}, which plots the 1.4-GHz continuum flux density of sources against the peak optical depth ($\tau_{\rm peak}$) for detections and non-detections (taken from \citealt{Gereb14a}), shows that there is no evidence for any trend of optical depth with source brightness. This means that with more sensitive observations  fainter  source populations will be detected in \HI\ absorption.  The figure  also shows that some upper limits are very tight and deeper than the level of typical detections. This shows that there is a population of radio sources that does not contain \HI, which can be valuable information about the nature of these sources. However, many   upper limits   are at the same level as most of the detections, thus they are not very constraining about the absence or presence of absorption in those sources. Higher sensitivity will allow to better separate these different populations.

\section{AGN seen through \HI\ absorption: an overview}
\label{sec:absorbingStructures}

Over  the years, \HI\ absorption studies have given important new insights into the  structure of the gaseous medium in the vicinity of an AGN, into the distribution and kinematics of gas in the nuclear regions of active galaxies,  into the role gas may play in the evolution of AGN, and vice versa into the influence AGN can have on the gas properties of galaxies. In the following sections we present this in detail, but here we briefly summarise the main results and put them in the context of earlier work.

\begin{figure}
\centering
  \includegraphics[angle=0,width=0.45\textwidth]{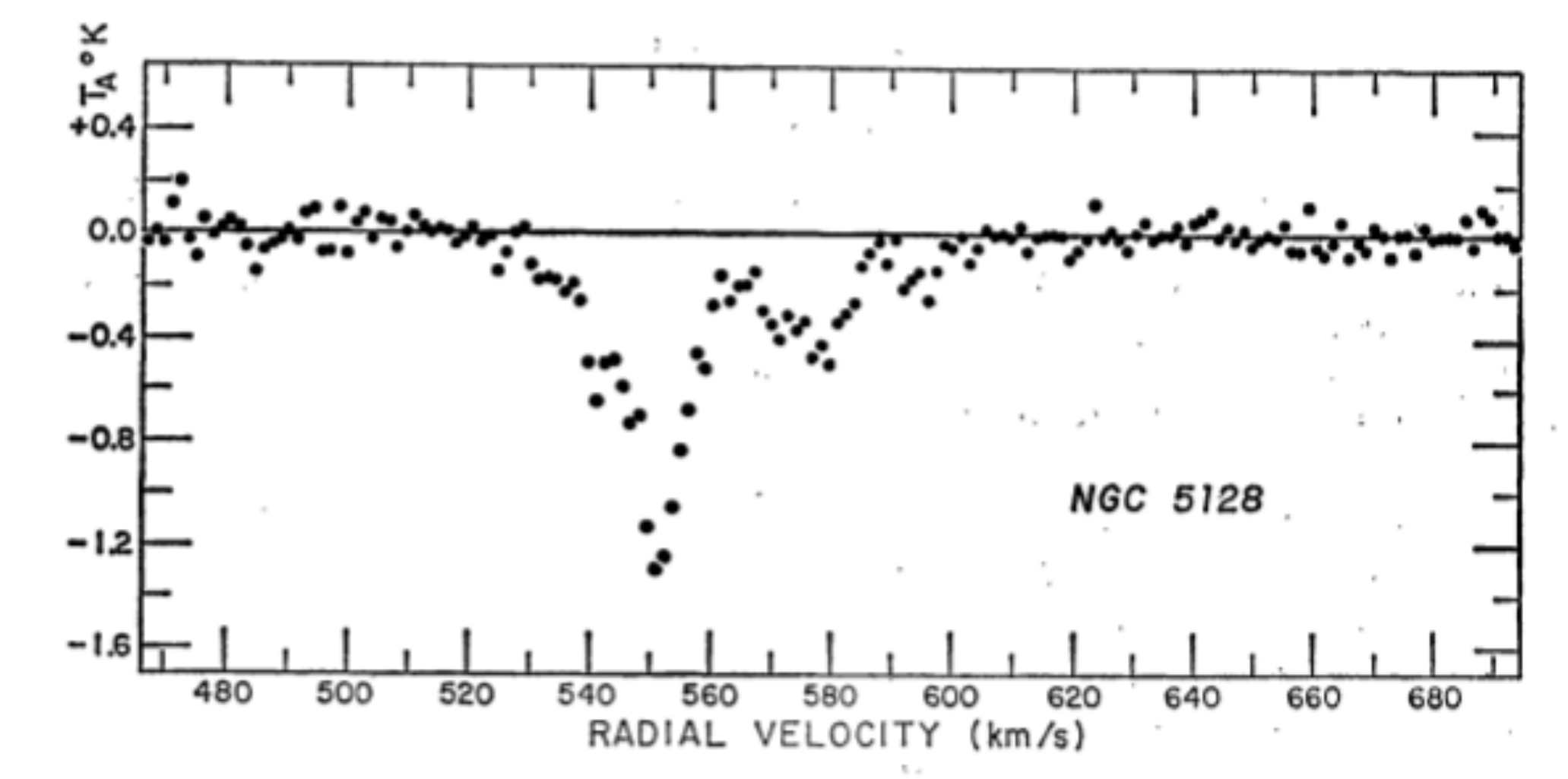}  
  \includegraphics[angle=0,width=0.45\textwidth]{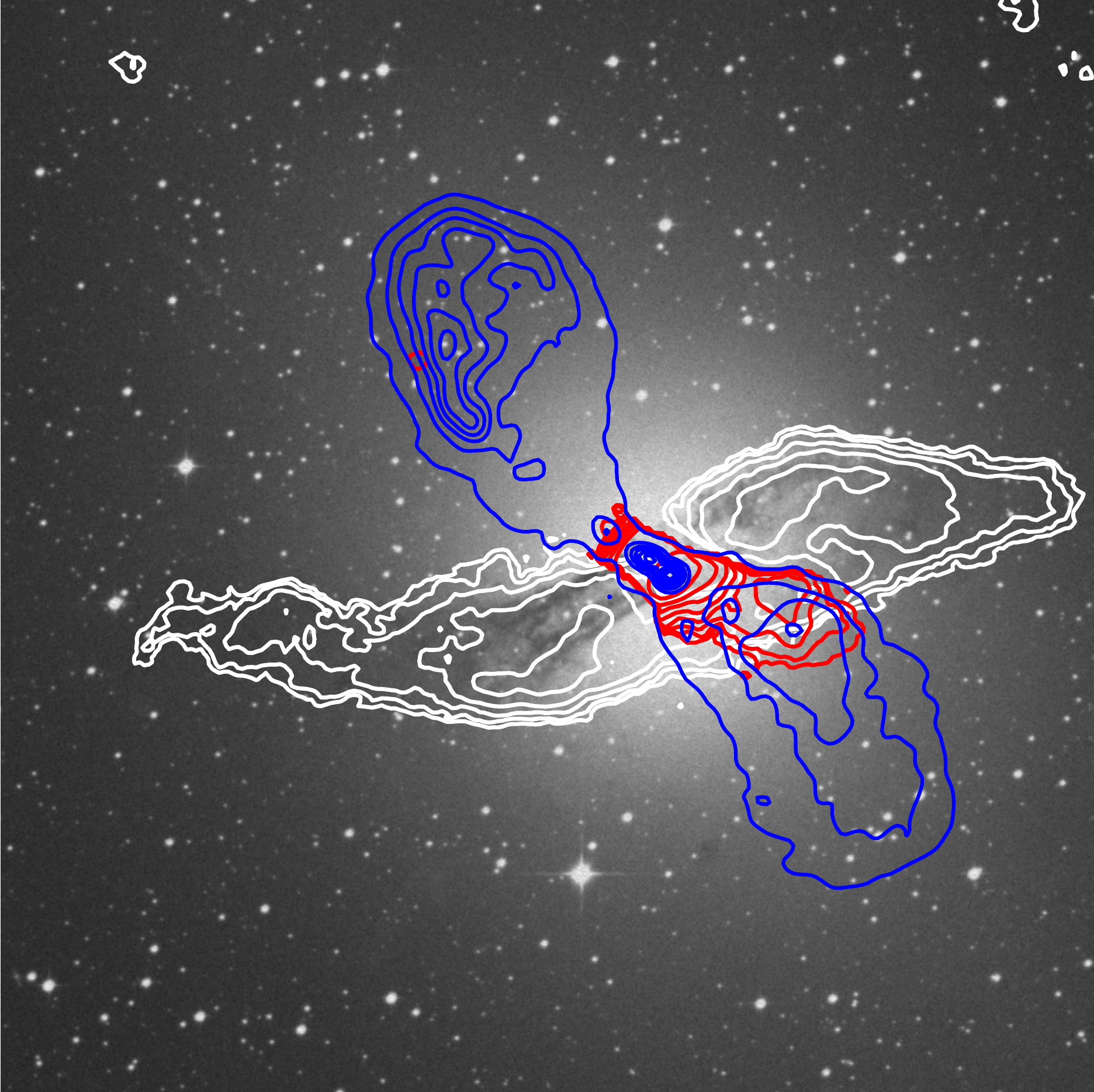}
\caption{Profile of the first \HI\ absorption detection which was obtained against the central region of Centaurus~A with the NRAO 140-foot telescope \cite{Roberts70}. On the right, we illustrate the geometry of the \HI\ absorption using the resolved image of the same region made with the Australia Telescope Compact Array \citep{Struve10b}. Superposed to an optical image, the blue contours represent the radio continuum coming from the inner lobes of Centaurus~A, the white contours trace the \HI\ emission from the warped gas disk of the host galaxy,  while the red contours represent the \HI\  detected in absorption against part of the continuum emission. In this object, most of the gas observed in absorption is that part of the large-scale disk that happens to pass in front of the radio continuum. This allows to  constrain the geometry of this disk relative to the continuum  lobes. }
\label{fig:cena}       
\end{figure}

One important aspect of the early studies has been that they   established the presence of \HI\ even in the circumnuclear regions of AGN, and that they  demonstrated the feasibility and the potential of \HI\ absorption studies for better understanding many AGN related phenomena. 
The first detections of \HI\ absorption were made against the central region of the iconic radio galaxy Centaurus~A by \citet[][see Fig.\ \ref{fig:cena}]{Roberts70}  and in the nearby spiral galaxy NGC~4945 \citep{Whiteoak76}.  
This was followed by the detection of \HI\ absorption in another famous radio galaxy, 3C~84 (Perseus~A), where - similar to what is seen in  optical emission lines - two absorption systems were found at very different velocities by \citet{Young73} and \citet{Crane82}; see also \S \ref{sec:clusters}.    

\begin{figure}
\centering
\includegraphics[angle=0,width=0.95\textwidth]{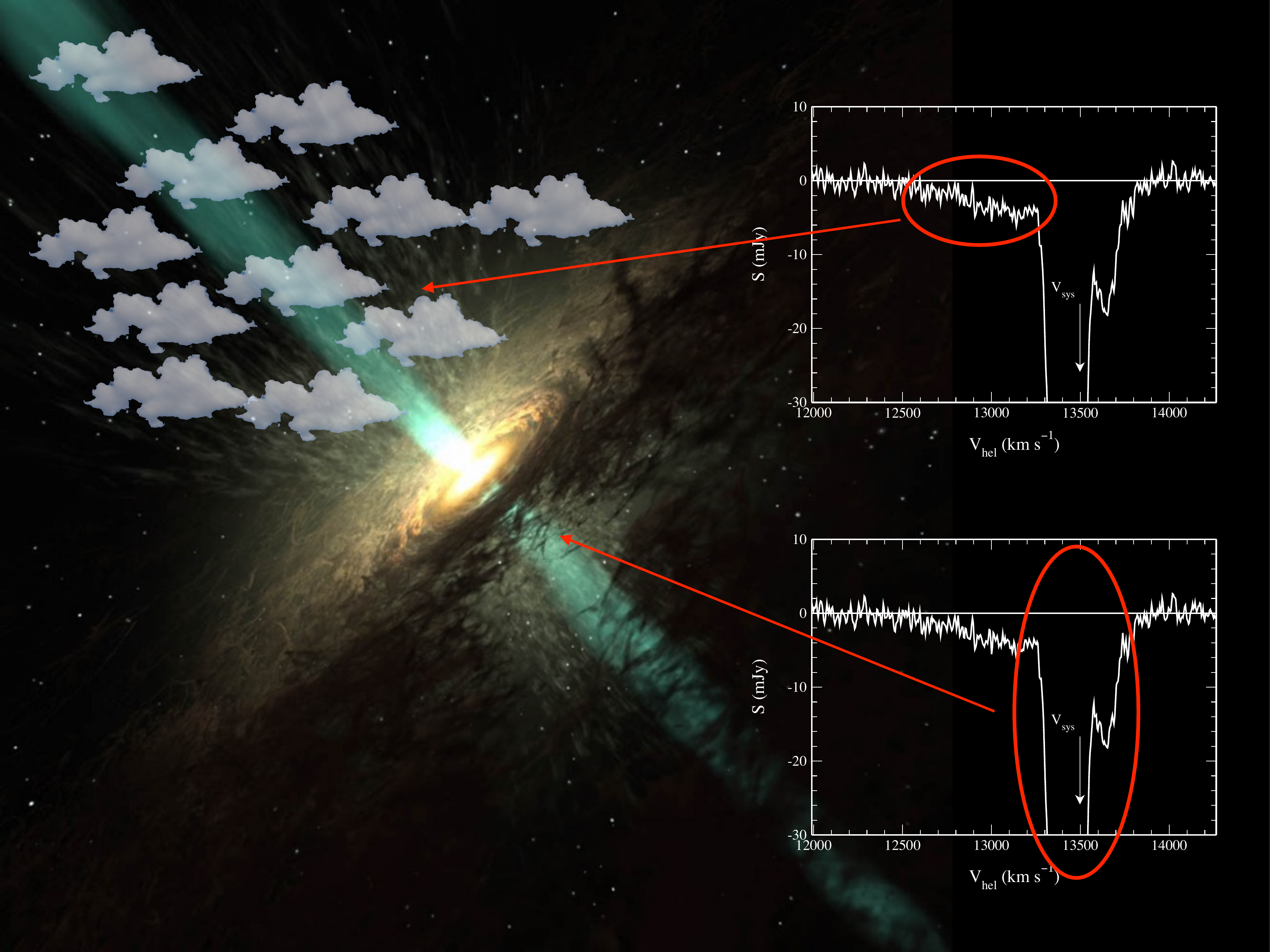}  
\caption{Illustration of  common structures of  absorbers and the related \HI\ profiles. Deep absorption profiles with widths of up to a few hundred \kms\  centred on the systemic velocity of the galaxy originate from regularly rotating gas disks which are projected on the bright inner radio continuum emission. Asymmetric wings to the absorption profile are in general associated with unsettled gas structures, such as gas outflows driven by the radio jet, or tidal gas streams. 
}
\label{fig:absorbers}       
\end{figure}

Since then, the earlier studies of associated \HI\ absorption in extragalactic sources were mostly  motivated by tracing the presence of nuclear disks and the presence of infalling gas fuelling the AGN.  Indeed, the first \HI\ observations of radio-loud galaxies \citep[e.g.][]{Dickey82,Mirabel82,Shostak83,Heckman83}  have shown that the \HI\ was likely  associated with disk structures with column densities about a factor 10 higher than of the \HI\  in the Milky Way and having a thickness of $10^2 - 10^3$ pc which was derived from a trend between column density and inclination of the disk derived from  optical images \citep{Heckman83}. 
Typically, these  spectra have deep absorption  with a full-width half maximum (FWHM) up to a few hundred \kms,  symmetric with respect to the systemic velocity of the host galaxy.  This suggests the \HI\ gas is tracing a regular rotating structure  (see the bottom panel in  Fig.\ \ref{fig:absorbers} for an example).  
The conclusions of these initial studies were mostly focused on showing the widespread presence of gas disks relevant for understanding the structure of the central regions and for unified schemes for AGN. 
Many current studies still focus on the larger-scale properties of the absorbing gas structures, whether they are regular or not, and their relation to the evolution of the AGN and of the galaxy that hosts the AGN.
The group of recently born, young radio sources, i.e.\ Compact Steep Spectrum (CSS) and Gigahertz Peaked Spectrum (GPS) sources (as well as restarted radio sources) are those having the highest \HI\ detection rate  and they will appear prominent in this review (see more in \S \ref{sec:young}). 
Interesting  is that recent studies have uncovered more examples of circumnuclear disks and tori on scales smaller than a kpc (i.e.\ tens to  a few hundred pc). Such gas close to the BH is typically characterised by broader absorption profiles with widths of several hundred \kms\ (e.g.\  Cygnus A, \citealt{Conway95,Struve10b,Peck01}, see more in \S \ref{sec:vlbi}).

Following the initial studies, \citet{Gorkom89} expanded the samples and found evidence that \HI\   absorption lines redshifted relative to the systemic velocity occur relatively more often than blueshifted absorption lines. This suggests  that the \HI\ could be involved in the fuelling of the AGN. For quite a period, largely based on this work, the perception in the community has been that redshifted absorption occurs quite often. However, separating infalling gas from the one regularly rotating has proven to be more challenging than expected because of the similar amplitudes of both infalling and rotational  velocities, as also shown by some of the more detailed studies (e.g., \citealt{ODea94,Taylor99,Maccagni14,Tremblay16}, see also \S \ref{sec:fuelling}).

Subsequent larger surveys have indeed confirmed that the situation is more complex and, in particular in the last 15 years, several studies have shown that the situation may actually be the reverse, namely that blueshifted absorption occurs relatively more often \citep{Vermeulen03a}.   In fact, after the first discovery \citep{Morganti98,Morganti03}, many examples are now known of very broad, shallow  ($\tau \ll 0.01$) blueshifted absorption components which indicate fast \HI\ outflows driven by the AGN   or by a cocoon inflated by the AGN (see \S \ref{sec:outflows}). In some cases  these outflows are located off-nucleus (so far they have been found up to a few hundred pc from the nucleus) at the location of bright radio components. Together with the high occurrence of outflows in newly born (or reborn) sources (see \S \ref{sec:young}), this suggests that, at least in some objects, the radio jets can play a role in driving them.  
This discovery has changed many ideas about the relation between an AGN and the gas in its immediate vicinity. The possible relevance  of these outflows in connection with  the role of AGN feedback in the context of galaxy evolution was recognised immediately. The discovery of \HI\ outflows can have a similar outflowing component of molecular gas \citep{Feruglio10} has further confirmed the relevance of the cold  AGN outflows.

\section{The nature of the absorbing structures in the various types of radio sources}
\label{sec:kinematics}


For a large fraction of the detections, the \HI\  absorption in radio sources is  seen against their central  regions.  Because of this, it allows to probe the physical conditions of the gas in the inner regions of radio sources and the nuclear structure of the various types of AGN.
The ``zoo'' of AGN is very rich (see \citealt{Padovani2016} for a review) and as described in several papers (see e.g.\ \citealt{Heckman14,Ramos17} for recent reviews), the nuclear regions of AGN of different (radio) luminosity are expected to show differences in the distribution and physical conditions of the gas and dust. 

In particular, two main types of AGN have been identified and, following \citealt{Heckman14}, referred to as radiative-mode and jet-mode AGN. 
Early studies proposed (e.g.\  \citealt{Krolik88}) for the first group, where the energy released is dominated by radiation, that the SMBH is surrounded by an accretion disk and, on larger scale, an obscuring structure, usually referred to as {\sl torus}, is present. The physical state of the gas near an active nucleus has been described by, for example, \citet{Maloney96}. The torus will be molecular if its pressure exceeds a critical value which depends on the luminosity of the central source, the distance from the source, and the attenuating column density between the central source and the point of interest in the torus. Under certain conditions, the torus can also have a significant component of atomic hydrogen. If the pressure is below this critical value, the gas is warm ($T \sim10^4$ K) and atomic\citep{Maloney96}.  

\begin{figure}
\centering
  \includegraphics[angle=0,width=0.8\textwidth]{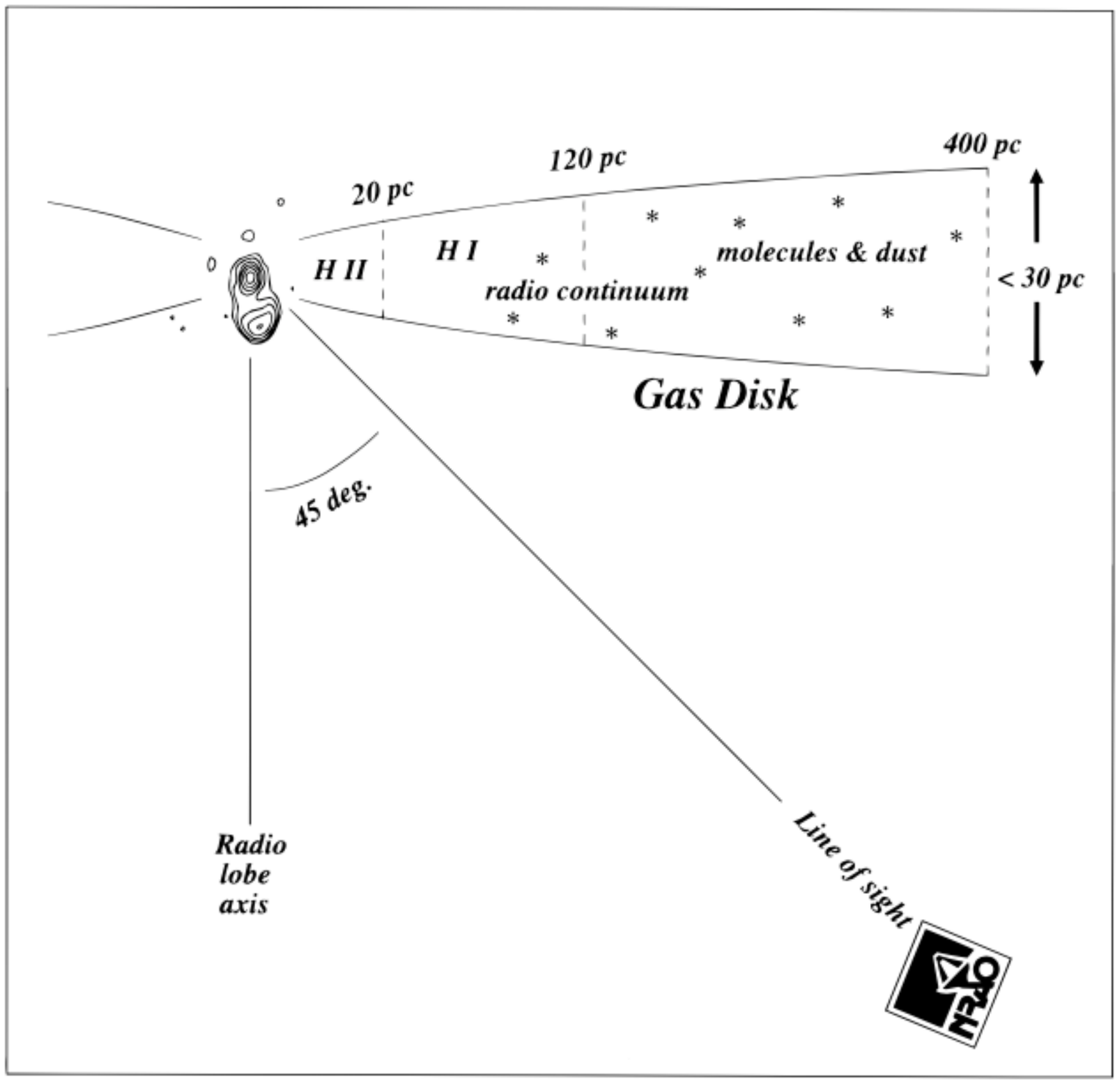}  
\caption{A sketch of the conditions of the gas in the circumnuclear disk of Mrk~231 as proposed by \citet{Carilli98a}, illustrating the general models for the structure of circumnuclear disks of, e.g., \citet{Maloney96}. The inner part of the gas disk is ionised while the outer regions consist of cooler molecular gas and dust. In between is a warm transition zone containing atomic hydrogen.}

\label{fig:mrk231}       
\end{figure}

These models predict that at the inner edge of the disk, likely between 0.3 pc and 1 pc from the AGN, a region of depth $\sim 0.1$ pc is fully ionised at a temperature of $\sim 10^4$ K and with a density of a few $\times 10^4$ cm$^{-3}$.  Such  ionised gas will radiate thermal emission and will cause free-free absorption of nuclear radio components viewed through the torus. 
Going outward, a warm transition zone is present, traced by H$_2$O maser emission and by \HI\ absorption. Further outward, a cooler molecular zone exists.  The conditions of the gas also change  with the vertical distance from the mid-plane of the disk/torus. The most recent theories suggest that the tori/disks are likely to be clumpy and that they even can be warped \citep[e.g.][]{Ramos17}, features which could be detectable using the high spatial resolution provided by VLBI observations.
A sketch of  a model of this kind, as applied to the structure of the inner disk/torus of Mrk 231, is given in  Fig.\ \ref{fig:mrk231}.

On the other hand, jet-mode AGN are instead associated with low accretion-rate and radiatively inefficient processes. Thus, an accretion disk and/or torus are likely absent, but the reservoir of gas can be in the form of larger circumnuclear disks (on scales of tens to hundred of pc).

Apart from their role in determining the differences between the physical properties of the AGN, these inner gas structures are also relevant for how we see and classify AGN, depending on to what extent they obscure our view of the AGN and its immediate surroundings (as in the unified schemes). In addition, they are also thought to play a key role in  feeding  the SMBH. Because of all this, a large effort has been dedicated also to trace the presence of  \HI\ in these structures. 

In the case of radio AGN, powerful radio galaxies (i.e.\ the so-called Fanaroff-Riley type II, FR-II, radio galaxies) and low-power, edge-darkened radio galaxies (Fanaroff-Riley type I, FR-I, radio galaxies)
follow, to first order, the separation between radiative-mode and jet-mode AGN (but see 
\citealt{Heckman14} and \citealt{Tadhunter16} for more details and exceptions).

Signatures of the presence of tori in FR-II has been found by observations in various wavebands (see \citealt{Tadhunter16}). However, the search for \HI\ in these structure has given mixed results, as described below.  

On the other hand, in low-power, edge-darkened radio galaxies (Fanaroff-Riley type I, FR-I, radio galaxies), the observations support the absence of a thick torus and indicate that the gas and dust is distributed in larger, thinner circumnuclear disks. For example, unresolved optical cores (e.g.\ seen in HST images, \citealt{Chiaberge99}) are commonly present in FR-I radio galaxies. The flux of these optical cores appears to correlate with their radio core flux, arguing for a common non-thermal origin (i.e.\ synchrotron emission from the relativistic jet) and supporting the absence of a pc-scale geometrically thick, obscuring torus  in FR-I radio  galaxies.  The presence of \HI\ in these circumnuclear structures is confirmed by the results described below.

In order to derive information about the kind of \HI\ structures that are detected in absorption, it is useful to know more about the larger \HI\ structures seen in emission, often unrelated to the AGN activity, in nearby early-type galaxies, i.e. the typical host of radio galaxies.   
Such early-type galaxies have historically been  considered red-and-dead, but in recent years  it has become clear that many early-type galaxies have significant amounts of gas, including atomic hydrogen  \citep[see][for some statistical results on the large-scale \HI\ emission in early-type galaxies]{Morganti06,Oosterloo10,Serra12}.
The best sample to  use as a reference point is the \atlas\ sample  \citep{Cappellari11} for which a broad suite of multi-wavelength observations are available, including deep  imaging  of the \HI\ emission \citep{Serra12}. This sample  consists of nearby early-type galaxies, most of them without a strong (radio) AGN \citep{Nyland17}. Based on \HI\ observations of a large fraction of the \atlas\ galaxies, \citet{Serra12} find that in about 40\% of those residing in the field, \HI\ emission is detected. In many cases, the \HI\ is found to form a large-scale \HI\ disk, but also  irregular structures, like tails and filaments, were observed in many  galaxies. This suggests that for  non-cluster radio sources there can be components in the absorption spectra that are not related to the circumnuclear environment. In a few members of the \atlas\ galaxies \HI\ absorption is detected and these spectra suggest that the profiles produced by these large, regular structures are typically relatively narrow (between 70 and 100 \kms\ FWHM) and centred on the systemic velocity of the host galaxy \citep{Serra12}.  

Only a small fraction of the absorption profiles in radio galaxies  are as narrow as those   found in radio-quiet  galaxies (i.e.\ like those in \atlas) and these narrow lines are mostly found at low radio power \citep{Maccagni17}.  In most   other cases, the absorption is typically several hundred \kms\ wide. This suggests that  in addition to possible absorption from large-scale disks, radio galaxies also have  \HI\ distributed in other components, for example  much closer to the central radio source.  If a rotating disk is closer to the centre, a larger range of velocities project on a background continuum source of a given size  and hence produces a larger profile width.

Below we go in more detail in describing the relevant results on central gas structures obtained for different groups of AGN. Seyfert galaxies and radio galaxies are presented separately because they typically (albeit with some exceptions) are  hosted by different type of galaxies:  Seyferts in spiral galaxies (where large-scale gas disks are expected to be present) and radio galaxies in earlier-type galaxies.

\subsection{\HI\ absorption in nearby galaxies and Seyfert galaxies}
\label{sec:seyferts}

For nearby galaxies, the study of \HI\ in absorption has been focussed on Seyfert galaxies and on interacting, FIR-bright and some peculiar galaxies.   
 
\begin{figure}
\centering
    \includegraphics[angle=0,width=0.55\textwidth]{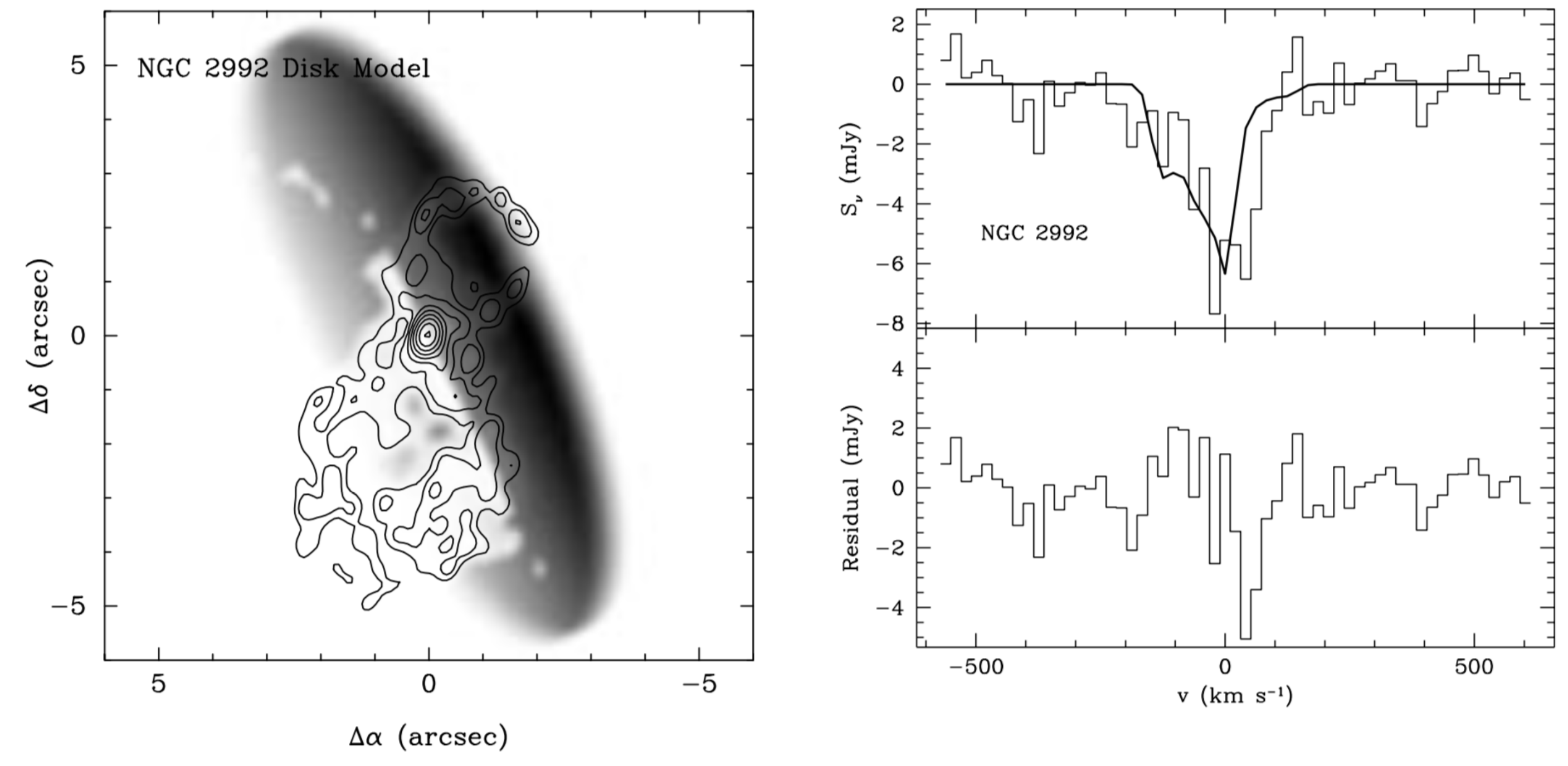}  
  \includegraphics[angle=0,width=0.4\textwidth]{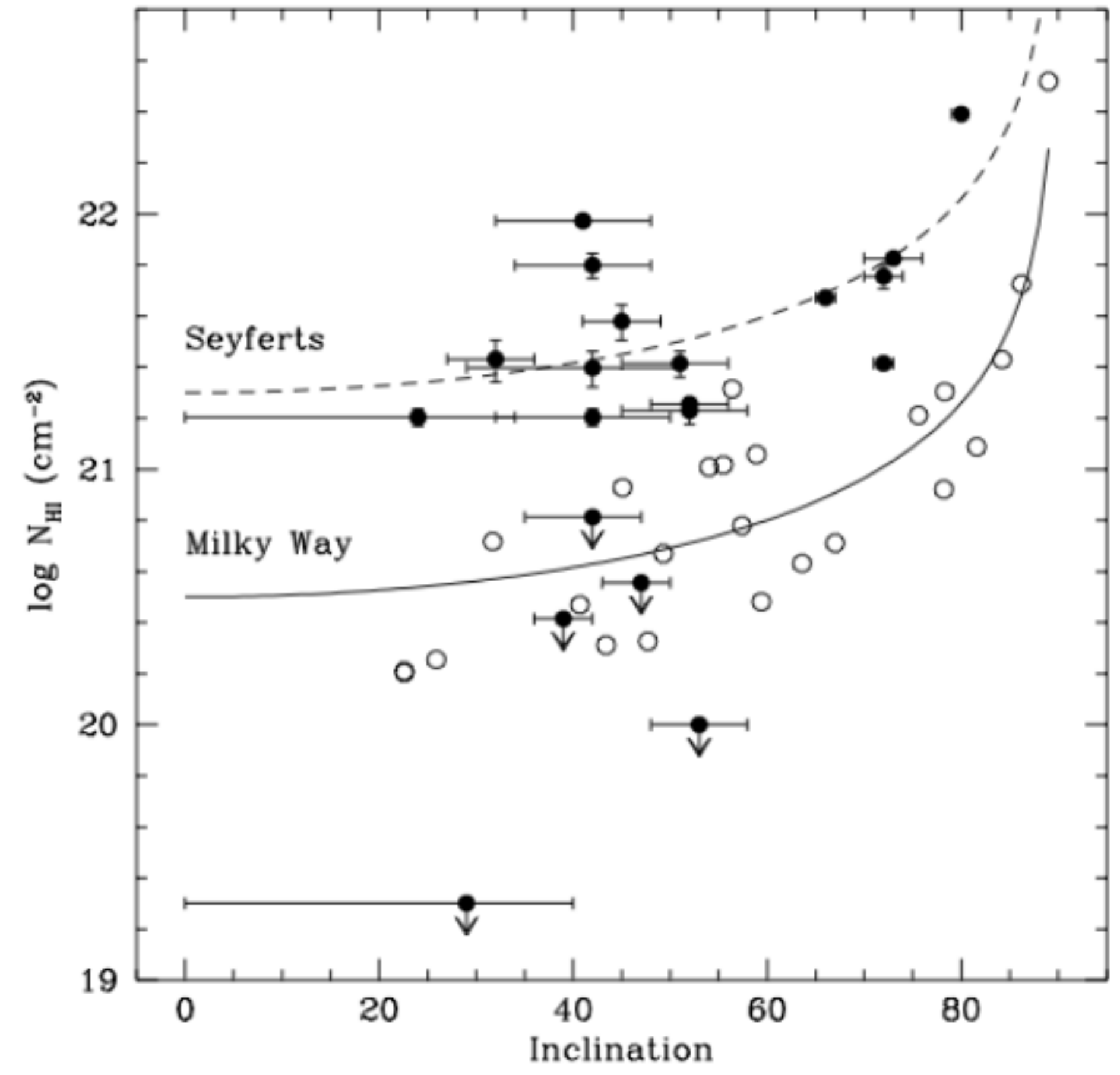}  
\caption{{\sl Left} The \HI\ absorption in NGC~2992 and the model of the absorbing structure and orientation respect to the continuum (from \citealt{Gallimore99}). {\sl Right}
 Distribution of column density for various groups of sources. Seyfert galaxies are  from \citet{Gallimore99}. The columns density derived for the Seyfert galaxies assume a characteristic \tspin = 100 K. 
}
\label{fig:Seyferts}       
\end{figure}

Early studies of \HI\ in the nuclear regions of Seyfert galaxies were carried out by \citet{Dickey82} while the most complete  study up to now has been done by \citet{Gallimore99}. In the latter work,   13 galaxies were observed which were selected as a flux-limited sample of spiral galaxies with evidence for nuclear activity and having a radio brightness $S > 50$ mJy. These observations show a high detection rate, with 9 detected objects having a column density ranging between $5 \times 10^{20}$ and $10^{22}$ cm$^{-2}$ assuming  \tspin\ = 100 K. Furthermore, there is a weak correlation between the probability to find \HI\ absorption and the  inclination of the host galaxy and similarly for the observed \HI\ column density. 
Figure \ref{fig:Seyferts} shows the \HI\ absorption from one of the galaxies, NGC~2992.  

A main result of \citet{Gallimore99} is that no \HI\ absorption is detected against the central, pc-scale, radio sources, even in known obscured Seyfert galaxies like Mrk~3 or NGC~1068.  In general, the absorption is detected against extended radio jet structures and appears to avoid central compact radio components.  Extensive modelling  by \citet{Gallimore99} suggests that the \HI\ absorption appears to trace rotating gas disks on the 100 pc-scale, which are aligned with the outer, larger disks of the host galaxy, rather than gas associated with the very central (pc) regions of the AGN.  The trend between the \HI\ column density and the inclination of the disk (see Fig.\ \ref{fig:Seyferts}, right) is consistent with this.   Among the possible explanations for the result of \citet{Gallimore99} is that free-free absorption suppresses the central background source so that \HI\ absorption cannot be detected against it. 


Another example  of this  is seen in high-resolution  observations of the nearest quasar, Mrk~231, studied in detail by \citet{Carilli98a} using both VLA and VLBA.  They found a rotating disk with an east-west velocity gradient of about 110 \kms\ traced by \HI\ absorption against a diffuse component on scales of hundred parsecs. However, this absorption disappears when full resolution spectra are taken only from the core. 
Thus, also in this object the absorption is not against the pc-scale radio core, but against a more diffuse radio continuum component seen on  hundred pc scales.  A circumnuclear disk similar to those used by \citet{Gallimore99} in their modelling can be invoked to explain these findings, 
where in this case the \HI\ is located at a few tens of parsec from the BH. 
The only Seyfert galaxy where, so far,  \HI\ absorption associated with a small (pc-scale) disk/torus is NGC~4151 as reported by \citet{Mundell95}.

However, the presence of at least some \HI\ gas located close to the AGN cannot be excluded by observations at high (tens of mas) spatial resolution. This is the case in particular in Seyfert galaxies that show evidence of a particular rich and dense nuclear ISM by the presence of water masers  as, for example,  in NGC~5793 \citep{Pihlstrom00} and TXS~2226--184 \citep{Taylor02}.
Eight galaxies have been reported to exhibit both water masers and \HI\ absorption \citep{Taylor02}. There appears to be a fair amount of overlap between the presence of water megamasers  and \HI\ absorption, although the search in five sources with \HI\ optical depth larger than 0.1 did not bring new detections. Whether this is due to the structure of the central disk/torus or on the conditions of the medium  is still unclear \citep{Taylor02}.

The column densities of the \HI\ absorption is often compared with the column densities derived from X-ray observations. This will be discussed in more detail in \S \ref{sec:OtherPhasesGas}. In the case of Seyfert galaxies, \citet{Gallimore99}  did not find a correlation between the column densities derived in these two ways, with those estimated from the X-ray absorption systematically higher. The mean column density from \HI\ absorption is $\sim 10^{21}$ cm$^{-2}$ (with cases up to $10^{22}$ cm$^{-2}$) assuming  \tspin\ = 100 K, while it is $\sim 10^{22.5}$ cm$^{-2}$ for the mean hydrogen column density derived from soft X-ray observations.  This is possibly indicating that the radio and the X-ray sources are not co-spatial (see \S \ref{sec:OtherPhasesGas}). 

No evidence for substantial infall/outflow of neutral hydrogen was found by \citet{Gallimore99} and they concluded that in the Seyfert galaxies  they  studied, the atomic gas is mainly in  rotating structures.  
However, this result is likely affected by the limitations of the telescopes used at the time, in particular the limited bandwidth. Fast \HI\ outflows  are now known in a number of Seyfert galaxies and their detection was only possible due to the broader observing bands available currently. Tentative evidence of \HI\ outflows  was already presented for two objects (NGC~1068 and the starburst/Seyfert galaxy NGC~3079) where broad ($\geq 600$ \kms) absorption features were found by \citet{Gallimore94}.    The presence of a faint blueshifted feature in the spectrum of NGC~3079 has been confirmed by more recent observations \citep{Shafi15}.
Furthermore, clear cases  of broad, blueshifted wings of \HI\ absorption indicating  gas outflowing at high velocities ($\sim$1000 \kms) have been found in the Seyfert 2 galaxy IC~5063 and in the nearby quasar Mrk~231. 
IC~5063  is  the clearest example of an \HI\ outflow and one of the best studied cases of AGN-driven outflows   so far \citep{Morganti98,Oosterloo00} and for which now also a strong molecular counterpart has been detected (see more in \S \ref{sec:OtherPhasesGas}).  More about \HI\ absorption as a powerful way to trace gas outflows will be discussed in \S \ref{sec:outflows}.

\HI\ absorption can also  be used to disentangle the complex distribution of the gas in the centre of interacting systems \citep[e.g. Mrk 1;][]{Omar02,Srianand15}) and in large merging systems, like  Arp~220 \citep{Mundell01,Allison14} and  NGC~6240 \citep[][ and refs.\ therein]{Beswick01,Baan07}. Such systems typically show broad \HI\ absorption profiles that are tracing disturbed disk-like structures. The presence of broad profiles in such cases is confirmed for the major-merger systems included in the survey of radio sources selected from cross-correlating SDSS and FIRST \citep{Gereb15,Maccagni17}, which show absorption profiles with full widths up to 800 \kms. 

A final remark is about the peculiar galaxy NGC~660 which is well-known for its prominent large, warped disk and  in which in the central regions \HI\ absorption is detected  \citep{Baan92}. Interestingly, the AGN in the galaxy has recently gone through an outburst which has produced a new, bright  radio continuum structure in the core of the galaxy.
The appearance of this new continuum component has affected  the shape and depth of the \HI\ absorption profile. Changes have been observed in some of the components  of the profile, for example a slightly broader redshifted wing, suggesting  inflow of material along our line of sight \citep{Argo15}. Whether there are more objects where changes in the profile shape occur due to changes in the \HI\ (on \HI\ coverage) it is still an open question. Campaigns to investigate this are still too time consuming, but will be done as result of the deep observations  in upcoming surveys. 

\subsection{\HI\ absorption in radio galaxies}
\label{sec:radiogalaxies}

A fairly large number of studies have presented observations of \HI\ absorption in radio galaxies    \citep[e.g.,][for some examples]{Morganti01,Vermeulen03a,Gupta06a,Curran08,Chandola11,Allison12,Allison14,Gereb15,Aditya16,Glowacki17,Maccagni17}. In addition, an analysis of a compilation of all  data published so far has been presented by \citet{Curran18}. These studies   typically show  a detection rate of \HI\ absorption   around 20-30\%. To first order, the detection rate appears to be independent of  radio power throughout the more than 5 orders of magnitude of radio power sampled in recent surveys.  
This holds also for low radio powers ($P < 10^{23}$ W Hz$^{-1}$) where the separation between radio galaxies and radio-quiet, normal galaxies becomes fuzzy. The similarity of the detection rate of \HI\ absorption and \HI\ emission suggests a continuity between what found (in emission) for radio-quiet, nearby early-type galaxies and radio galaxies. The only study were a combination of \HI\ in emission and in absorption has been traced in a sample of radio galaxies has been done by \cite{Emonts10}. The results of this study appear to support this continuity, albeit that the statistics and the depth of the observations will need to be improved to confirm the results. 

The column densities of the \HI\ absorption in radio galaxies are typically  between a few times $10^{20}$ and $10^{21}$ cm$^{-2}$ for \tspin\ = 100 K. Barring uncertainties in \tspin,  it appears that the typical column densities in radio galaxies are somewhat lower compared to those seen in Seyfert galaxies where column densities up to $10^{22}$ cm$^{-2}$ are observed.  However, some of the absorption lines may come from gas located quite close to the AGN where  \tspin\ can be much higher and, as a consequence, also the column density. 
 
Below we  discuss how these different structures can be identified thanks to statistical consideration or high-spatial resolution observations. 

\begin{figure}
\centering
  \includegraphics[angle=0,width=0.95\textwidth]{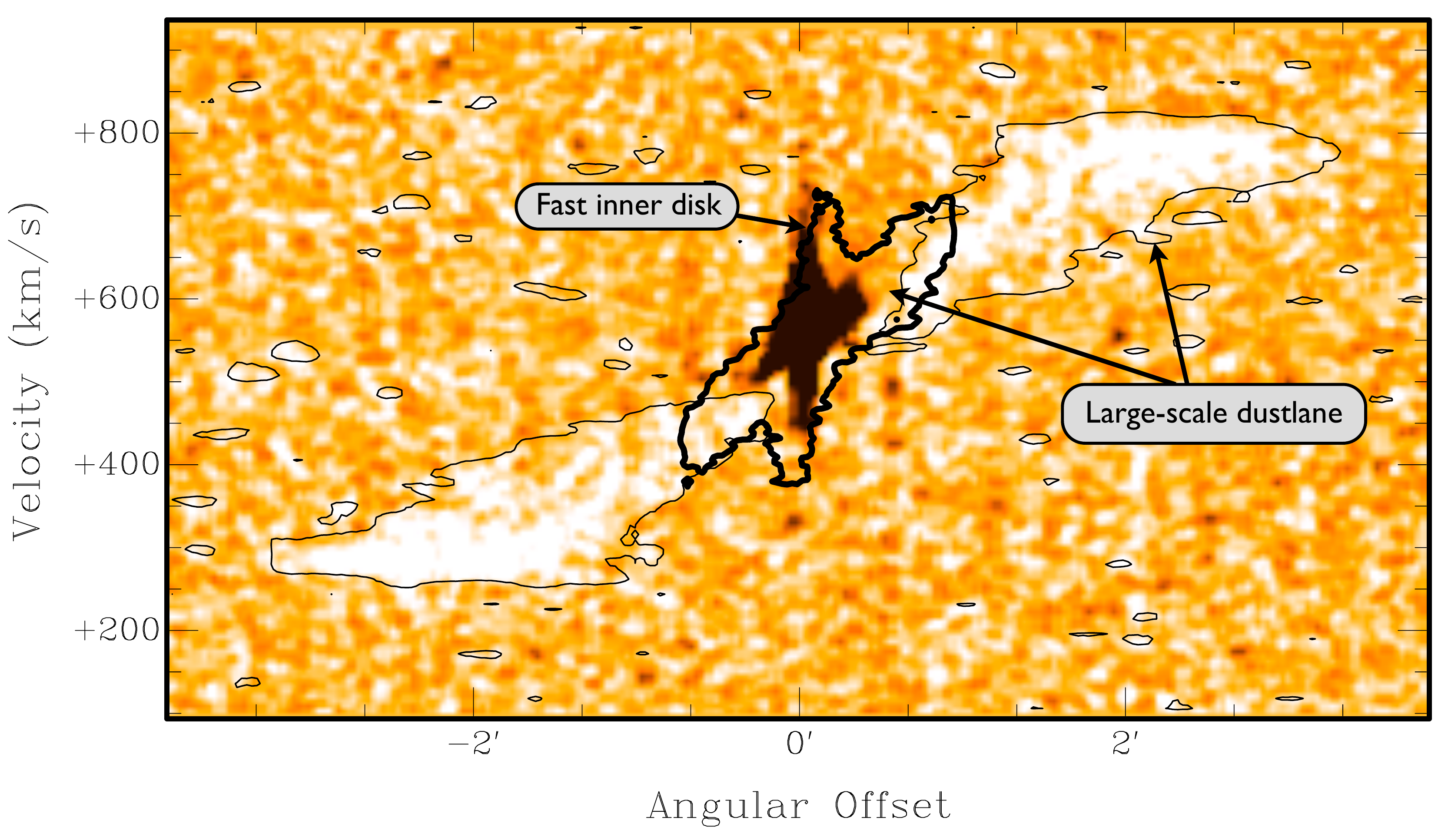}  
\caption{Position-velocity plot of the \HI\ (grey-scale and thin contours) in Centaurus A, taken along position angle $139^{\circ}$, with superimposed the CO emission \citep[thick contours; from][]{Liszt01}. The grey-scale represents the high-resolution \HI\ data (beam 8 arcsec) while the thin contour is from the same dataset smoothed to 15 arcsec. Note that the CO observations do not extend beyond a radius of about 1 arcmin. From \citet{Morganti08}.}
\label{fig:PVcena}       
\end{figure}

\subsubsection{Circumnuclear disks and tori in radio galaxies}

As described above, the inner gas disks of powerful radio galaxies (FR-II)  and of low-power radio galaxies (FR-I) are thought to be very different in size, thickness and density. The AGN in an FR-II galaxy is surrounded by  a  small, thick torus while in FR-I objects the central gas and dust is distributed in a large, thinner circumnuclear disk (see Fig.\ \ref{fig:mrk231}).
Based on this picture, and assuming \HI\ is present in all these structures, the detection rate and properties of \HI\ absorption should be different for the two classes of radio galaxies, for example a higher detection rate would be expected in FR-II objects compared to FR-I sources. Moreover, the detection rates should  also depend on the orientation of the nuclear disk.
Testing the hypothesis of differences in the structure of the central region of FR-II and FR-I is, however, not entirely straightforward and the observations are affected by a number of limitations. For example, performing \HI\ absorption observation of powerful radio galaxies is not always easy. The radio flux of their core is typically relatively low (see e.g.\ \citealt{Giovannini88,Morganti97}) making difficult to reach, with standard observations, sufficient optical depth sensitivity.

Most of the evidence for circumnuclear structures comes from statistical studies which we summarise below. However, some exceptions like the one from high-resolution, down to pc-scale, VLBI  observations (\S \ref{sec:vlbi}) and e.g. the observations of the nearest radio galaxy Centaurus~A allow to trace the circumnuclear structures in detail.

Given its proximity, Centaurus~A represents a special case for the detection and study of a circumnuclear disk in an  FR-I radio galaxy.  Not only the \HI\ absorption, but also the \HI\ emission of this galaxy can be studied in detail. The position-velocity diagram of Fig.\ \ref{fig:PVcena} illustrates how a circumnuclear disk can be identified from the \HI\  observations \citep{Morganti08}. The position-velocity plot shows how the kinematics of the large dust lane and disk (seen in emission, grey scale) cannot explain the large velocity range seen in the very centre (region in black representing the absorption). Instead, the nice correspondence of the inner, broad \HI\ absorption with the 160 pc sized inner H$_2$ disk (in emission, thick black contours) detected in Centaurus~A \citep{Liszt01}, strongly suggests that  the central \HI\ absorption also traces this circumnuclear disk.  This can be used as template for other radio AGN where such resolution cannot be reached.

The expected differences between the nuclear regions of FR-I and II have been mainly  addressed by statistical studies. For example, using a flux limited sample of radio galaxies, \citet{Morganti01} found that in powerful FR-II  radio galaxies, no \HI\ absorption was detected in the few galaxies observed that have broad (permitted) optical lines, while three out of four galaxies with narrow optical lines were detected (the one non-detection having quite a high upper limit). This is consistent with what is expected according to the unified schemes which predict that the presence of broad optical lines indicates an unobscured view of the AGN, while narrow optical lines means that the torus is blocking the view of the AGN and hence can also cause  \HI\ absorption. 
The same study suggested  that the detection rate of \HI\ absorption in FR-I objects is low: only one of the 10 FR-I galaxies observed was detected in \HI\ absorption,  consistent with the idea that their cores are relatively unobscured.
Similar results have been obtained by \citet{Gupta06b} and by \citet{Chandola13}. The latter study, using new observations combined with literature data, found \HI\ absorption towards 4 out of 32 ($\sim$13 per cent) FR-I objects, and towards 3 out of the 15 ($\sim$20 per cent) FR-II sources observed. However, considering the small samples of all these studies,  it difficult to make firm conclusions about differences in the torus/disc structure between FR-I and FR-II sources.

The distribution of dust in the central regions of FR-I radio galaxies has been further investigated in the study of \citet{Bemmel12}, by combining HST images, tracing the dust, with \HI\ absorption observations.  As expected, they found that the occurrence of \HI\ absorption is linked to the presence of dust. They also found that the absence of \HI\ absorption correlates (to first order) with the presence of an optical core (hence an unobscured view of the AGN), while an optical core is not detected if $N_{\rm \HI} > 10^{21}$ cm$^{-2}$ .
These results are again consistent with the neutral gas and the dust being associated with  the kiloparsec-scale disks and do not require parsec-scale dust tori to be present in FR-I radio galaxies.

However, it is important to keep in mind that the detection rate is also related to the more general conditions of the interstellar medium.  For example, a link between presence of \HI\ and dust has been seen using IR colours. \citet{Maccagni17}, using data from the WISE satellite, report a detection rate of almost 40\% for sources bright at 12 and 4.6 $\mu$m  and that are thus rich in AGN-heated dust. On the contrary, for dust-poor galaxies the detection rate is as low as 13\%. From shallower observations of a sample of compact-core radio galaxies,
\citet{Glowacki17} confirm the detection rate in early IR type radio galaxies but found a lower detection rate (i.e.\ $6\pm4$\%) in late IR type objects. The low number statistics together with the different type of radio sources and depth of the observations, make difficult to explain this difference. These statistical studies are still at the beginning and they will highly benefit from the upcoming large surveys (\S \ref{sec:future}). 
Finally, \citet{Chandola17}  find significant differences in the distributions of WISE W2(4.6$\mu $m)-W3(12$\mu$m) colour between sources with \HI\ absorption detections and the non-detections. They also report a very high detection rate of $70\pm20$ per cent for low-excitation radio galaxies (LERGs) with W2-W3 $ >2$ and having compact radio structure (possibly connected to young radio sources). The higher specific star-formation rate  for these galaxies  suggests that the \HI\ absorption may be largely due to star-forming gas in their hosts.

The presence of obscuring structures in the circumnuclear regions can also be explored  by comparing the detection rates of \HI\ absorption in type-2 (in this case radio galaxies) and type-1 (i.e.\ quasars) radio AGN.
A group of objects for which this has been done are  the young radio sources, such as Compact Steep Spectrum (CSS) and Gigahertz Peaked Spectrum (GPS) sources. The general properties of the \HI\  absorption in these objects  will be discussed in   \S \ref{sec:young}, but in the context of studying the effects of orientation and obscuration, it is interesting to note that CSS/GPS objects classified as radio galaxies have an \HI\ detection rate ($\sim$40\%) significantly higher than the detection rate ($\sim$20\%) observed towards CSS/GPS quasars \citep{Gupta06a,Gupta06b}. Similarly, from a study of 49 GPS and CSS sources, \citet{Pihlstrom03} find that \HI\ absorption is more likely to arise in objects classified as galaxies, rather than in quasars. This is expected if most of the \HI\ is distributed in a circumnuclear structure that is obscuring a type-2 AGN (i.e.\ radio galaxies), while type-1 AGN (i.e.\ quasars) are  unobscured. 
The situation seems to be different for red quasars. These objects have shown higher detection rates of \HI\ absorption \citep{Carilli98b,Yan16}, consistent with these sources likely being in a crucial, early phase of their lives when the AGN emission is still obscured by gas and dust (see also \S \ref{sec:disturbed}).

As a final remark, it is interesting to note that, as for Seyfert galaxies, the conditions of the gas close to a more powerful AGN  ($r<$ few tens of pc)  are strongly affected by the presence of the AGN. The presence of circumnuclear disks and tori can be traced by ionised, atomic neutral and molecular gas. Warm ionised gas can be traced by  free-free absorption, while the hot ionised gas can be identified by Compton thick structures \citep[see e.g.,][]{Risaliti03,Ursini18}. 
The confirmation of the presence of largely ionised gas (likely distributed in a rotating structure) in the immediate proximity of the AGN and, consequently, of a ``central hole'' (inside $\sim 2$ pc) of atomic gas, has been confirmed by the combination of free-free and \HI\ observations in a few cases. One of the best studied objects where the observations have shown that these structures are actually present, is the nearby galaxy NGC~1052 \citep{Vermeulen03b} but other examples with similar properties will be discussed in \S \ref{sec:vlbi}.

\subsubsection{Evidence of disks and tori from VLBI observations}
\label{sec:vlbi}

The final test to pin down the properties of circumnuclear-disks and tori is to spatially resolve them and trace the kinematics of the gas. Such  studies  require extremely high (i.e.\ milli-arcsecond) spatial resolution which can be reached using VLBI observations. This can be done in the cases where the structure and extent of the radio continuum on such small scales provides a large enough screen so that the overall kinematics of the absorbing gas can be reconstructed. A useful compilation of VLBI and VLBA \HI\ observations published up to 2010 is presented in \cite{Araya10}. Some more publications have appeared since and will be described below.

\begin{figure}
\centering
  \includegraphics[angle=90,width=1\textwidth]{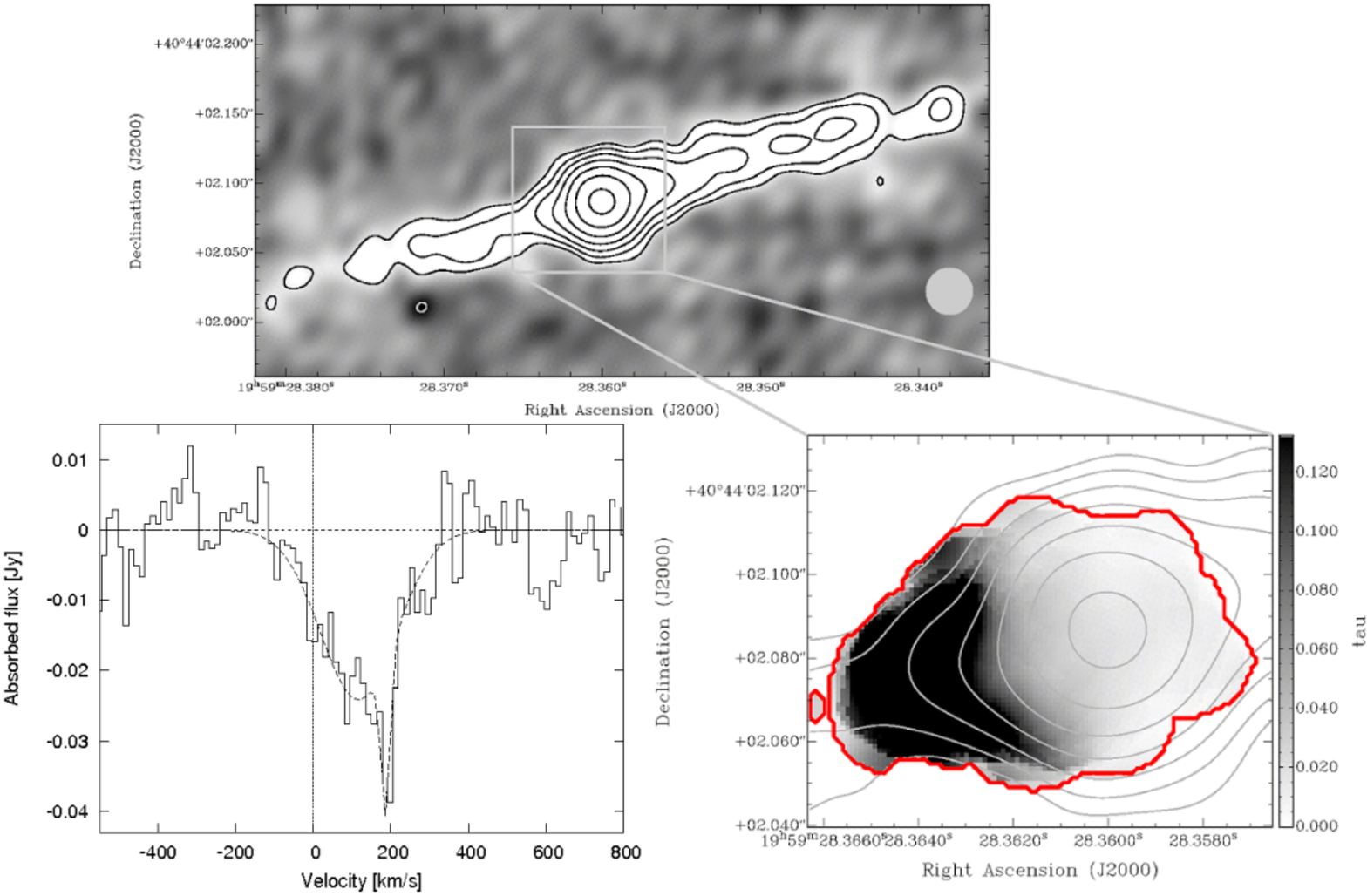}  
\caption{Top panel: continuum image of the nuclear jets of Cygnus A at 1340 MHz. The lowest contour is at 2~\mJybeam\ with subsequent contours increasing by factors of 2. The effective beam FWHM (see Sect. 2) is indicated in the lower right corner. Bottom left panel: integrated absorption spectrum from the blanked cube with the velocities shifted to the rest frame of Cygnus A. The dashed line shows the two component Gaussian fit. Bottom right panel: contours show continuum. Grayscale shows the mean opacity over the rest frame velocity range $-80$ to $+170 $\kms. The thick dark line shows the un-blanked region over which the integrated absorption spectra (shown in the bottom left panel) is calculated. Figure taken from \cite{Struve10c}.}
\label{fig:CygnusA}       
\end{figure}

One of the first sources where the \HI\ absorption was attributed - already from low spatial resolution observations - to a circumnuclear absorbing structure was the prototype FR-II radio galaxy Cygnus~A (\citealt{Conway95}).
The large width of the \HI\ absorption profile ($379\pm62$ \kms)  from the nuclear region  and the offset from the systemic velocity argue against the absorption arising at large (kpc-scale) distances from the nucleus because one would expect the profile to be much narrower and centred on the systemic velocity. Instead, the large line width indicates the absorption is produced by a geometrically thick torus or a circumnuclear disk close to the active BH, since the velocity of the torus/disk must be comparable to the orbital velocity in this case.  This scenario was confirmed by  VLBA observation of Cygnus~A \citep{Struve10c}. The data showed that the \HI\ absorption comes from a region perpendicular to the radio axis and as well as from the counter jet. This is exactly what one would expect for a nuclear disk. A velocity gradient over the absorbing region was seen,  as illustrated in Fig.\ \ref{fig:CygnusA}.  From these data one can estimate that the \HI\ absorbing gas lies at a radius of $\sim 80$ pc and that the disk has a scale height of about 20 pc, density $n > 10^4$ cm$^{-3}$ and a total column density in the range $10^{23}$ -- $10^{24}$ cm$^{-2}$ with \tspin\ $ < 740$ K. 

\begin{figure}
\centering
  \includegraphics[angle=-90,width=0.9\textwidth]{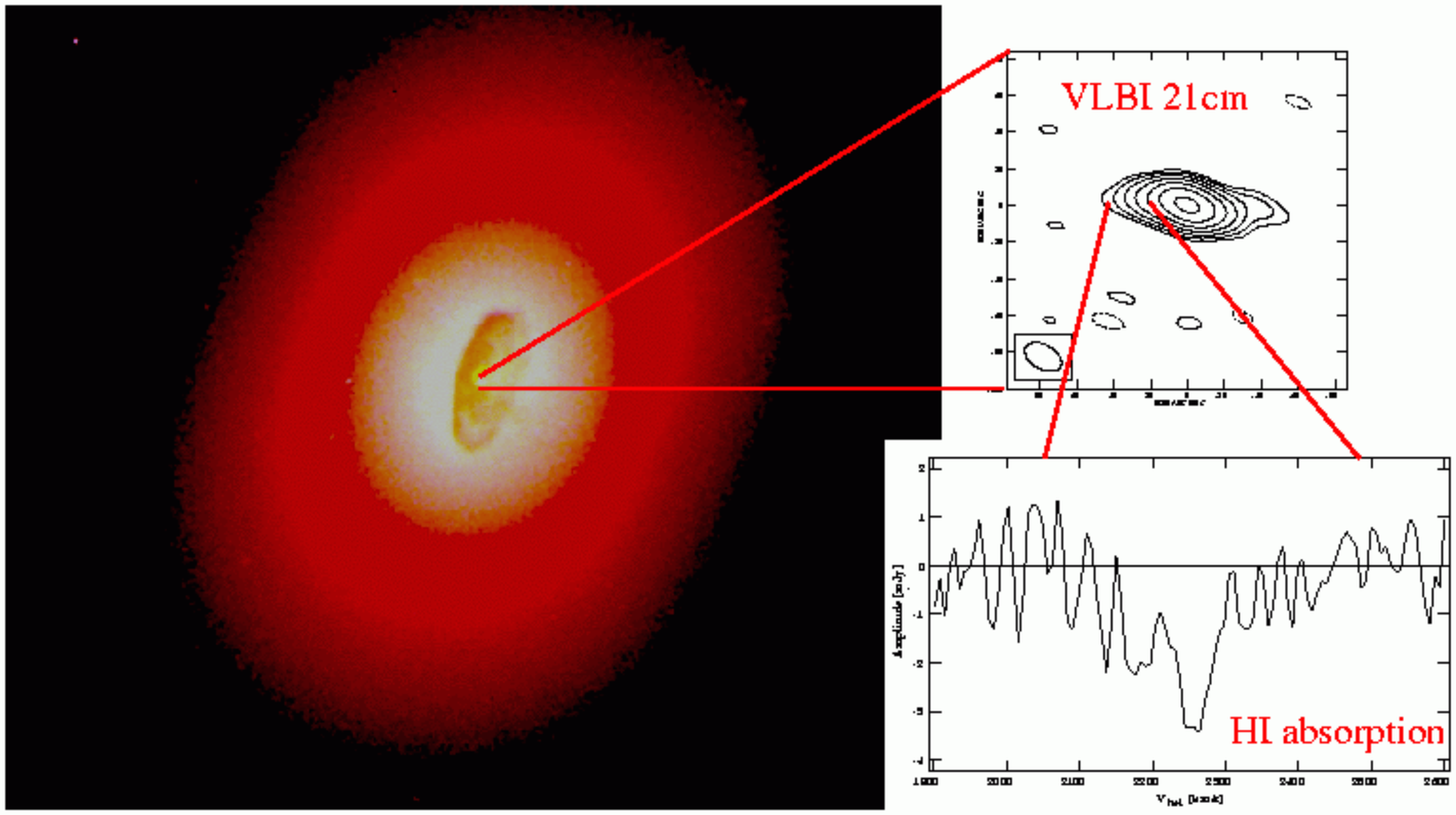}  
\caption{The background represent the HST image of NGC~4261. The inset are the
VLBI images (from  \cite{Langevelde00}) of the nucleus at 21 cm (right) and the \HI\ absorption spectrum
observed slightly offset from the nucleus (left). Courtesy of Huib Jan van Langevelde.}
\label{fig:4261}       
\end{figure}

Even in a powerful radio galaxy like Cygnus~A, the counter-jet is only just bright enough for what is needed for detecting \HI\ absorption. 
This is why detailed studies of \HI\ absorption using VLBI data have been done for only a limited number of extended sources, some of them low-power radio galaxies. 
The best examples are NGC~4261 \citep[][see Fig.\ \ref{fig:4261}]{Langevelde00}, Hydra~A \citep{Taylor96} and Centaurus~A \citep{Espada10}.  In the first two objects,  the evidence for a nuclear disk is based on the fact that the \HI\ absorption is relatively broad ($\sim$80 \kms) for the small background source  ($< 10$ pc) sampled by the VLBI data.  In Hydra~A, the torus/disk could be quite edge-on, allowing to estimate the vertical extent ($\sim$30 pc).  In NGC~4261, the \HI\ absorption is detected against the counter-jet (see Fig.\ \ref{fig:4261} taken from \citealt{Langevelde00}) at only 6~pc from the nucleus. 
For Centaurus~A, the VLBA observations of \citet{Espada10} have explored the pc-scale inner regions and identified a number of narrow absorption components seen along the jet, and a broader ($\sim$50 \kms, also detected by \citealt{Sarma02}) \HI\ line  which is more prominent towards the central and brightest 21-cm continuum component, base of the nuclear jet but not the nucleus itself, i.e. at a radius of $\sim$20 mas (or 0.4 pc) further from the from the nucleus. Whether this component is part of the circumnuclear disk described in \S \ref{sec:radiogalaxies} still needs to be confirmed. Another FR~I radio galaxy, the giant NGC~315, has been studied in detail with VLBI focusing on the role of \HI\ for the fuelling process (see \S \ref{sec:fuelling}).

As mentioned above, a group of sources that is suitable for VLBI observations are the young as well as restarted radio galaxies. In these objects, the radio continuum typically extends on  scales of a few hundred pc to about a kpc
and this offers the best background for tracing the complex distribution and kinematics of the multiple structures that can be present.

In the case of the restarted radio galaxy 3C~293, at least two structures with different velocity widths, both having large-scale velocity gradients, are seen against the kpc-sized continuum \citep{Beswick04}. One is seen against the eastern radio jet co-spatial with the location of dust lanes observed by the HST and which has a small velocity gradient of $\sim$50 \kms\ arcsec$^{-1}$,  consistent with the velocity gradient observed for the ionised gas. The second structure, with possibly a larger velocity gradient, could be due to a circumnnuclear disk on scales of $\sim$400 pc.   Also in the restarted, giant galaxy 3C~236 (see \citealt{Tremblay10} for an overview of its properties) a velocity gradient is observed which, also in this case,  seems to correspond to the dust-lane, the ionised gas disk and the molecular gas disk. Clouds of \HI\ are also detected with masses of a few $\times 10^4$ \msun\   \citep[see][recently confirmed by Schulz et al. 2018]{Struve12}. In both 3C 236 and 3C 293 the overall distribution and kinematics of the gas is complicated by the presence of fast \HI\ outflows, see \S \ref{sec:outflows}.

\begin{figure}
\centering
  \includegraphics[angle=0,width=1\textwidth]{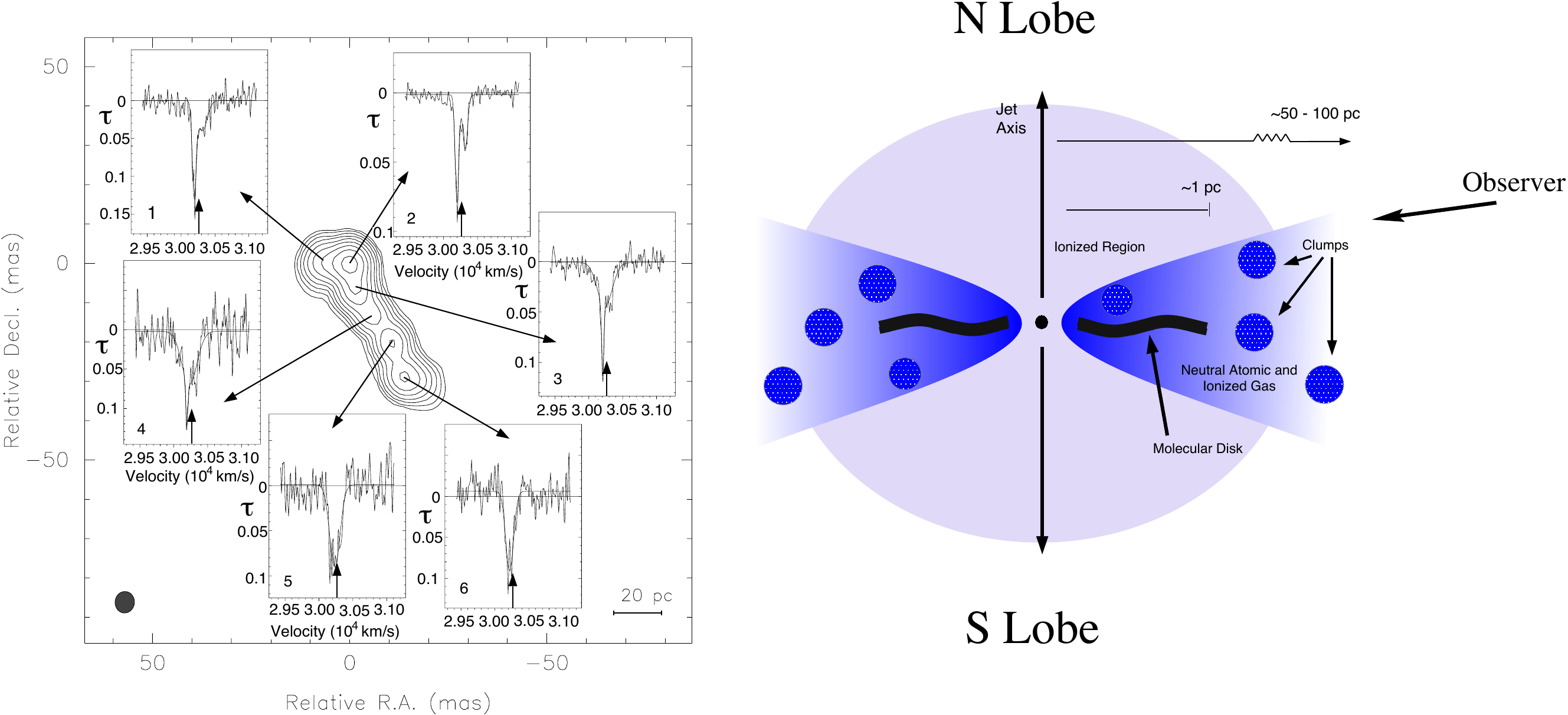}  
\caption{{\sl Left:} \HI\ absorption profiles toward the radio source PKS~1946+708 overlaid on the 1.3~GHz continuum contours.  {\sl Right:} cartoon illustrating the possible distribution of the gas along the  tine of sight to the jet components.  From \cite{Peck01} and \cite{Peck99}. }
\label{fig:1946}       
\end{figure}

The groups of objects most commonly studied in VLBI are the CSS and GPS radio galaxies. The \HI\ absorption in these objects appears to trace a variety of absorbing structures (including infalling and outflowing gas), but  a number of these radio galaxies have also shown, on  pc-scales, \HI\ distributed in torus-like structures with evidence of rotation. 

One of the best examples is 4C~31.04. In this compact, double-lobed radio galaxy, a sharp edge in optical depth has been observed against the western lobe, while for the eastern lobe the opacity is quite uniform \citep{Conway96,Struve12}.  This has been taken as a signature of an almost edge-on torus/disk which extends to radii $r < 200$ pc.   HST observations of 4C~31.04 \citep{Perlman01} have confirmed the presence of a dust disk of the  dimensions as derived from the \HI\ absorption. 

Another example is the compact, double-lobed radio galaxy, PKS~1946+708, where \HI\ absorption was detected  towards the entire $\sim$100 pc extent of the continuum source  \citep{Peck99,Peck01}. The absorption against the core is broader ($\sim$300 \kms), but has lower optical depth than the absorption against the rest of the source (see Fig.\ \ref{fig:1946} and \citealt{Peck01} for details).  According to \citet{Peck01}, this is in agreement with the thick torus scenario in which gas closer to the central engine rotates faster and is much warmer, thus lowering the optical depth.  The narrow lines  are explained as coming from gas further out in the torus, possibly related to an extended region of higher gas density and having a diameter of the order of at least 80 - 100 pc.  
In an other double-lobed radio galaxy, B2352+495,  \HI\ absorption with a width of $\sim$100 \kms\ has been found located against the core,  and tentatively showing a  velocity gradient perpendicular to the jet direction \citep{Araya10}. Also for this case, a component  tracing circumnuclear material is suggested to cause the absorption. In addition, a very narrow and redshifted (about 130 \kms) absorption is  detected and likely tracing an infalling cloud at larger distance from the nucleus. 

Finally, in  PKS~1814--637,  a radio source  $\sim$400 pc is size and a rare example of a powerful radio source  hosted by a disk galaxy,   one can identify at least two separate disks producing the \HI\ absorption \cite{Morganti11}. The deep absorption seen in this object  likely originates from  cold gas located at large distances from the nucleus and is probably associated with the large-scale disk  (which has a thickness of 400 -- 500 pc) of the host galaxy. A shallower, broader component could, instead, trace a circumnuclear disk located closer to the radio source (and having a thickness of the order of 100 pc). The large width of the shallow, shifted absorption features would be due to unresolved rotation  projected along the line of sight. 
The VLBI observations are confirming the presence of circumnuclear structures containing \HI, likely playing an important role in the fuelling of the SMBH.

Most VLBI studies are based on sources having redshifts up to $z\sim 0.12$, corresponding to the frequency of the redshifted \HI\ to be larger than 1270 MHz. This frequency represents the limit of the receivers of many of the VLBI/VLBA antennas. Being restricted to very low redshifts is one of the major limitations for the high-resolution study of the properties of \HI\ in radio galaxies.
For a few years, a number of antennas which are part of the European VLBI Network (EVN) were equipped with UHF receivers covering the frequency range  800 -- 1300 MHz. This has allowed to reach redshifts up to  $\sim 0.8$ and to explore the \HI\ absorption in a number of more distant sources, albeit the relatively poor sensitivity of these receivers has limited the observations to  bright sources. 
Examples are the compact, double-lobed radio galaxies 2050+364   \citep[$z=0.35$;][]{Vermeulen06}, 3C~49, and 3C~268.3 \citep[$z=0.62$ and $z=0.37$ respectively;][]{Labiano06}. Also in these objects the VLBI data show a variety of situations. 
In 3C~49 and 3C~268.3, the radio lobes against which the absorption occurs are brighter, are closer to the core and, in the case of 3C~268.3,  are more depolarised. The observed asymmetries suggest that the \HI\ gas is involved in an interaction with the radio plasma.  
In 2050+364, like in some of the  objects described above, two absorption systems - one very narrow and one shallow and broader - have been found. The narrow component covers the entire source, perhaps indicating that also in this case the origin is gas at larger distances from the core. The origin of the broad absorption is more difficult to establish, because of a difference between the redshift derived from the optical emission lines (\OIII\ and H$\beta$, \citealt{Vermeulen06}).

\begin{figure}
\centering
  \includegraphics[angle=0,width=0.99\textwidth]{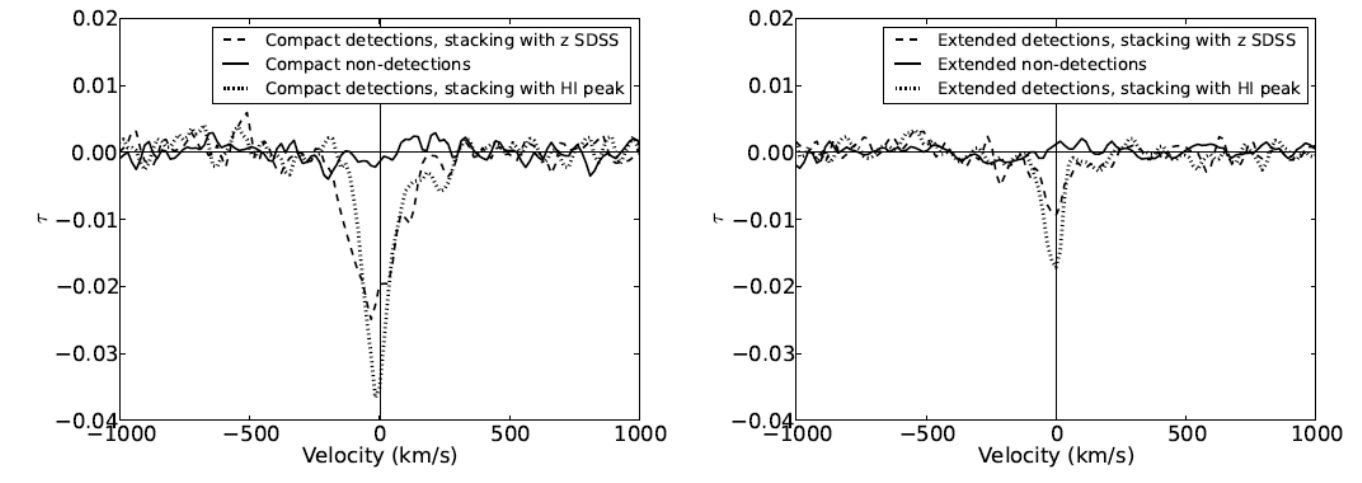}  
\caption{Stacked \HI\ profiles for compact (left) and extended (right) sources \citep[from][]{Gereb14b}. The profiles show the higher optical depth and the broader width of the stacked profiles of the former, suggesting differences in the medium in which the two groups of objects are embedded.  Both panels show that for both compact and  extended sources, even if a large number of spectra are stacked of sources where no absorption was detected in the individual spectra, the stacked spectrum still does not show a detection, indicating that a population of gas poor galaxies exists and/or where the conditions of the \HI\ (e.g.\ spin temperature) are different.  From \cite{Morganti15a}}
\label{fig:Stacking}       
\end{figure}

\subsubsection{Nuclear structures in young radio galaxies}
\label{sec:young}

The detection rate, column density and line width of \HI\ absorption in radio galaxies appears to depend on the characteristics of the sources. For studying \HI\ absorption, intrinsically compact radio sources like GPS and CSS have  gained particular attention. These sources are considered to be recently born (or re-born) radio  AGN, although their properties may also be the result of a strong effect of the surrounding ISM. 

A number of studies have shown that these compact sources, compared to more extended ones, have a higher detection rate of \HI\ absorption, reaching  40-45\% \citep{Veron00,Vermeulen03a,Pihlstrom03,Gupta06a,Chandola11,Gereb15}.
Interestingly, although the statistics is not as good, such a high detection rate is also observed for restarted radio galaxies, where a newly born radio source is embedded in a large, much older object (see the cases of 3C~236 \citealt{Mirabel89,Morganti05a}; CTA 21 \citealt{Salter10}; 3C~321 and 4C~29.30 \citealt{Chandola10,Chandola12}; 4C~12.50 \citealt{Morganti13} and \citealt{Saikia07} for some examples).
The \HI\ absorption detected in CSS/GPS and restarted sources has also been found to trace gas with higher column density compared to extended radio sources, and the spectra more often indicate gas with disturbed kinematics, traced by the asymmetric, often highly blueshifted profiles (\citealt{Gereb15,Glowacki17} see also Fig.\ \ref{fig:Stacking}). 

\begin{figure}
\centering
  \includegraphics[angle=0,width=0.6\textwidth]{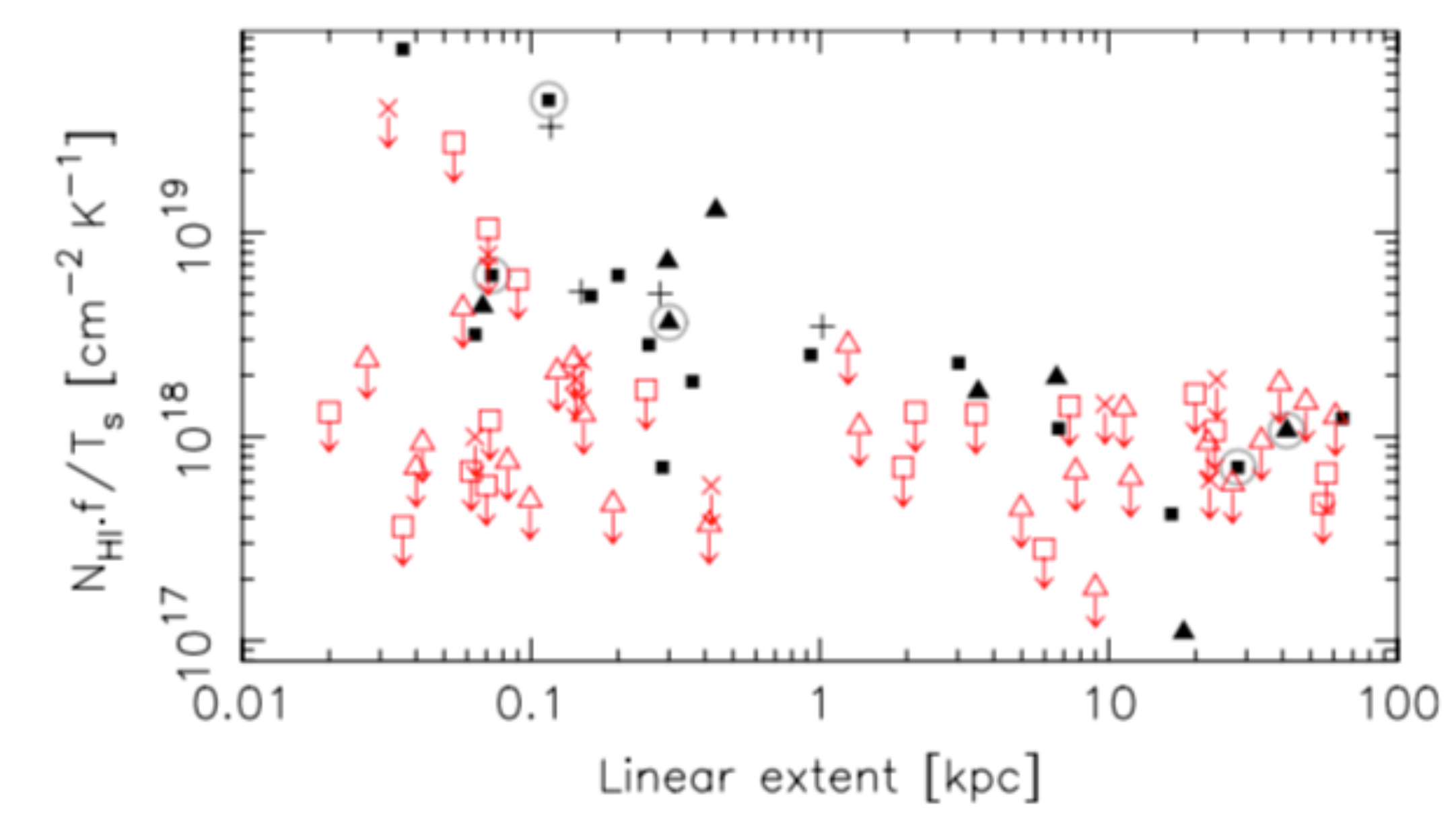}  
   \includegraphics[angle=0,width=0.3\textwidth]{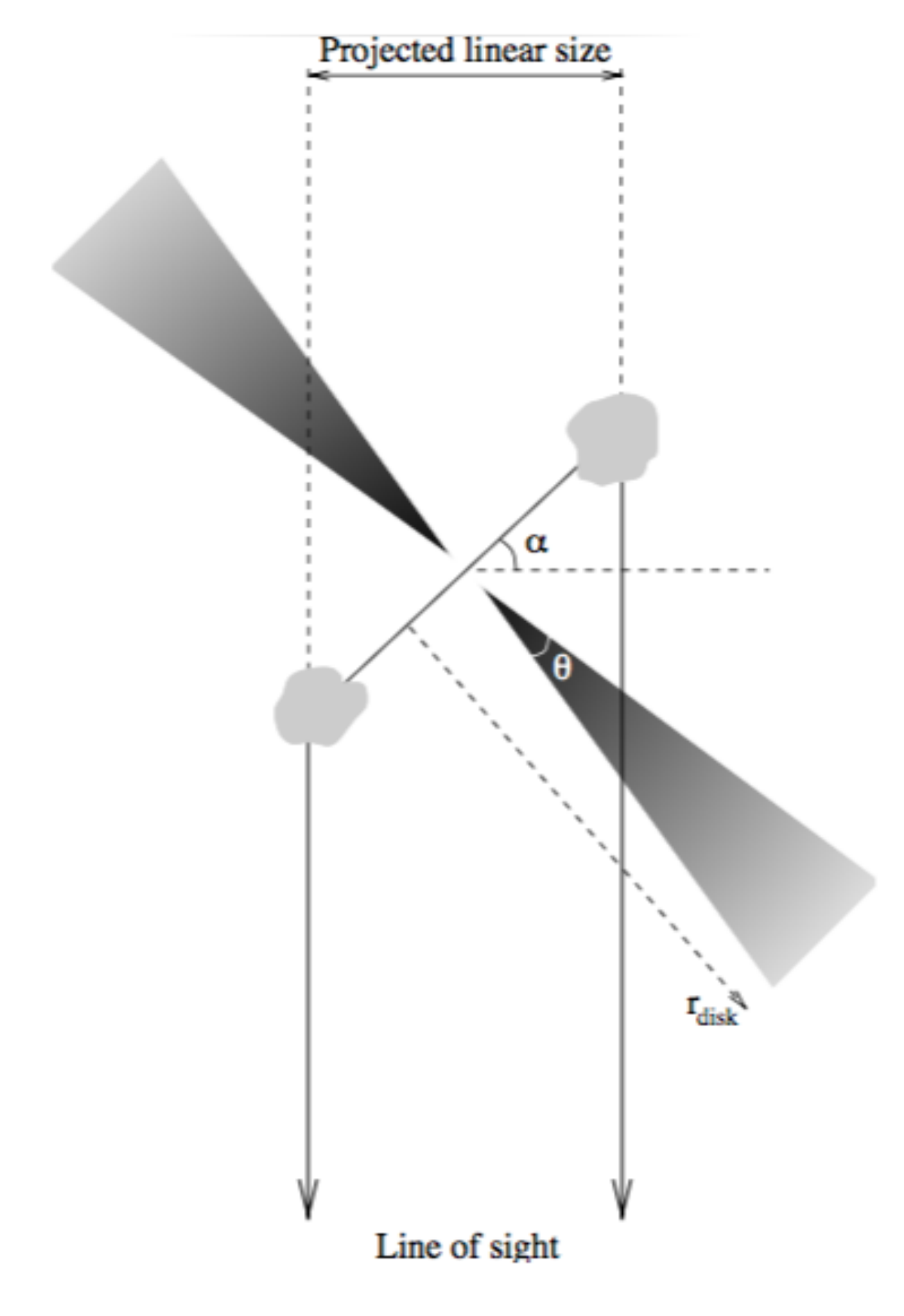}  
\caption{{\sl Left:} scaled velocity integrated optical depth ($1.823 \times 10^{18}  \tau dv$) of the \HI\  absorption vs. the linear extent of the radio source size. Black symbols are detections while red symbols are upper limits (taken from \citealt{Curran10}). {\sl Right:} cartoon of the possible  geometry of the disk structure producing the \HI\ absorption in CSS/GPS sources. Suggested by \cite{Pihlstrom03}}
\label{fig:sizecolumn}       
\end{figure}

The high detection rates could suggest that these radio sources are, on average, more often embedded in a rich ISM. This would have  implications for the impact these AGN may have on the surrounding medium which, in turn, may be important for the role of AGN feedback in  galaxy evolution (see \S \ref{sec:young} and \S \ref{sec:outflows}).
\citet{Pihlstrom03} have found an anti-correlation (albeit with a large scatter) between the source linear size and the \HI\ column density.
They derive a relation $N_{\rm \HI} = 7.2 \times 10^{19} L^{-0.43}$ cm$^{-2}$, with $L$ the linear size in kpc, using the data from their survey (see also \citealt{Vermeulen03a}). 
The  explanation put forward by these authors is that the relation is probing a surrounding medium  whose density decreases with radial distance from the centre. Under reasonable assumptions, a  spherical distribution of the absorbing gas does not produce the observed relation. A better model is a disk-like gas distribution in a plane roughly perpendicular to the radio source and with opening angle of 20$^{\circ}$. However, this explanation (described by the cartoon in Fig.\ \ref{fig:sizecolumn} right) is likely not the full story. The continuum morphology of CSS/GPS sources is not always this symmetric, the disk itself is likely not necessarily uniformly filled and is not always perpendicular to the radio source axis.
Furthermore, Fig.\ \ref{fig:sizecolumn} (left) shows the integrated optical depth  vs. linear size  for all objects (i.e. not only CSS/GPS) collected by \cite{Curran10}, including the upper limits. It can be seen that, deviations from the correlation do exist. For example, for a sample of High Frequency Peakers (HFP), which are extremely small sources   (i.e.\ with linear sizes $\leq 20$ pc) and considered to be radio AGN in their very first stages of evolution, only a low fraction of  \HI\ absorption detections was found \citep[see][]{Orienti06}. This is at odds with the expectations  from the above relation for such small scales, where high \HI\ column densities ($\geq 10^{21}$ cm$^{-2}$) would have been expected. 
On the other hand, and because of the small sizes of the sources, these findings could be explained if the line of sight is passing  mainly through the inner regions of the disk where the gas is likely to be mostly ionised, as discussed above. 
The correlation may also be affected by the limited sensitivity of some of the \HI\ absorption observations. For example, in the case of the ``baby'' source PKS~1718--63 having a size of only 2 pc (see \citealt{Veron00,Maccagni14}, more in \S \ref{sec:fuelling}), two narrow absorptions with $N_{\rm \HI} = 7 \times 10^{17} T_{\rm s}/c_{\rm f}$  cm$^{-2}$. This is lower than  expected from the anti-correlation between the source linear size and the \HI\ column density for a source of such a small size (see Fig.\ \ref{fig:sizecolumn}). 

An alternative explanation for this anti-correlation has been put forward by  \citet{Curran13a} who suggested that the higher detection rate in compact objects could result if the size of the sources and the size of the absorbers are very similar. This would indicate that the typical cross section of cold absorbing gas is of the order 100 - 1000 pc.
They remark on the possibility that the correlation is then the by-product of indirect relations between the column density and the covering factor which in turn is suggested to be related to the source size.  
The large scatter in the correlation suggests that, although this can affect the correlation, it is unlikely that this effect applies to every source.

Despite the high detection rate of \HI\ absorption and the high column densities observed, it is interesting to note that among compact/young objects, and the same is true for extended radio sources, there is a population  that appears to be poor in \HI. This is shown in Fig.\ \ref{fig:Stacking},  taken from \citet{Gereb14b}, where the stacked spectra are shown for compact (left) and of extended (right) sources. The result from the stacking of non-detections is indicated, in both plots, by a solid line.  It can be seen that, even if a large number of spectra of such non-detections are stacked, no \HI\ absorption is seen even among the compact sources, Fig.\ \ref{fig:Stacking} (left).
It appears, therefore, that there is a significant group of galaxies where either there is little \HI, or in which the conditions of the gas are different, for example, that the spin temperature is higher. Orientation effects play a role, but does not  seem to be fully able to explain the effect \citep{Maccagni17}. 
To further study this, \citet{Maccagni17} have stacked the central emission spectra of the  those early-type galaxies from the \atlas\ sample that were undetected in emission and found a detection of \HI\ emission  equivalent to a column density of $N_{\rm \HI} \sim 2.1 \times 10^{19}$  cm$^{-2}$. This is quite low, confirming the presence of a population of early-type galaxies with genuinely a limited amount of \HI. \HI\ at these low column densities likely has  \tspin\ well above 100 K \citep{Kanekar11} which makes it even harder to detect these low column densities in absorption. Indeed, the limits reached by the stacking of \HI\ absorption spectra discussed above is not yet low enough to detect such columns in absorption and this is one of the tasks for the future large surveys.

\subsection{\HI\ absorption in galaxy clusters}
\label{sec:clusters}

Some of the radio galaxies where \HI\ absorption has been found  (e.g.\ Hydra~A and Cygnus~A) are located in the centre of a galaxy cluster.  Cold gas is now often detected in clusters, in particular thanks to the improved sensitivity and resolution provided by ALMA. This is especially the case in the central regions of cool-core clusters (see  e.g. \citealt{Edge01,Salome06,Russell17} and refs. therein). In some cases the detected molecular gas is extended and likely represents the end product of hot gas lifted by the radio jets and cooling rapidly as result of instabilities. 

The information on \HI\ absorption in cluster galaxies is still limited and heterogeneous. As a result, very little is known about to what extent the observed properties depend on the dynamical state of a cluster or whether \HI\ absorption more often in, for example, cool-core clusters. The only attempt of a more systematic study of \HI\ absorption in clusters has been done by \cite{Hogan14}. This has addressed the presence of \HI\ absorption in samples (albeit small) of  Bright Central Galaxies in clusters. In general this study finds that many \HI\ absorption profiles of galaxies in clusters show the presence of a narrow (often redshifted) absorption line and of a broad absorption component at the systemic velocity.
The absorptions tend to have column densities covering a not too broad range, i.e.\ $N_{\rm \HI} < 3.0 \times 10^{21}$ cm$^{−2}$, for \tspin = 100 K. 
Following \cite{Hogan14}, the findings are broadly consistent with the majority of the broad absorptions being attributable to a central toroidal structure and the narrower components due to infalling clouds on their way to feed the AGN or to replenish the torus.

A number of studies focusing on single objects are also available. \HI\ absorption has been detected against the central radio galaxy of large clusters, for example in  Abell~2597 (PKS2322--123, see \citealt{ODea94,Taylor99}), in the Perseus cluster (3C~84/NGC1275, \citealt{Sijbring89,Jaffe90} and refs.\ therein), in Abell~1795 (4C~26.42, \citealt{Bemmel12}), and in Abell~407 (4C~35.06, \citealt{Shulevski15}). In addition, other cases of absorption in cluster galaxies are known like PKS~1353--341 \citep{Veron00}, Hydra~A \citep{Taylor96} and Cygnus~A \citep{Conway95,Struve10c}. 
For some of these objects, high spatial resolution, VLBI observations are also available giving more detailed information on the structure of the absorber (see \S \ref{sec:vlbi} for more details). 

In the extreme case of NGC 1275 in the Perseus Cluster, two absorbing systems have been found, at very different velocities. The so-called {\sl high velocity system} (discovered by \citealt{Young73}) is a relatively narrow \HI\ absorption line redshifted $\sim$3000 \kms\ from the systemic velocity of NGC~1275. Based on  optical narrow-band images \citep{Lynds70} and  recent VLA observations \citep{Momjian02}, this system is likely caused by a foreground galaxy, projected onto NGC 1275. 
The {\sl low velocity system} (discovered by \citealt{Crane82} and studied in some detail by \cite{Sijbring89,Jaffe90}) is, instead, centred on the systemic velocity of NGC~1275, at 5300 \kms. This absorption is broad, with FWHM $\sim$500  \kms\ \citep{Jaffe90}, larger than what is typically found in radio galaxies outside clusters.

The case of PKS~2322-12 in the Abell~2597 is similar \citep{Taylor99}. At low spatial resolution, the \HI\ absorption shows  a broad and a narrow  velocity component \citep{ODea94}. Using VLBI, \citet{Taylor99} showed that the narrow component is seen against both  jets and  against the nucleus, whereas the broad component is observed only against nucleus. They interpreted the broad component as a circumnuclear torus and the narrow component as inward streaming (cold) gas. This has been recently confirmed by the study of the molecular gas with ALMA, which finds a similar kinematics \citep{Tremblay16}. 

Thus, studies of \HI\ absorption in cluster galaxies, although they are still sparse in number, can provide complementary information to the molecular gas to complete the picture of the gas in these complex environments.

\section{Absorbing structures: fast outflows and kinematically disturbed  \HI\ gas}
\label{sec:kinematicallydisturbed}

Perhaps the most surprising result in recent years in the field of \HI\ absorption has been the discovery of  fast ($> 1000$ \kms) AGN-driven outflows of atomic hydrogen.  Such outflows were identified through the discovery of very broad, blueshifted and typically shallow absorption wings in the spectra of active galaxies. Such velocities are much larger than  typical rotational velocities and thus cannot be associated with rotating gas structures. The fact that these fast components  are blueshifted  relative to the systemic velocity clearly implies that the absorbing gas is outflowing. No redshifted components of similar widths have been detected so far (see also \S \ref{sec:fuelling}). 

AGN driven gas outflows are one of the manifestations of the effects of AGN on the  gas surrounding them. They are a key component in current models of  galaxy evolution because of their suspected role in regulating star formation as well as the growth of the central super massive black hole. The presence of  \HI\ (and also cold molecular gas)  in such outflows is perhaps counter-intuitive because of the large amounts of energy emitted by the AGN. The origin of the atomic gas (and more in general of the cold gas) associated with outflows is still a matter of debate.  The preferred explanation appears to be that this cold gas is the result of in-situ fast cooling after the gas has been shocked and accelerated \citep[e.g.,][]{Mellema02,Fragile04}, but the possibility of pre-existing  gas being accelerated has  also been put forward (see  \S \ref{sec:OtherPhasesGas} for more details).   Outflows of ionised gas have been known for a long time to be relatively common in AGN \citep{Tadhunter08,King15}. However, the discovery of the role of the \HI\ (and molecular gas) in these phenomena has broadened our view of AGN and of the conditions of the gas around them. All phases of the gas take part in the outflows and therefore, in order to get the full picture of their physical properties, multi-wavelength observations are needed, including those of  \HI\ \citep[see e.g.,][]{Morganti17}.

\begin{figure}
\centering
  \includegraphics[angle=0,width=0.45\textwidth]{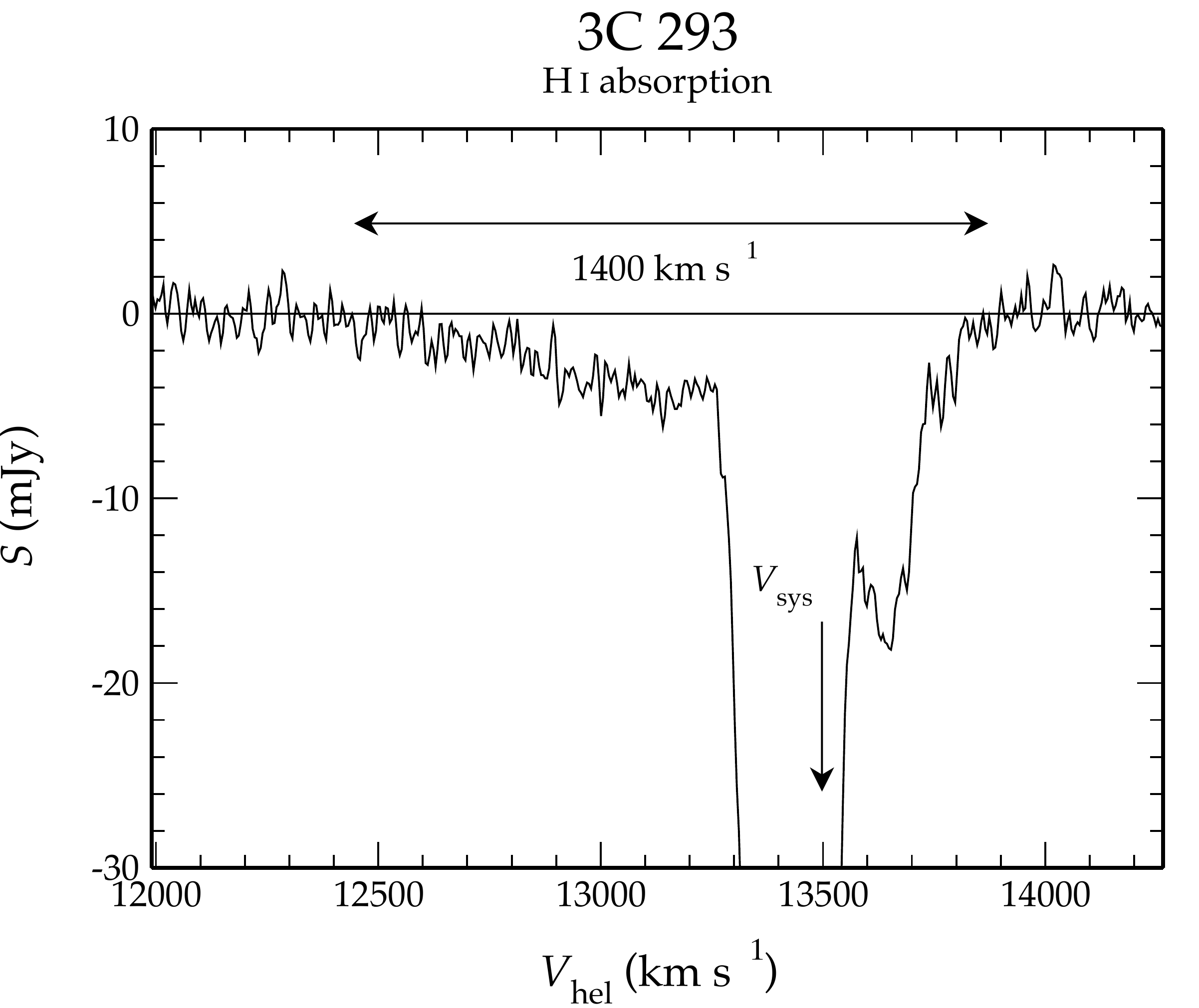}  
  \includegraphics[width=0.45\textwidth]{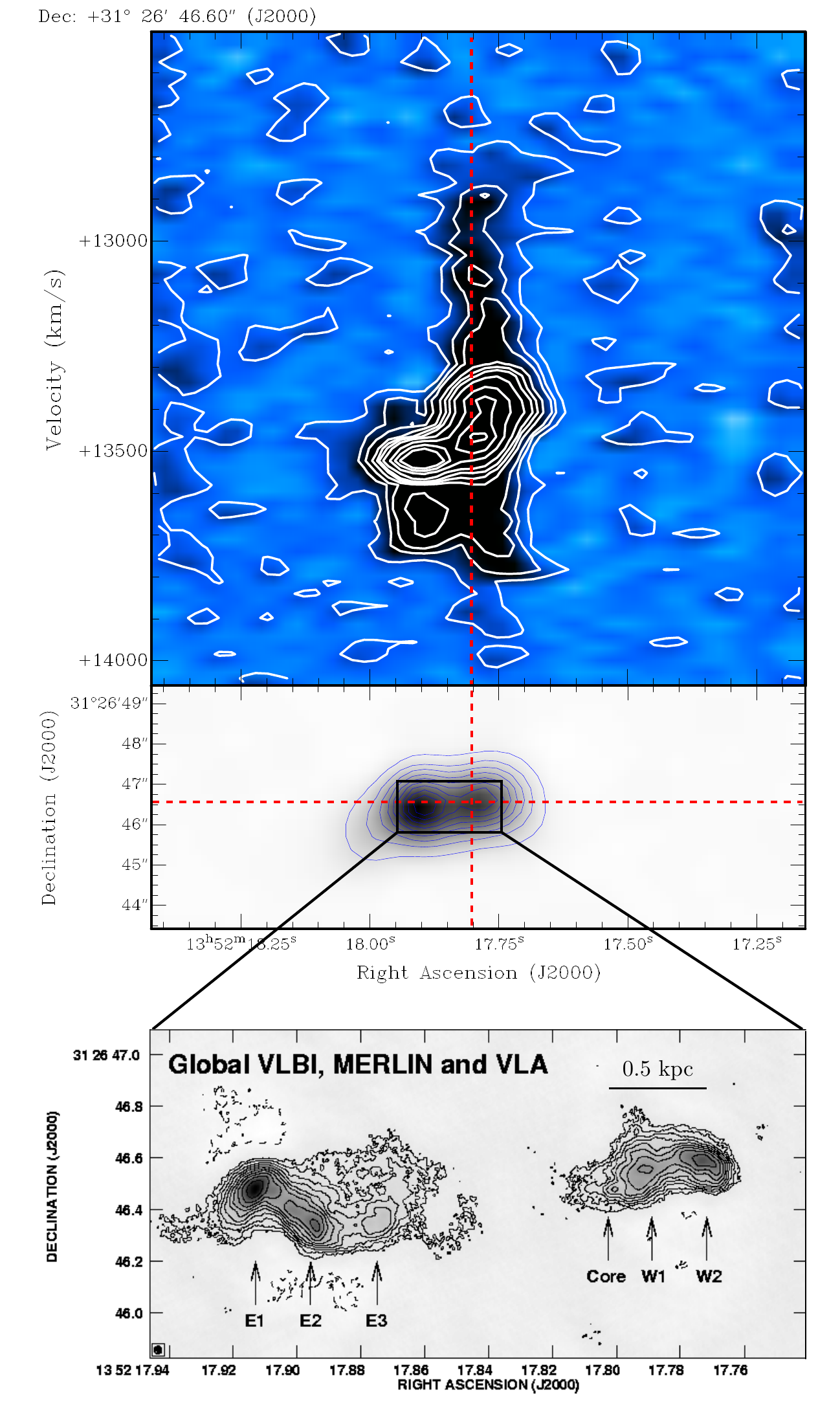}
\caption{
{\sl Left} Integrated \HI\ absorption profile obtained from the WSRT  \citep{Morganti03} showing a very broad, blueshifted wing, indicating a fast \HI\ outflow, in addition to the deep absorption profile near the systemic velocity due to the large-scale gas disk in the object. 
{\sl Right} 
Top: Position-velocity diagram extracted along the radio axis of the continuum source of 3C293 as obtained with the VLA, showing that the outflow is associated with the western lobe of this radio galaxy   \citep[from][]{Mahony13}. Middle: The continuum image of 3C293 from these VLA observations with a resolution of $~$1 arcsecond. The horizontal red line marks the axis along which the position-velocity diagram was extracted while the vertical red line indicates the position of the core as determined from the VLBI image in the bottom panel. Bottom: the combined 1.4-GHz global VLBI, MERLIN and VLA image of the central regions of 3C293 \citep[from][]{Beswick04}.
}
\label{fig:3C293outflow}       
\end{figure}

\subsection{Occurrence of fast \HI\ outflows }
\label{sec:outflows}

The discovery of the fast \HI\ outflows has been possible due to the larger instantaneous observing bandwidths (covering at least several thousands of \kms, see e.g.\ Fig.\ \ref{fig:3C293outflow} left) and the better spectral stability which have both become available as result of  upgrades of several radio telescopes. Earlier observations  failed to detect these fast outflows, partly because the velocity range covered  was simply not large enough, and partly due to  lack of sensitivity and/or stability. However, some indications for the AGN having an effect of the surrounding ISM had been found earlier as, 
for example, in the study by \citet{Vermeulen03a} of a sample of CSS/GPS  in which     the prevalence of blueshifted absorption in the \HI\ profiles was already emphasised, although  the  widths of these blueshifted components were smaller than those of the fast outflows found later.

\begin{figure}
\centering
\includegraphics[angle=0,width=0.9\textwidth]{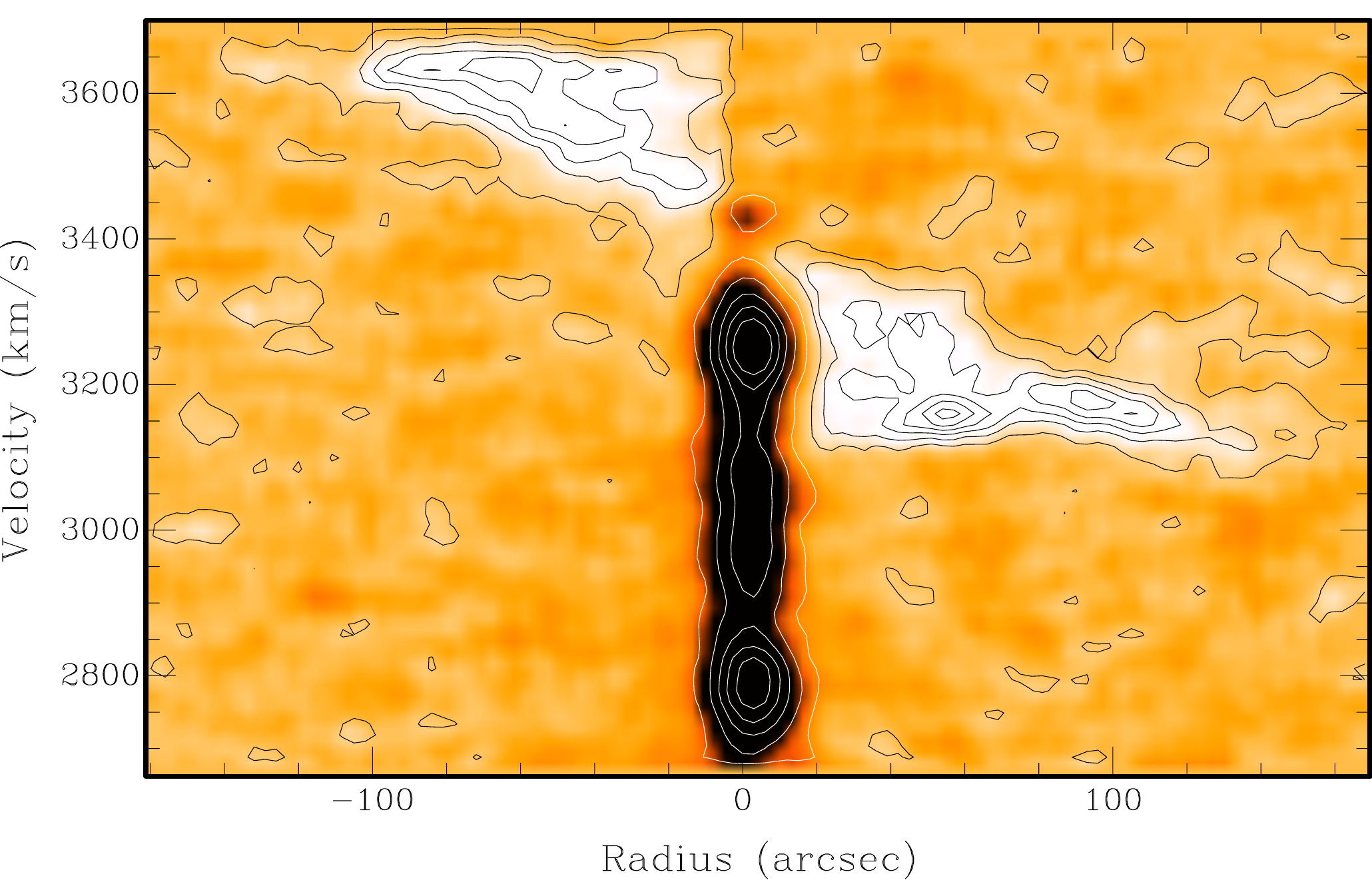}  
\caption{
 The position-velocity plot from the ATCA \HI\ data of IC~5063 taken along the  axis of the radio jet showing the first fast \HI\ discovered \citep{Morganti98}. The black contours represent the \HI\ emission coming from the  large-scale, rotating \HI\ disk of IC 5063 and indicate the range of velocities associated with normal rotation. The grey contours represent the absorption due to the \HI\ outflow and show that the blueshifted velocities of the outflowing gas extend to well outside the velocity range of    the regularly rotating gas. }
\label{fig:IC5063ATCA}       
\end{figure}

Since the first discovery of a fast \HI\ outflow \citep[][see Fig.\ \ref{fig:IC5063ATCA}]{Morganti98,Morganti03}, a growing number of cases have been reported. A summary of the   \HI\ outflows known  so far  is given in Table \ref{tab:TabOutflows}. Studies of \HI\ outflows have focused on three main issues: the occurrence of fast outflows, the localisation and physical parameters of the outflows, and the mechanisms driving them. 
Fast \HI\ outflows can only be observed using \HI\ absorption  due to the fact that, so far, these outflows appear to be limited to the central kpc, meaning they have an apparent size of at most one arcsecond. The column density sensitivity of current (and future) radio telescopes is insufficient for detecting the emission from these outflows at the required resolutions (as described in \S \ref{sec:TelescopesLimitations}).

An example of one of the first \HI\ outflows detected in a radio galaxy is 3C293, shown in Fig.\ \ref{fig:3C293outflow}. On the left is shown the integrated absorption profile detected against the unresolved central radio source of this object, using the WSRT \citep{Morganti03}. The profile shows the broad blueshifted wing with velocities which clearly exceed the velocity range of the regularly rotating gas disk indicated by the extent of the deep absorption. The large blueshift of the wing is an unambiguous signature that the gas is outflowing. The position-velocity diagram from higher spatial resolution observations (VLA, $\sim$1 arcsec resolution, \citealt{Mahony13}) shows that much of the outflow is occurring against the western radio lobe of 3C293 which implies that the outflow extends to at least 0.5 kpc from the nucleus.

\begin{figure}
\centering
\includegraphics[angle=0,width=0.6\textwidth]{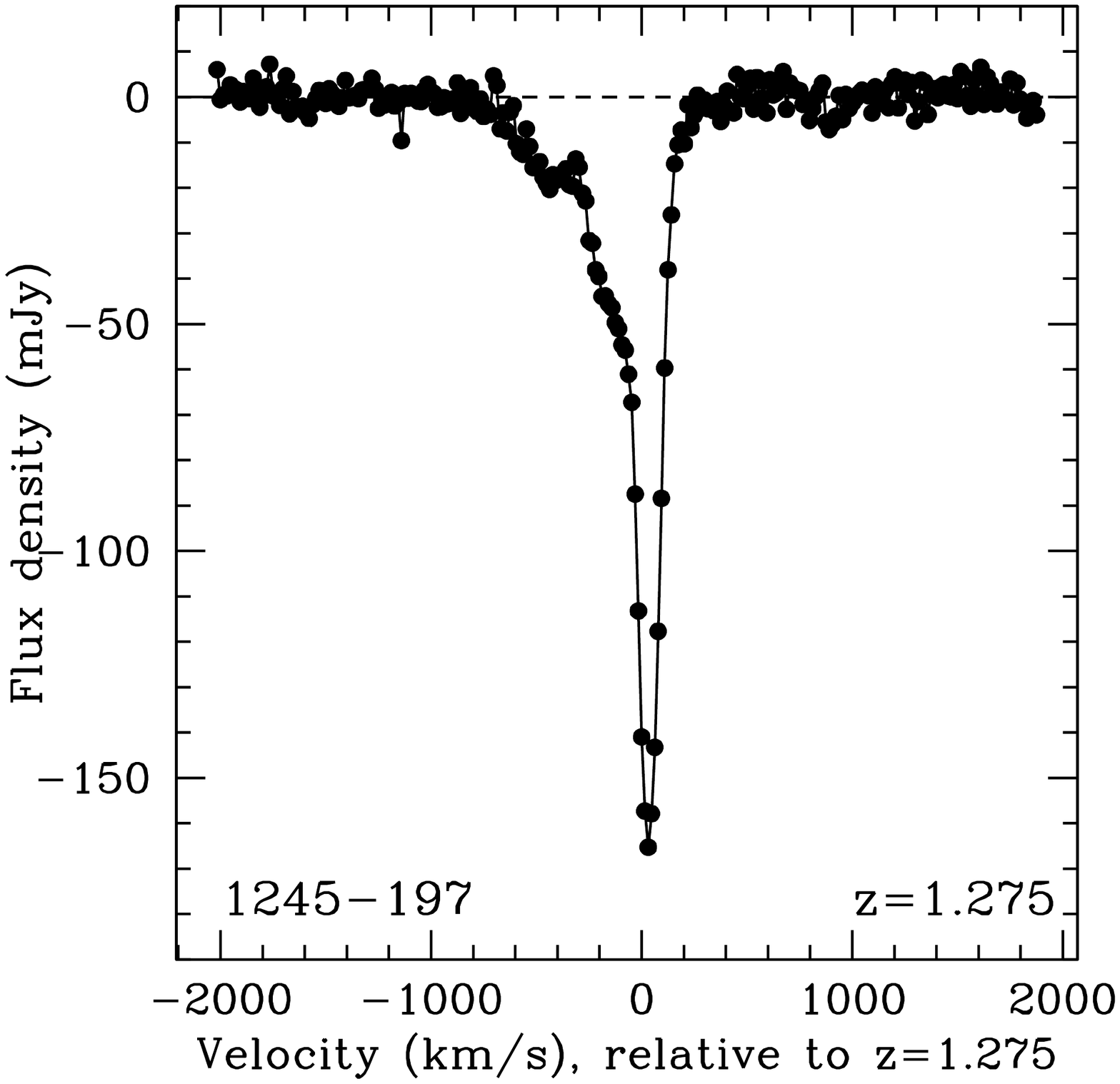}  
\caption{the GMRT \HI\ 21 cm absorption spectrum towards TXS 1245−197 at z = 1.275, 
showing a broad, blueshifted component extending more than 800 \kms, taken from \cite{Aditya18}.}
\label{fig:Aditya1245}       
\end{figure}

The only attempt done so far to quantify how common \HI\ outflows are has been based on  the survey  presented in \citet{Gereb15} and \citet{Maccagni17}. Of the galaxies detected in \HI\ absorption, at least 15\% show fast outflows which, given the overall detection rate of \HI\ absorption of 30\% in their sample, translates to about 5\% of all radio sources having fast outflows. This is a lower limit because of the limited sensitivity of the survey, but also because not every outflow will be located in front of the radio continuum and  will not be detected in absorption. From the above survey, but also from other  studies \citep[e.g.,][and refs. therein]{Morganti05a,Aditya18}, it is now clear that in particular young radio galaxies, as well as restarted radio galaxies, are the objects where  \HI\ outflows are more likely to occur.  This is expected given the small sizes of such radio sources and their dense ISM. Interestingly, the highest redshift outflow observed so far (from \cite{Aditya18}, see Fig.\ \ref{fig:Aditya1245}) is also found in a GPS source.

In addition, a possible correlation  between the velocity of the outflow and the radio power of the AGN  has been found \citep{Maccagni17}.  This is consistent with the observation that most of the \HI\ outflows are found among relatively high-power radio sources  (\citealt{Gereb15}; 4C~31.04 from the WSRT Morganti et al.\ in prep.). However, also low-power radio sources can provide a driver for gas outflows and examples are NGC 1266 \citep{Alatalo11,Nyland13} NGC 1433 \citep{Combes13}, IC 5063 \citep{Oosterloo00} and Mrk 231 \citep{Morganti16}. Interestingly, these objects are commonly classified as radio-quiet objects, but  IC 5063 is now nevertheless one of the best cases known showing the relevance of the jet for driving the outflow. 

Interesting are also the results derived from \HI\ absorption for interacting  and FIR-bright galaxies and ULIRG objects.  A high fraction of \HI\ outflows (with velocities  exceeding 1500 \kms\ in some cases) have been found using the GBT in  nearby ULIRGs and quasars studied by \citet{Teng13}. 
Very intriguing is that they find  polarisation-dependent \HI\ absorption  which they  attribute   to absorption of polarised continuum emission    by foreground \HI\ clouds. This result  still needs confirmation from interferometric observations. 

\subsection{Location and other parameters of the \HI\ outflows}
\label{sec:location}

In a  (unfortunately still limited) number of cases,  the spatial resolution of the data  is insufficient to determine the location of the outflow. The first object where an \HI\ outflow was discovered is the Seyfert 2 galaxy IC~5063 \citep{Morganti98}. This object is one of the few Seyfert galaxies with  relatively strong radio continuum. The radio morphology  shows a core and two lobes, one of which is ending in a particularly radio bright hot spot, about 500 pc from the nucleus.
In this galaxy,  highly blueshifted absorption ($\sim 700$ \kms, see Fig.\ \ref{fig:IC5063ATCA}) was detected using the ATCA. The spatial resolution of these ATCA data were not sufficient to determine the location of the outflow, but later VLBI observations allowed to establish that the location of the fastest part of the outflow  is coincident with the bright hot spot mentioned above, around 0.5 kpc from the nucleus \citep{Oosterloo00}. This observation was the first indication that  the jet, and not the central source,  could play a role in driving a fast \HI\ outflow. Later observations of the molecular gas in this object have shown that, although the fastest outflow occurs at the bright hotspot at the end of the jet, the region with outflowing gas exactly coincides with the entire radio jet, reinforcing the idea that the jet is driving the outflow (see also \S \ref{sec:OtherPhasesGas}).

\begin{figure*}
\centering
  \includegraphics[angle=0,width=1\textwidth]{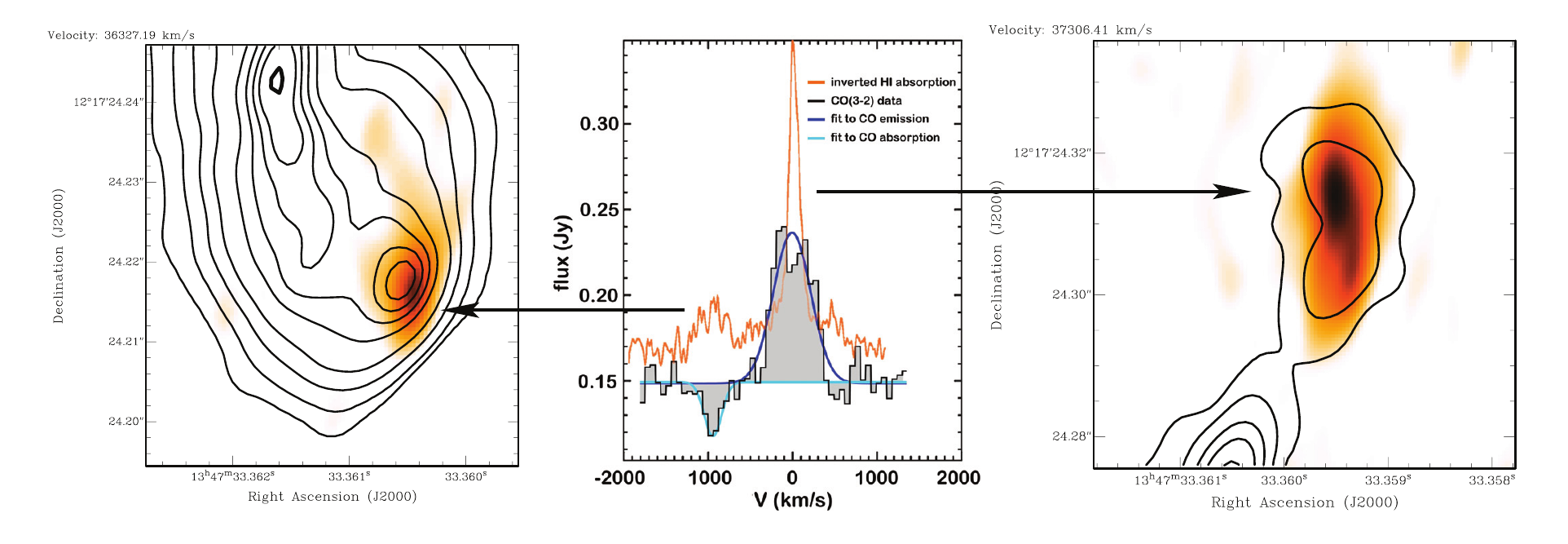}  
\caption{The distribution of the \HI\ absorption  (orange-white) in two velocity channels for the source 4C 12.50, superimposed to   the continuum emission (contours) showing the location of the two   \HI\ clouds detected in absorption (from \citealt{Morganti13}). The inverted integrated \HI\ absorption profile is shown in the central panel and clearly shows a strong, blueshifted \HI\ outflow of $\sim 1000$ \kms\ (from \citealt{Morganti05a}).  The cloud shown in the right panel has a narrow absorption profile near the systemic velocity, while the left panel shows the location of \HI\ absorption associated with the fast \HI\ outflow,  co-spatial (in projection) with the bright radio hot spot. Also shown is the   CO profile (taken from \citealt{Dasyra12} which  shows that a very similar outflow is detected in CO absorption.  }
\label{fig:4C12.50outflow}      
\end{figure*}

Also in the case of 3C~293  (see Fig.\ \ref{fig:3C293outflow} right), the strongest outflow of $\sim 1000$ \kms\ is seen against the western inner radio lobe, about 0.5 kpc in size \citep{Morganti03,Mahony13}. Similarly, the outflow observed in 3C~305 is  located about 0.5 kpc from the core  \citep{Morganti05b}.
 A very  convincing example for the outflows being off-nucleus  is the case of 4C~12.50, a young (possibly restarted, see \citealt{Stanghellini05}), IR-bright radio galaxy. In this object, the fast outflow \HI\ has been found to be located at the end of a radio lobe, and is wrapped around a bright hot spot in this lobe (\citealt{Morganti13}, see Fig. \ref{fig:4C12.50outflow}).

Thus, in a significant number of cases,  the spatial coincidence of the outflow with bright, non-nuclear continuum features suggests that the outflow  likely originates from the interaction between the expanding radio plasma and the ISM.  This is in agreement with what is now predicted by numerical simulations \citep{Wagner11,Wagner12,Mukherjee16,Mukherjee18a}. Such  simulations describe the effects of a newly formed radio jet  trying to find its way through a dense and clumpy gaseous  medium. Such a porous medium with dense clumps forces the jet to find the path of least resistance. While doing so, the jet interacts with, and gradually disperses, the dense clouds away from the jet axis, forming a turbulent cocoon of over-pressured gas. In this way, clouds can be accelerated to high velocities and over a wide range of directions, not only in the direction of the jet movement, but also away from the jet axis.

However, not in every radio source  the jet is the main driver of the outflow. For example, in the case of Mrk~231  an \HI\ outflow is present which reaches velocities up to 1300  \kms\ blueshifted compared to the systemic velocity and which has properties very similar to outflows found in other  radio galaxies. However,  the energetics of the radio source, and the lack of a clear kpc-scale jet, suggest that the most likely origin of the \HI\ outflow is a wide-angle nuclear wind, as earlier proposed for this object to explain the neutral outflow traced by Na I and by molecular gas. This suggests that different mechanisms can produce an \HI\ outflow, similarly to what is  found for  other phases of the gas (see \S \ref{sec:OtherPhasesGas}, see also \citealt{Tadhunter14}).  

The mass outflow rate is an important parameter to be able to judge the impact the outflow may have on the host galaxy.  The estimate of this parameter can be quite uncertain, in particular if we don't have location and extent of the \HI\ outflow. This is in any case always limited by the size of the radio continuum, which can be smaller than the scale of the outflow, setting a lower limit to the outflow rate derived. 
The \HI\ mass outflow rate can be
estimated following \citet{Heckman02} and \citet{Rupke02}:
\begin{equation}
\dot{M} = 30\cdot  {{\Omega}\over{4\pi}}\cdot {{r_*}\over{\rm 1\, kpc}}\cdot
{N_{\rm HI}\over{10^{21}\, {\rm cm}^{-2}}}\cdot { v \over 300\, {\rm km
s}^{-1}} \ M_\odot\, {\rm yr}^{-1}
\end{equation}
 and for the known cases is listed in Table 1. Here it is assumed that the gas is flowing at a velocity equal to the FWZI/2 of the blueshifted component,  at an observed radius $r_*$ and 
into a solid angle $\Omega$ assumed to be $\pi$ steradians.  

The  mass outflow rates of the \HI\ outflows range between a few and a few tens \msunyr\ (see Table \ref{tab:TabOutflows}). For radio galaxies the number of outflows known is sufficient to draw some statistical conclusions and their \HI\ outflow rates typically appear to be  higher than those associated with the outflow of  ionised gas. For the very few objects where this comparison can be done directly, the  outflow rate of the \HI\ outflow appears  to be lower (or at most comparable) with those derived  for the molecular gas, but such comparisons will need to be done for  more objects before solid conclusions can be drawn.  
 
Interestingly, the study of outflows on pc-scales using VLBI observations have shown also other interesting features of the \HI\ associated with the outflow. First of all, in some objects  outflowing gas is also found very close to the active BH (a few tens of pc in projection; 3C~236 and 3C~293, \citealt{Schulz18}).  
Furthermore, the detected outflows appear often to be clumpy with cloud sizes of tens of pc  and \HI\ masses of the order of $10^4$ \msun. Such a clumpy medium has been traced in a few cases (e.g.\ 4C~12.50, 3C~236, 3C~293) and it supports the predictions made by numerical simulations where it is indeed the clumpiness of the medium that  enhances the impact of the jet because the jet is meandering through the ISM to find the path of minimum resistance and so creating a cocoon of shocked  medium from where the outflow originates.

\begin{table}
\begin{center}
 \caption{Table listing known (and confirmed)  broad \HI\ absorption associated with outflows (first part of the list) and broad absorption associated with mergers and binary black holes (second part of the list).  }
\label{tab:TabOutflows}       
\begin{tabular}{llll}
\hline\noalign{\smallskip}
Source  name & FWZI & $\dot{M}$ & Ref.  \\
             &   \kms        & $M_\odot$
yr$^{-1}$ & \\
\noalign{\smallskip}\hline\noalign{\smallskip}
IC~5063 & 750 & 35 & \cite{Morganti98} \\
3C~190 & 600  & - & \cite{Chandra03} \\
3C~293 & 1400 & 8-50 &\cite{Morganti03,Mahony13} \\
3C~305 & 800 & 12&\cite{Morganti05a,Morganti05b} \\
3C~236 & 1400 & 47&\cite{Morganti05a} \\
4C~12.50 & 1400 & 8-21&\cite{Morganti05a,Morganti13} \\
NGC~1266  & 400  & 13$^1$ &\cite{Alatalo11,Nyland13}  \\
OQ~208  & 1200  & 1.2&\cite{Morganti05a} \\
4C~52+37  & 674$^2$  &  4 & \cite{Gereb15}, Schulz et al. in prep. \\
3C~459  & 600  &  5.5&\cite{Morganti05a} \\
4C~31.06   & 700  & - & unpublished (Morganti in prep) \\
1504+377  & 600  & 12 & \cite{Kanekar08} \\
TXS~1200+045    &  600 & 32 &  \cite{Aditya18} \\
TXS~$1245-197$ & 1100   & 18 & \cite{Aditya18}  \\
Mrk~231  & 1300  &  8-18 &\cite{Morganti16},Teng13 \\
NGC~3079  &  600 &  0.2-2.5 & \cite{Gallimore94,Shafi15} \\ 
NGC~1068  & 300$^3$    & - &  \cite{Gallimore94} \\
\noalign{\smallskip}
\hline
\noalign{\smallskip}
    4C37.11 & 1200 & - &candidate binary BH \cite{Maness04}\\
     & & &\cite{Morganti09b,Rodriguez09} \\
    NGC~6420 & 780 & - & \cite{Baan85,Beswick01} \\
\noalign{\smallskip}\hline
\\
\end{tabular}
\end{center}
$^1$ M$_{{\rm \HI}+{\rm H}_2,outflow}$ from \cite{Alatalo11}\\
$^2$ FW20, from \cite{Gereb15} \\
$^3$ Tentative, see \cite{Gallimore94} 
\end{table}

\subsection{Irregular  \HI\ structures }
\label{sec:disturbed}

Finally, there are situations where  \HI\ absorption is not expected to occur, but  is nevertheless detected. In such cases, even if \HI\ is not outflowing, it is likely not in a settled configuration and is  indicating the presence of extreme conditions.

An example of this is the presence of \HI\ absorption in  quasars.  Despite the fact that these objects are expected to be pointing close to the line-of-sight and, therefore, not  highly obscured, associated absorption has been reported for a few cases (e.g. \citealt{Pihlstrom99,Chandra03}) including red-quasars (see e.g. \citealt{Carilli98b,Yan16}). In the red quasar 3C190, the detected \HI\ absorption appears to be also broad (with a width of 600 \kms, \citealt{Chandra03} suggesting an ongoing interaction between the jet and the ISM. This is not the only case where the  quasars can still be  embedded and interacting with a cocoon of gas perhaps originating from the event that has fuelled their central SMBH. This scenario has also been proposed  for the radio source PKS~1549--79 \citep{Holt06}. For this object, an \HI\ profile with only moderate width ($\sim 300$ \kms) is observed. However, the presence of a fast outflow (with velocities exceeding 1000 \kms) is traced by the ionised gas \citep{Holt06} and, more recently, by the molecular gas observed with ALMA (Oosterloo et al.\ in prep).   
Interesting is also the case of, although not extremely broad ($< 100$ \kms),  \HI\ absorption  blueshifted by about 300 \kms, observed in a blazar object (TXS 1954+513, \citep{Aditya17}), likely the result of a recent realignment of the jet axis possibly connected with restarting activity. 

A major implication of these results is that the simplest version of the unified schemes may not always hold for young, compact radio sources and confirms that in these objects the  jet is interacting with the surrounding medium.

Finally, some other examples of gas structures with extreme kinematics have been found, often with  extremely broad ($\geq 800$ \kms) profiles and detected in objects either undergoing major mergers and, in some cases, showing double nuclei like   NGC~6240 (see also \S \ref{sec:seyferts} for more details), or candidate hosts of binary black holes such as 0402+379 (see \citealt{Morganti09b,Rodriguez09}). 

\section{Tracing the fuelling of the AGN}
\label{sec:fuelling}

Given that \HI\ absorption spectra can sample the immediate environment of a SMBH,  the search for signatures of  gas fuelling the SMBH has  been a central theme  since the earliest work on \HI\ absorption. If clouds of cold gas are involved in 'feeding the monster', these clouds could, in principle, be observable as redshifted \HI\ absorption seen against the central radio core.
Initial results were encouraging as they seemed to indicate that redshifted \HI\ absorption was more common than absorption at the systemic velocity and even more so than blueshifted absorption    \citep{Gorkom89}. However, subsequent studies on much larger samples have now shown a clear prevalence of blueshifted  (i.e.\ outflows) over redshifted absorption (i.e.\ infall). This is particularly evident in samples of young CSS/GPS sources (e.g.\ \citealt{Vermeulen03a}). More recent studies (e.g.\ \citealt{Gereb15,Maccagni17}) also show that if the profiles show  \HI\ gas deviating from regular rotation, this 'anomalous gas' is mostly seen as blueshifted absorption. The prevalence of blueshifted absorption is likely connected to AGN-driven outflows. Cases of \HI\ absorption with large redshifted velocities do exist, but they are rare (see e.g.\ Fig.\ 5 in \citealt{Maccagni17}) and if they are detected, they are deviating less from the systemic velocity than many blueshifted absorptions. The reason for this is still not completely understood. Perhaps \HI\ is not involved in  feeding the SMBH, but even if it is, it may not be easy to identify redshifted absorption as infalling. Contrary to gas outflows (i.e.\ blueshifted absorption), for which the velocities involved are determined by the energy input of the AGN and which well exceed the typical gravitational velocity, infall velocities should only slightly exceed the rotational velocities of a regular circumnuclear  disk so that the absorption of infalling clouds blends with the absorption of the circumnuclear disk.

\begin{figure*}
\centering
  \includegraphics[angle=0,width=1\textwidth]{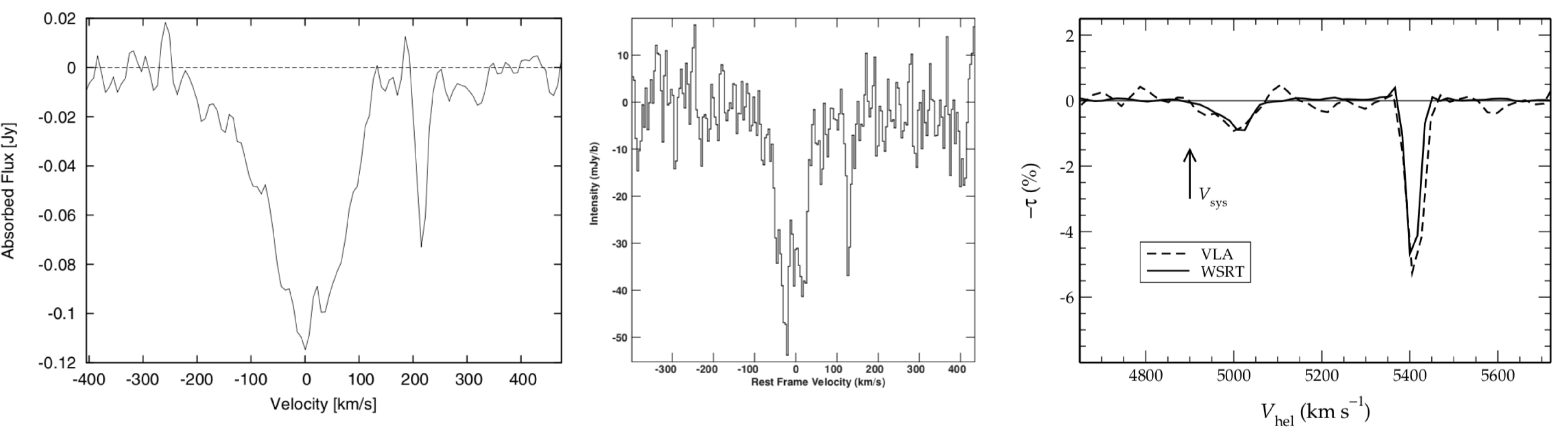}  
\caption{Examples of radio galaxies where narrows, redshifted \HI\ absorption are seen likely pinpointing the presence of infalling clouds connected with the feeding of the SMBH.  Left: integrated absorption spectrum of 4C 31.04 with the velocities shifted to the rest frame of the host galaxy taken from \cite{Struve12}.
Middle: 
\HI\ absorption detected toward the CSO B2352+495 after continuum subtraction. The systemic velocity of the system (V$_{\rm Hel} = 71321 \pm
48$ \kms) is almost at the centroid of the broad absorption feature.  Taken from \cite{Araya10}. Right: \HI\ profile (in optical depth) of the giant radio galaxy NGC~315 obtained from the WSRT (solid line) and the VLA (dashed) data. The two \HI\ absorption systems are clearly visible (see text for details). The systemic velocity is also indicated.  From \cite{Morganti09}.}
\label{fig:InfallingClouds}      
\end{figure*}

At least in some objects, \HI\ is likely connected to feeding the SMBH and infalling clouds of \HI\ have been detected. Often this is indicated by the presence of narrow absorption components redshifted compared to the systemic velocity of the target, some examples are showed in Fig.\ \ref{fig:InfallingClouds}.
An example is  observed in 4C~31.04 \citep{Conway96,Struve12} for which results from  VLBA observations (\citealt{Conway99}, see \S \ref{sec:vlbi}) confirm that its size (5-10 pc) and opacities  are similar to those of Galactic High Velocity Clouds which are located in the halo of the Milky Way.  Similar results have been found in data of the compact symmetric object B2352+495 \citep{ Araya10} which shows  redshifted narrow absorption. The VLBA observations reported by \citet{Araya10} were not able to localise this component, but the idea of an infalling cloud is the most likely. 

However, the situation can be more complex. One interesting case is seen NGC~315.  Like in 4C~31.04 and B2352+495, a deep, narrow \HI\ absorption line, redshifted by about 500 \kms\ from the systemic velocity, was discovered by \citet{Dressel83} using Arecibo against the inner regions of this giant radio galaxy. The presence of fainter, broader \HI\ absorption closer to the systemic velocity was later revealed by deeper observations \citep{Morganti09}. Interestingly, this component is also redshift, albeit only of only $\sim 80$ \kms\ compared to the systemic velocity (see Fig.\ \ref{fig:InfallingClouds}. When traced by the VLBI (i.e.\ parsec-scales), the two absorption systems have very different structure. The narrow, redshifted absorption smoothly covers most of the central continuum (the core and the inner 10 pc of the jet) and does not show any morphological or kinematic features. This suggests that this component is likely caused by clouds falling into NGC~315, not necessarily located  extremely close to the AGN.  Interestingly, the broad component centred on the systemic velocity  shows a velocity- as well as a column density gradient  (on the scale of only $\sim$10 pc)  suggesting that the gas has been influenced by the AGN and that it is located physically close to it. Because of these properties and other considerations, it has been suggested that this gas is {\sl also} involved in fuelling  the AGN. If NGC 315 is representative, these data indicate that only observations at high spatial resolution (pc-scales) are able to determine whether gas is close to the AGN and falling into the SMBH. 

A similar situation is seen in PKS~2322--123, where identifying the infalling gas could be done in even more detail by the combination of \HI\ and of  molecular gas \citep{Tremblay16}. In this object, both a narrow (110 \kms\ FWHM) and a very broad (735 \kms\ FWHM) line are seen  redshifted ($\sim 220 \pm  100$ \kms) with respect to the systemic velocity. Thanks to  VLBI observations, the location of the components could be determined  with the very broad line seen only against the core while the narrow is found about 20 pc from the core  \citep{ODea94,Taylor99}. These changes on such a small scale suggest that we are looking at gas very close to the nucleus and that the kinematics of the gas can be explained as the result of an atomic torus centred on the nucleus with considerable turbulence and inward streaming motions suggesting infalling gas. This has been nicely confirmed by the ALMA observations of the same object showing absorption due to molecular clouds with similar properties (width and velocities) to the one of the \HI\ (see \citealt{Tremblay16}). The combination of the two diagnostics gives strong support to the scenario of inward-moving, clumpy distribution of molecular clouds within a few parsec.

Another extremely interesting case is the very young ($\sim$100 yr) radio galaxy PKS~1718--63 where very narrow absorption lines (both redshifted and blueshifted) were detected. This radio source is only 2~pc in size and the presence of narrow absorption components at different velocities  suggests  they trace unsettled gas clouds in the inner regions \citep{Veron00,Maccagni14}. Extra constraints about their nature come from the detection of redshifted absorption and emission due to molecular clouds in this object \citep{Maccagni16,Maccagni18}.  Thus, also in this object,  both atomic and molecular gas show signatures  of clouds that can be involved in  feeding  the recently activated SMBH. 

The discovery of such small clouds with unsettled velocities supports the scenario of AGN fuelling known as chaotic cold accretion, now predicted by  several numerical simulations \citep{King15,Gaspari13,Gaspari17a,Gaspari17b}. Such models
 suggest that small clouds of multi-phase gas can condense out of turbulent eddies in the hot gas   of the halo of the host galaxy.
Angular momentum cancellation occurs via tidal stress and cloud-cloud collisions,  resulting in significant inward radial motions of these clouds.

\section{Evolution with redshift, effect of UV radiation}
\label{sec:highredshift}

Given the importance of gas for the evolution of galaxies, it is not surprising that observing and understanding the changes of the properties of \HI\ absorption  with redshift has been a key  goal of many absorption studies. 
Exploring the high-$z$ Universe using \HI\ absorption has  so far been difficult because it is hampered by RFI and by the relatively poor performance  of  radio telescopes at low frequencies (sensitivity, but also bandpass stability and the presence of standing waves which makes obtaining flat spectral baselines difficult). Published studies have  mostly used the GMRT and the GBT.
In contrast to the over 150 detections at $z \leq 1$,  there are only 7 detections of associated \HI\ 21-cm absorption above this redshift   \citep{Chandra03,Curran13c,Aditya17,Aditya18} and only two of those are  at  $z > 2$ (one of which is even a tentative detection; \citealt{Uson91,Aditya16}). 

The few published studies of small samples of objects at high redshift use a variety of selection criteria for the targets. Observed targets include compact radio sources and GPS \citep{Gupta06a, Aditya18}, radio quasars and flat spectrum sources \citep{Curran08,Aditya16}, and radio sources selected for their blue magnitude or optical faintness \citep{Curran13b,Curran13c}.
The heterogeneity of the samples, and the limited number of objects, makes it particularly difficult to derive strong conclusions about the properties of \HI\ absorption at $z>1$. Furthermore, the quality of the spectra available is, in most cases, lower than what can be obtained at low redshift. Therefore,  a fair comparison with findings at low redshift is still difficult. 
  
Despite these limitations, \citet{Curran08} were the first to suggest a lower detection rate for  sources at  $z>1$ compared to sources below that redshift. This was based on obtaining no detections for all of the ten $z > 3$ sources searched. This study has been followed by other work  which, nevertheless, has not been able to fully clarify whether an evolution of the detection rate with redshift is actually present. 
While a low detection rate was found in the high-$z$ samples by \citet{Curran13b} and \citet{Curran13c}, the significance of this effect was not fully confirmed by the study of compact flat-spectrum sources - at redshift around 3 - selected from the Caltech-Jodrell Bank sample by \citet{Aditya16}.  
They compare their findings with a lower redshift sample of similarly selected objects reporting  marginal evidence for a lower detection rate for the higher redshift sources.  The significance (or not) of the trend relies on a tentative detection found at redshift 3.53. In the study of GPS of \citet{Aditya17}, no trend is found when they compare the detection rate with similar studies at low redshift. The final word is likely to come from the new surveys that will be done with the SKA pathfinders (see \S \ref{sec:future}). 
 
Nevertheless, it is worth discussing whether a possible trend with redshift is due to either limitations of the observations, or due to real physical differences in the high-$z$ objects.
One issue is that the high-$z$  samples typically contain objects of higher rest frame luminosities than the sources of the comparison samples at low redshift, in both the radio and the UV wavebands. The population of radio sources studied at high redshift is dominated by very powerful sources ($>10^{26.5}$ W/Hz) while at low redshift the studied sources are at much lower radio power (see Fig.\ 4 in \citealt{Curran18}). One should also keep in mind that  the more powerful sources tend to have relatively faint radio cores, making it difficult to reach low optical depths if the absorption is  against the core. 

\citet{Curran08} noted that a high AGN luminosity in either the radio- or the UV band might result in a lower strength of \HI\ absorption. They pointed out that near an AGN  with high UV luminosity, the excitation, as well as the ionisation of the hydrogen gas,  may be affected such that it results in a lower incidence of \HI\ absorption.  Both redshift evolution and  a dependence of local conditions on the AGN UV luminosity were found to be a viable explanations for the observed low  occurrence of associated \HI\ absorption in high-$z$ AGN \citep{Curran08,Curran13b}. 
They also suggested that for high redshifts there is another selection effect, such that  the faint optical targets ($B \geq 19$) were in fact UV bright in the source rest-frame, thus ionising/exciting the gas to below the detection limit. They suggest that a critical UV luminosity of $L_{\rm UV} \sim 10^{23}$  W Hz$^{-1}$ exists above which these effects occur. 
Interestingly, the GPS at redshift 1.275 where recently \HI\ absorption was detected by \cite{Aditya18} has a high UV luminosity ($> 10^{23}$ ) as well as an \HI\ outflow with properties similar to  found for objects at low redshift.

If the evidence for redshift evolution in the strength of associated \HI\ absorption is confirmed, alternative effects should also be considered as discussed by \citet{Aditya16}. For example, one possibility is that ratio CNM to WNM in  high-$z$ objects is typically lower resulting in  a   higher spin temperature on average.  Evidence for an increase of \tspin\ with redshift has been obtained by the study of intervening damped Lyman-$\alpha$ systems (e.g.\ \citealt{Kanekar14} and refs.\ therein) suggesting a larger fraction of the warm neutral medium. Although the regions of the ISM sampled by damped Lyman-$\alpha$ systems are typically at larger galactic radii and hence not near an AGN, a similar trend could also be present in the environments of high-redshift AGN \citep{Aditya16}.

\begin{figure}
\centering
\includegraphics[angle=0,width=0.95\textwidth]{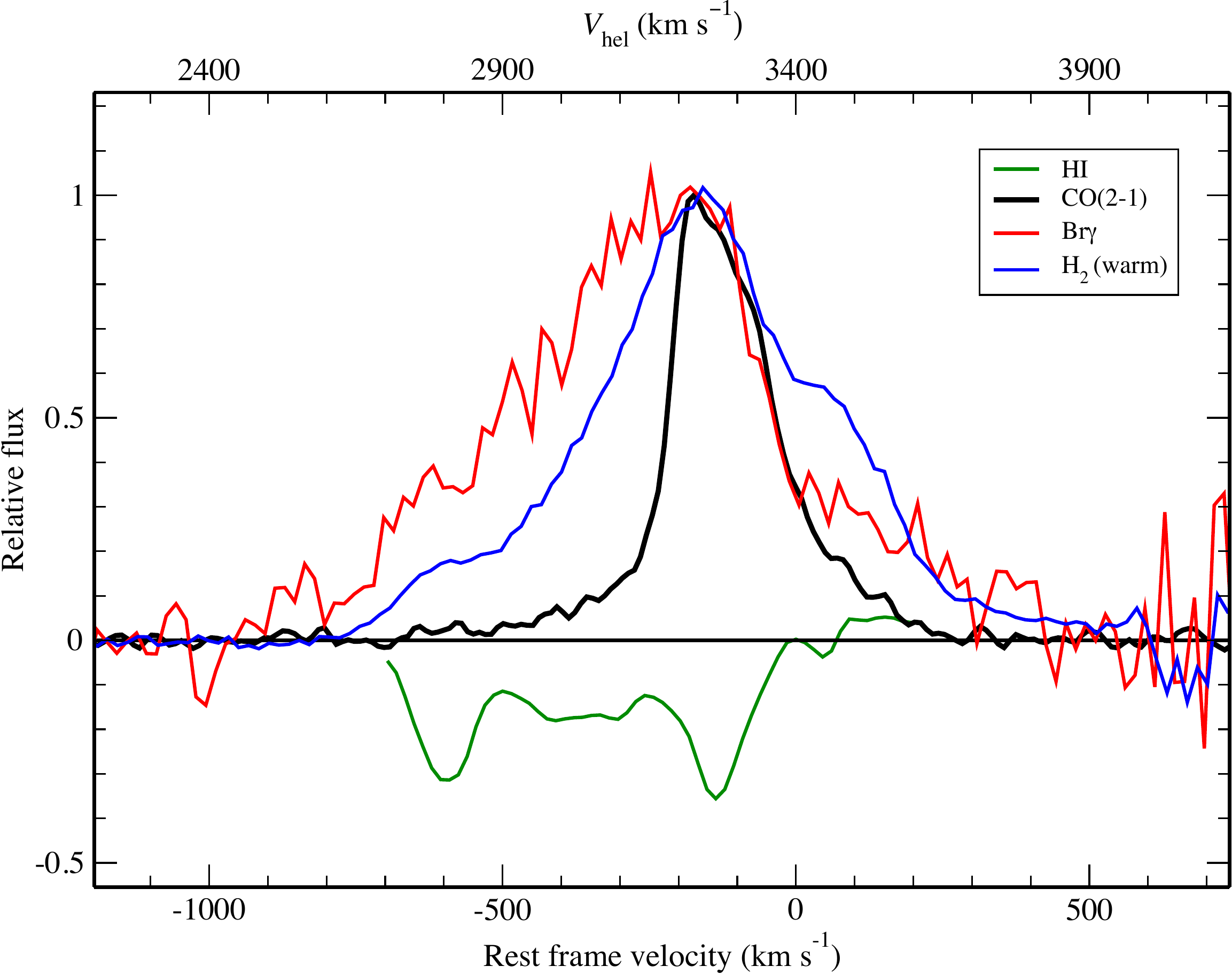}  
\caption{
Velocity profiles of the  multiphase outflow in IC~5063. Shown are H$_2$ 1-0 S(1) (blue line) and Brackett$\gamma$ (red  line) velocity profiles for the western lobe of IC 5063 (from \citealt{Tadhunter14}), the spatially integrated \HI\ absorption (green line, \citealt{Morganti98}) and integrated CO(2-1) spectrum (black line, \citealt{Morganti15b}). Although the detailed shapes differ greatly,  all phases show the same extent for their blueshifted component indicating they participate in the same outflow. The flux scaling between the different profiles is arbitrary. }
\label{fig:IC5063}       
\end{figure}

\section{Connection with other phases of the gas}
\label{sec:OtherPhasesGas}

Obtaining a complete view of the ISM near  AGN requires  information for all phases of the gas. For this reason, the results obtained from \HI\ absorption are, in most cases, used in combination with other diagnostics, such as   ionised gas, and warm and cold molecular gas. There is  ample literature on these other gas phases  in AGN, far too broad to be covered in this review. Here we describe a few  cases where the \HI\ observed in absorption has played a particularly relevant role.

In order to characterise the  properties of the gas in the circumnuclear regions, it of interest to compare  the column densities derived from  \HI\ absorption with those obtained from absorption of the soft X-ray emission. The X-ray emission from AGN is thought to mostly come  from the immediate vicinity of the central BH and resulting from thermal emission from the accretion disc which is  reprocessed by the hot plasma in the corona (inverse-Compton scattering). However, in the case of radio galaxies (and especially intrinsically compact CSS/GPS sources),  thermal emission from the ISM shocked by the expanding radio lobes, and  non-thermal emission from compact lobes produced through inverse-Compton scattering of the local radiation fields, can also play a role. 

This X-ray emission can be affected by photoelectric absorption by the foreground consisting of not-fully ionised material in the circumnuclear region. In principle, this absorbing material can be (partly) the same as that producing the \HI\ absorption.
However, there are several factors that complicate a comparison between column densities derived from \HI\ absorption with those derived from X-ray observations. While  \HI\ absorption provides the  column density of  neutral hydrogen,  X-ray absorption provides the equivalent {\sl total} hydrogen  column density $N_{\rm H}$ (i.e., atomic, molecular, and ionized hydrogen), of the column of material obscuring the central engine. Therefore differences in location may play a role, as well as the primary X-ray absorber being  dust, molecular or ionised gas, i.e.\ other phases of the ISM. In addition, the background source which is being absorbed may have a different spatial distribution resulting in differences in the absorption properties, even if the absorber is the same. Moreover,  spin temperatures for \HI\ close to an AGN are very uncertain. Also relevant is that the absorption of the X-ray photons is mainly due to  elements other than hydrogen so that assumptions on, for example, metallicity have to be made in order to be able to derive the column density of hydrogen \citep[see, e.g.,][]{Wilms00}.

Studies of a possible a relation between \HI\ column densities derived from \HI\ absorption and  hydrogen column densities derived from X-ray observations  have brought mixed results. 
In the case of Seyfert galaxies, \citet{Gallimore99} find no measurable correlation between \HI\ absorption and  absorption columns derived from X-rays (giving a reasonable match only for Mrk~231). For most (eight out of 11) of the sources, the total hydrogen column exceeds the \HI\  column by at least an order of magnitude (see Fig.\ 14 of \citealt{Gallimore99}). Even in Seyfert 2 objects, where  obscuration is known to play a major role, the \HI\ absorption column densities are not correlated with the hydrogen column derived from X-ray spectroscopy. This agrees with the result of  \HI\ absorption avoiding the very central regions in most  Seyferts  (see \S \ref{sec:seyferts}) . 
 
Similar to what is seen for Seyfert galaxies, for radio galaxies, in particular GPS and CSS objects,  $N_{\rm H}$  is systematically larger than  $N_{\rm HI}$  by 1--2 orders of magnitude (e.g., \citealt{Vink06,Tengstrand09}). On the other hand, trends have been found between $N_{\rm H}$  and $N_{\rm HI}$  suggesting that GPS/CSOs with increasingly large X-ray absorption have increasingly larger probability of being detected in \HI\ absorption observations \citep{Vink06,Tengstrand09,Ostorero10,Ostorero17,Glowacki17,Moss17}. 

The detailed analysis of \citet{Ostorero17} suggests that  a correlation  exists of the form $N_{\rm \HI} \sim N_{\rm H}^b$, with $b$ ranging between 0.35 and 0.47 depending on the sample used. However, the correlation displays a large scatter which may originate from a scatter of $T_{\rm s}/c_{\rm f}$,  the ratio between the spin temperature and the covering factor of the radio continuum  needed to estimate $N_{\rm \HI}$.  
The observed relation would imply \tspin/\cf\ to be in the range 12--832 K.
This is consistent with the limits  on \tspin\ derived by \citet{Glowacki17} and \citet{Moss17}, obtained under the assumption that all of the hydrogen gas producing the X-ray absorption is atomic. They find an upper limit of $T_{\rm spin} \leq 1130\pm 380$ K assuming a covering factor of 1.

Thus, the gas responsible for the X-ray obscuration and for the \HI\ absorption in compact radio galaxies may be part of the same  (settled or unsettled) structure, likely  larger than the typical torus, as discussed in \S \ref{sec:vlbi} and \ref{sec:outflows}.  This result may suggest that the dominant contribution to the X-ray emission of GPS/CSOs does not come from the accretion disc, but from the larger-scale jet and lobe components.
  
Optical spectra  can provide a further way to compare \HI\ with ionised gas and with other estimates of the neutral gas. Such comparisons have  mostly focussed on characterising gas outflows. 
Many outflows of ionised gas  are known and many of these also have  an \HI\ component. In particular, most of the \HI\ outflows found in a sample of SDSS galaxies seem to have a counter-part of ionised gas (\citealt{Santoro18}, Santoro et al.\ in prep). The converse statement, i.e.\ outflows detected in ionised gas also showing  an \HI\ outflow, is not true. However, this could be, at least partly, due to the limitations of the available data on \HI\ absorption for the optical samples used and deeper data are needed to further investigate this.

Optical spectra can also be used to trace the neutral gas. The most common diagnostic is    absorption by the Na D doublet of \NaI\ at 5890 and 5896 \AA\ which can be detected against the optical stellar continuum of the host galaxy. The \NaI\ line is a  tracer of cold neutral gas because of its relatively low ionization potential (5.1 eV), quite a bit lower than that of hydrogen (13.6 eV). Thus, depending on the conditions, \NaI\ and \HI\ absorption can trace the same regions of the ISM. 
However, a direct comparison needs to be done  with care. The two tracers are observed in absorption against very different backgrounds, in terms of nature and of extent:  the stellar light of the galaxy for \NaI\ and the radio continuum for the \HI. A major complication is that the \NaI\ absorption due to the ISM is superposed on \NaI\ absorption occurring in the photospheres of the stars of the stellar background. Moreover, given the relatively low ionisation potential of Na,  Na is mostly ionised (i.e.\ \NaII) under many common conditions in the ISM and the conversion from \NaI\ column density to \HI\ column density  is not straightforward (see, e.g., \citealt{Rupke02} for a discussion). These complications may explain some of the differences seen in some studies where the two column densities are compared in  detail. 

Despite all this, consistent results are obtained in a number of cases. In Mrk~231  it is clear that  \NaI\ traces the absorption associated with the fast outflow in this galaxy over larger distances from the nucleus compared to the outflow seen in \HI\ absorption against the central  radio continuum. Interestingly, the column densities derived from the \HI\ data (ranging from 5 to $15 \times 10^{21}$ cm$^{-2}$ for \tspin\ = 1000 K) are  consistent with the column densities derived from the \NaI\ ($7.5 \times 10^{21}$ cm$^{-2}$; \citealt{Morganti16}). This suggests that the \NaI\ and \HI\ absorption may indeed come from the same outflow. 

\NaI\ has been used in particular for tracing outflows  found in  ULIRGs \citep{Rupke02,Rupke05,Rupke13a,Cazzoli16,Rupke11}  and in radio galaxies \citep{Lehnert11}. The latter studied \NaI\ in radio galaxies selected from FIRST, using SDSS spectra. Interestingly, they find mass- and energy outflow rates of about 10 \msunyr\ and few $\times  10^{42}$ \ergs\ respectively,   consistent with those derived for comparable radio galaxies from \HI\ absorption line observations.

Thanks to  new instrumentation, the comparison of the properties of the \HI\ and of the molecular gas in AGN has become much more interesting and relevant. 
Unsurprisingly, this has been triggered by the  very large improvement  in observational capabilities provided by ALMA for  observations of cold molecular gas, and by  NIR  instruments for the study of the warm molecular components (e.g.\ SINFONI and X-Shooter at the VLT). An important aspect is that the boost in sensitivity provided by ALMA is allowing to perform observations of molecular gas   {\sl in emission} at the sub-arcsecond resolution needed to study the structures seen by \HI\ absorption.  

Particular attention has been given to molecular gas outflows, after the discovery that not only ionised and atomic neutral hydrogen, but also molecular gas can be found in fast and massive AGN-driven outflows \citep{Feruglio10,Cicone14}.   Moreover, the result that in several objects the outflows  contain a  possibly even dominant component of molecular gas  has given new insights into the physical conditions of the gas close to the AGN and into the physics of the outflows and how they are driven. 
The fact that molecular outflows appear to be massive and that they have  high mass outflow rates, possibly higher than what typically found for the  \HI\ outflows (and much higher than the mass outflow rate of the ionised gas) indicates that the molecular gas may represent the most important component for AGN-driven outflows. Clearly, the connection between molecular and atomic outflows has to be understood. 

In this context, an important point is that, although the statistics are still limited by small sample sizes,   striking similarities have been found between the location and kinematics of the \HI\ and of the molecular gas in  outflows  in several radio galaxies \citep[see e.g.][]{Alatalo11,Dasyra12,Morganti15a,Morganti16}. One of the first examples of this was the CO outflow detected in the radio source 4C12.50 (see Fig.\ \ref{fig:4C12.50outflow}). Earlier \HI\ observations \citep{Morganti05a} had shown that a fast \HI\ outflow is present in this object which is located at a  bright radio hot spot in one of the radio lobes \citep{Morganti13}. The CO data show  absorption over the same velocities as where the \HI\ outflow occurs \citep{Dasyra12}.

One of the objects illustrating the similarity between \HI\ and molecular gas kinematics is  IC~5063  (see Fig.\ \ref{fig:IC5063}). This is also one of the best examples of the effect of a radio jet entering a clumpy medium, affecting the physical conditions of this medium and producing a fast multiphase outflow  \citep{Oosterloo17,Mukherjee18b}. Figure \ref{fig:IC5063} shows the velocity profiles of the cold and the warm molecular gas, of the ionised gas and of the \HI. Although the detailed shapes of the profiles differ very much, all phases show the same extent for their blueshifted components, indicating they participate in the same outflow.

The similarity in properties of the atomic and the molecular outflows suggests that both phases are participating in the same processes and that they are driven by a common mechanism. Interestingly, fast outflows of \HI\ and cold molecular gas have been found to be driven by wide-angle winds as well as by radio jets.  The discovery of the similarities between the atomic and molecular outflows has triggered large interest in what could be the scenario for accelerating  cold gas to the high velocities observed. The model that is currently favoured is that the cold  gas forms in situ. After being accelerated and ionised by fast shocks, the outflowing gas is able to cool radiatively on short time scales into clumps of cold molecular material. The \HI\ only represents  gas in an intermediate phase, cooling from warm ionised to cold molecular \citep{Zubovas12,Costa15,Richings18}. An alternative scenario  assumes that pre-existing molecular clouds from the host ISM are entrained in the adiabatically expanding shocked wind and  that they are accelerated to the observed velocities without being destroyed \citep[see, e.g.,][]{Scannapieco17}.

An interesting point is that the shared properties of the atomic and molecular outflows, if confirmed for more objects, suggests that the future large \HI\ absorption surveys planned for SKA1 and its precursors, and which are expected to identify many new cases of \HI\  outflows, will also provide many  interesting candidates for follow up with ALMA, an instrument not suitable for large, blind surveys. This synergy will greatly help in building up the statistics on the properties of molecular outflows and to help to understand AGN-driven outflows of cold gas in general.

\begin{figure}
\centering
  \includegraphics[angle=0,width=1.0\textwidth]{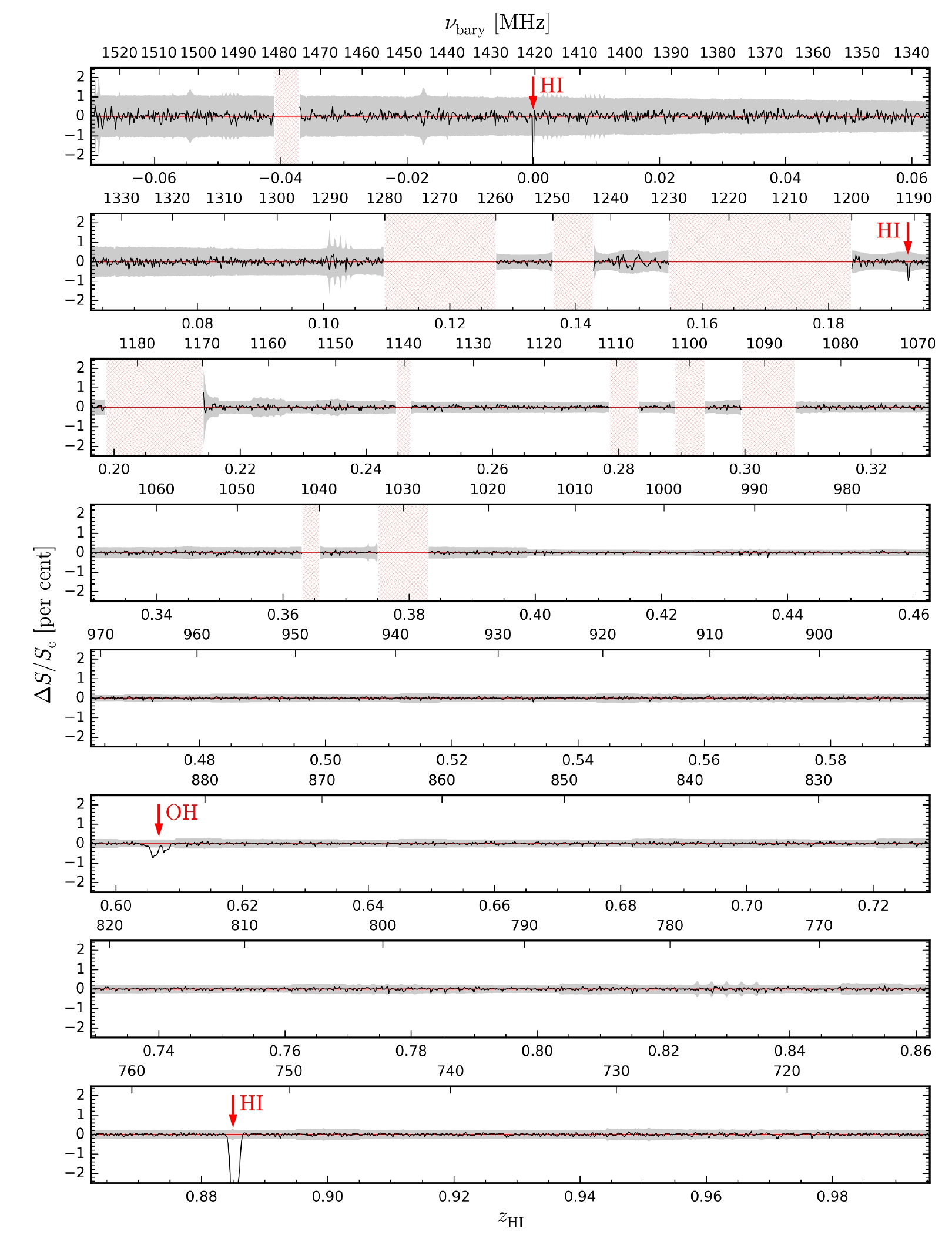}  
\caption{A spectrum obtained with the ASKAP test array BETA towards the gravitational lensed quasar PKS~B1830--211, taken from \citet{Allison17}. This figure illustrates the possibilities of the new generation of receivers. In the case shown here, the receiver can cover - with just three observations - a redshift range going from 0 to 1 and detect any \HI\ in this range located along the line of sight. The hatched regions mask data that were significantly contaminated by aviation and satellite-generated RFI. The red arrows indicate the positions of previously reported \HI\ and OH lines.
}
\label{fig:ASKAP}       
\end{figure}

\begin{figure}
\centering
  \includegraphics[angle=0,width=0.8\textwidth]{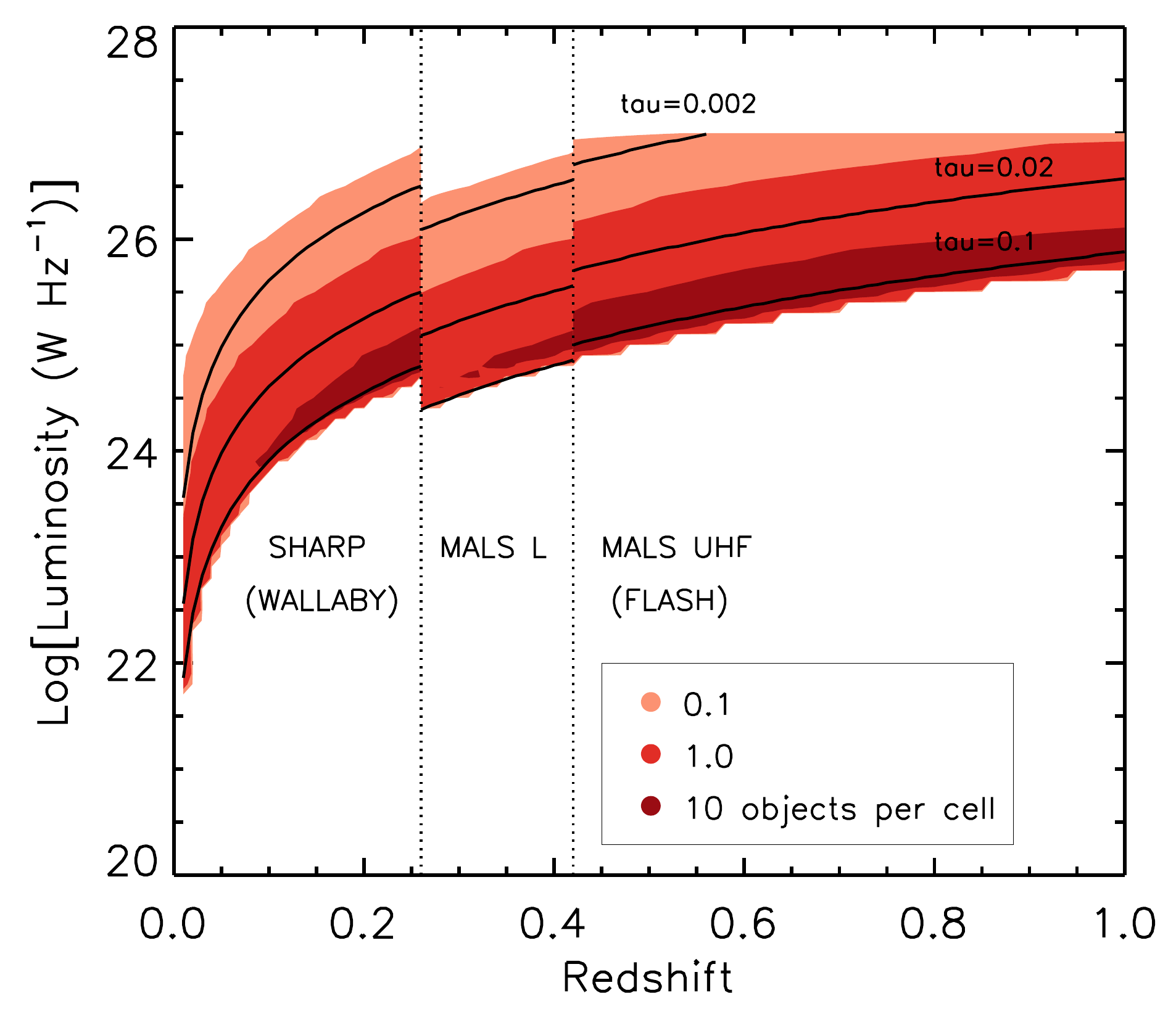}  
\caption{Coverage of the luminosity-redshift plane by upcoming absorption surveys. The parameter space is divided into cells of d$z = 0.01$ and d$L = 0.1$, and the colour coding indicates 0.1 (light), 1 (medium), and 10 (dark shading) objects per cell. As expected, the majority of objects lie close to the flux limits of the surveys. At $z \leq 0.26$, SHARP in the northern hemisphere and WALLABY in the south cover very similar parameter space. The full MALS L-band survey covers $0 < z < 0.58$, but is the only survey probing intermediate redshifts, $0.26 < z < 0.42$. MALS UHF targets intermediate redshifts, $0.42 < z < 1.44$. A similar redshift range is covered by FLASH, but with brighter flux limits. For clarity, we show the coverage of the plane out to $z = 1$, even though MALS UHF extends to $z = 1.44$ with the same behaviour. The black lines denote the minimum optical depth visible for sources of a given flux with the channel noise values as described in \citet{Maccagni17}.}
\label{fig:BlindSurveys}       
\end{figure}


\section{Future perspectives}
\label{sec:future}

Significant progress can be expected  in the coming years in a number of areas. Much of the work in the next few years will be focussing on exploiting the large survey speed of the SKA pathfinder and precursor telescopes, and later of SKA1 itself. For this purpose, a number of large surveys, which vary in depth and area covered, are being planned (see Table \ref{tab:surveys}; see also \citealt{Gupta17}). By utilising  the high survey speed, the broad frequency coverage, and the large instantaneous bandwidth of the new telescopes, these surveys will allow to create large, new samples of objects out to high redshift,  which will be used to explore new territory \citep[see][]{Morganti15a}. The expected number of detections will increase the  number of known associated \HI\ absorptions (currently about 150)  by almost two orders of magnitude with the majority of the detections in a redshift range that is poorly sampled currently (Table \ref{tab:surveys}). This will make it much easier to compare the absorption properties of various classes of objects. A very important difference  is that the planned surveys
will  carry out  {\sl blind searches} for \HI\ absorption in contrast to the targeted surveys done so far. This will  allow to probe the \HI\ and its properties, and the evolution thereof, in radio sources in a  much more systematic and unbiased way. 

A peek of things to come is shown in Fig.\ \ref{fig:ASKAP} which shows a  spectrum taken from commissioning observations done with the ASKAP test array BETA (\citealt{Allison17}; see also \citealt{Allison15}. The spectrum  covers the redshift range $0<z<1$ with  a velocity resolution of only a few \kms. For each  continuum source in the survey area, one will be able to obtain  such a spectrum covering such a large range in redshift.  This means that the presence of \HI\ can be recovered also for galaxies for which the redshift is not  known a priori. Thus, the planned surveys will provide a 'blind view' of a very large volume of the Universe.    Particularly exciting is the possibility   of having  a first proper census  at higher redshifts thanks to the higher sensitivity at lower frequencies and the low RFI levels at the  locations of the new radio telescopes. In the slightly more distant future, the higher sensitivity and the even larger frequency coverage of SKA1 will make sure that the above planned surveys can be expanded to fainter and to more distant  sources. The higher sensitivity of MeerKat and of SKA1 compared to available radio telescopes also means that fainter \HI\ absorption can be detected. This is  particularly relevant for studying fast \HI\ outflows, the study of which is now limited to relatively bright sources, given the low opacities involved. 
 
Figure \ref{fig:BlindSurveys} \citep[taken from][]{Maccagni17} illustrates another aspect of the upcoming surveys, namely their complementarity in terms of depth and redshift coverage (interestingly, this    came about largely by coincidence\ldots). The figure indicates the expected number of detections of the various surveys, as function of source luminosity and redshift. The parts of parameter space over which the different surveys will detect sources  match very nicely and in particular sources with a luminosity in the range $10^{26}$--$10^{27}$ W Hz$^{-1}$ can be studied over a large redshift range. This  will allow to construct well-selected samples of objects up to $z \sim 1.5$ and perform   studies of the evolution of gas in AGN probing optical depths of a few \% and \HI\ column densities of $10^{20}$ -- $10^{21}$ cm$^{-2}$.  

Observations to higher redshifts (up to $\sim 3$)  can be done, at present, only with the GMRT and with single dish telescopes, like the GBT. However, low frequency telescopes like LOFAR \citep{Haarlem13}, MWA \citep{Tingay13} and ultimately SKA-Low are now opening the challenging but exciting possibility to explore \HI\ in even higher redshift objects (i.e.\  $z \sim 5 - 10$). The fact that some of the very high-$z$ radio sources have been detected in molecular gas (see e.g.\ \citealt{Klamer05,Emonts18}), makes such absorption studies extremely relevant.

For these blind surveys, automated detection techniques  have already been developed \citep{Allison12,Allison14,Koribalski12}, along with techniques for stacking absorption spectra \citep{Gereb14b} and profile characterisation \citep{Gereb15} of large samples. The pathfinder and precursor surveys will allow us to expand the use of stacking techniques and derive general properties of groups of objects such as young vs old vs restarted radio sources, compact vs extended objects, and exploring trends with redshift, optical properties etc.. 

Since most of the planned surveys will be done on the southern sky, they  will provide  interesting targets for follow up with the excellent optical and mm facilities which are, or soon will be,  available in the southern hemisphere. In particular, the blind searches for \HI\ outflows will provide targets for observation with ALMA to investigate the molecular counterpart of the outflows. Future large-area optical spectroscopic surveys planned in both hemispheres will also be an important complement to the \HI\ data.

\begin{table}
\centering
\begin{tabular}{l c c c c c c c}  
\hline\hline                                                         
 Survey &  Redshift & Noise over  & Sky coverage   & Number of  & Detection \\
    & range  &   5 \kms      &            &detections & limit\\
    &        &[mJy beam$^{-1}$]  & [deg$^2$]  &        & [mJy ] \\
\hline
Apertif -- SHARP  & 0--0.26  & 1.3 & 4000  & 400 & 30 \\
ASKAP -- FLASH  & 0.4--1.0  & 3.8 & 25000  & 5500 & 90 \\
ASKAP -- Wallaby & 0--0.26  & 1.6 & 30000  & 2300 & 40 \\
MeerKAT -- MALS  & 0--0.57  & 0.5 & 1300  & 450 & 15  \\
(L-band)    \\
MeerKAT -- MALS  & 0.40--1.44 & 0.5--0.7& 2000  & 1500  & 15\\
         (UHF-band)   \\
\hline                           
\end{tabular} 
\caption{Summary of  upcoming large \HI\ 21-cm absorption line surveys (see also \cite{Morganti15a,Gupta17}. SHARP and WALLABY are  commensal \HI\ emission and absorption surveys, primarily investigating associated absorption.  FLASH is a blind survey of the southern hemisphere to detect \HI\ absorption in intervening and associated systems at intermediate redshifts. The MALS project is targeted survey focussing on relatively bright, high redshift background sources to search the line of sight for intervening absorption \citep{Gupta17}. However, with the more than $1\times1$~deg$^2$ field of view of MeerKAT, a substantial volume for each pointing is blindly, and commensally, probed both for associated and intervening absorbers. The expected number of detections has been estimated using the redshift distribution of radio sources from the S-cubed simulations \citep{Wilman08} and assuming a detection rate for associated absorption of 25\%.}
\label{tab:surveys}
\end{table}

\section{Conclusions}
\label{sec:conclusions}

In this review we have summarised the main results based on  \HI\ 21-cm absorption observations obtained in the last two decades about the cold ISM in the central regions of radio loud galaxies and about the interplay between this gas and the AGN. Given the specific nature of \HI\ absorption observations, they make it possible to study the cold gas in the environment of AGN at parsec scales also at higher redshift and therefore  can probe a completely different parameter space than observations of \HI\ emission.  Significant progress has been made in a number of important areas, in part due to technical improvements of radio telescopes which have allowed to obtain data over larger bandwidths and with higher spectral dynamic range. The main results can be summarised as follows:

\begin{itemize}

\item{Observations of various samples of various types of radio sources have shown that AGN  are surrounded by a regularly rotating gas disk which contains a detectable fraction of \HI, as expected from theoretical models. In low-power objects, these disks are thin and are typically a few hundred parsec in size. In more powerful objects, they extend to smaller radii and are thicker. In Seyfert galaxies, these disks connect to the larger scale galactic disks. Due to obscuration effects, these disks play a role in how the central AGN is detected in other wavebands.  In radio galaxies, the presence of circumnuclear absorbing structures containing \HI\ is by now well established and the kinematics of these structures has been traced by pc-resolution VLBI observations. Some groups of objects, and in particular young and recently restarted radio galaxies, appear to have a particularly high detection rate of \HI. This is interesting in connection with  the evolution of these AGN and their impact on the surrounding ISM.}

\item{An important discovery has been the occurrence of fast ($>1000$ \kms) \HI\ outflows in about 5\% of radio sources. They occur in particular in young radio sources. High-spatial resolution observations have shown that  most of these outflows are driven by the radio jet coming from the central AGN, but they can also be driven by  wide-angle nuclear winds. These outflows of atomic hydrogen  have a counterpart of molecular outflows. The preferred model is that the cold gas is the result of fast cooling of the gas after it has been shocked and accelerated by a passing jet, or by a cocoon of overpressured gas inflated by the jet. The observed mass outflow rates of the cold gas in these outflows are much larger than those of the ionised outflows.  This has provided a completely new way to look at the gas under the influence of the energy released by an active BH which is important input on the role of AGN feedback in models of galaxy evolution.}

\item{It has been more difficult than anticipated to detect evidence for \HI\ playing a role in  fuelling the central SMBH. The original consensus that redshifted absorption components, which would indicate accretion onto the SMBH, occur more often than blueshifted components has made place for  the reverse namely that blueshifted absorption is more common and also extends to more extreme velocities. 
It may not be easy to identify redshifted absorption as infalling given that infall velocities should be similar to the rotational velocities of a regular circumnuclear  disk and the absorption of infalling clouds may blend with the absorption of the circumnuclear disk. Nevertheless, a small number of high-resolution observations have shown evidence for fuelling. These observations also suggest that infalling \HI\ can only be identified as such in high-resolution data} 
\end{itemize}

The results summarised above show the  exciting progress of the last two decades  in the field of associated \HI\ absorption and the many significant contributions which have been made for better understanding the cold ISM near AGN and the role it plays in the evolution of AGN and of their host galaxies. It is very likely that a review on \HI\ absorption that will be written in two decades from now will contain even more exciting and important results. The new blind large-area surveys which will be done with several new radio facilities in the near future will have a major impact on \HI\ absorption studies. The number of detections of associated \HI\ absorption will increase by a very large factor ($\gg 10$) and the data will extend to fainter source populations and fainter absorption features. Very important is the fact that the new surveys are blind surveys. This will to a large extent remove the effects of many selection biases which now plague  most studies and which  makes drawing firm conclusions sometimes difficult. Perhaps the most exciting prospect is offered by the improved sensitivity at higher redshift. This will provide a major step forward for studying the evolution with redshift of the gas in AGN, something which has been very difficult to up to now.

\begin{acknowledgements} We would like to thank James Allison and Elaine Sadler for their insightful input.
RM gratefully acknowledges support from the European Research Council under the European Union's Seventh Framework Programme (FP/2007-2013) /ERC Advanced Grant RADIOLIFE-320745.

\end{acknowledgements}

\end{document}